\documentclass[aps,prd,preprint,11pt,a4paper,longbibliography,nofootinbib]{revtex4-1}

\usepackage{amssymb,amsmath,amsfonts}
\usepackage{graphicx}
\usepackage{slashed}
\usepackage{hyperref}
\usepackage{color}
\def\eq#1{{eq.~(\ref{#1})}}

\def\fig#1{{fig.~(\ref{#1})}}

\def\sec#1{{sec.~(\ref{#1})}}
\def\tab#1{{table~(\ref{#1})}}

\newcommand{\bea}{\begin{eqnarray}}
\newcommand{\eea}{\end{eqnarray}}
\newcommand{\GeV}{\mbox{ ${\mathrm{GeV}}$}}
\newcommand{\TeV}{\mbox{ ${\mathrm{TeV}}$}}
\newcommand{\ifb}{\mbox{ ${\mathrm{fb^{-1}}}$}}
\newcommand{\ra}{\rightarrow}

\newcommand{\gt}{\tilde{g}}

\begin{document}
\title{Custodial Vector Model}

\author{Diego Becciolini$^{1}$}

\author{Diogo Buarque Franzosi$^{1}$}

\author{Roshan Foadi$^{2,3}$}

\author{Mads T. Frandsen$^{1}$}

\author{Tuomas Hapola$^{4}$}

\author{Francesco Sannino$^{1}$}

\affiliation{$^{1}$CP$^{3}$-Origins and the Danish Institute for Advanced Study, University of Southern Denmark, Campusvej 55, DK-5230 Odense M, Denmark}
\affiliation{$^{2}$Department of Physics, University of Jyv\"askyl\"a, P.O. Box 35, 
FI-40014, University of Jyv\"askyl\"a, Finland}
\affiliation{$^{3}$Department of Physics \& Helsinki Institute of Physics, P.O. Box 64,
FI-000140, University of Helsinki, Finland}
\affiliation{$^{4}$Institute for Particle Physics Phenomenology, Durham University,
South Road, Durham DH1 3LE, UK}

\preprint{CP3-Origins-2014-033 DNRF90, DIAS-2014-33, IPPP/14/89,  DCPT/14/178}

\begin{abstract}
We analyze the Large Hadron Collider (LHC) phenomenology of heavy vector resonances with a $SU(2)_L\times SU(2)_R$ spectral global symmetry. This symmetry partially protects the electroweak S-parameter from large contributions of the vector resonances. The resulting custodial vector model spectrum and interactions with the standard model fields lead to distinct signatures at the LHC in the diboson, dilepton and associated Higgs channels. 
\end{abstract}


\maketitle


\section{Introduction}
The discovery of the Higgs-like particle at the LHC further supports the remarkable success of the Glashow, Salam and Weinberg  (GSW)  theory of electroweak interactions.

The GSW theory augmented with Quantum Chromodynamics (QCD) is known as the standard model of particle interactions (SM). 
Any extension of the SM must closely reproduce the GSW theory, including the Higgs sector.  
It is therefore natural to explore extensions where the low energy effective GSW theory is partially protected against contributions from new sectors via the presence of additional symmetries. 

One may either consider perturbative or non-perturbative extensions of the GSW theory. Here we consider the possibility that the new extension features massive spin-1 resonances in the TeV region. This is, for example, expected in any model of composite dynamics near the electroweak scale while many perturbative extensions also feature, via new Higgs mechanisms, massive spin-1 states, e.g. so-called $Z'$ states. 

Our model respects the custodial symmetry of the GSW  theory, i.e. $G = SU(2)_L\times SU(2)_R$  that protects the mass relation between the electroweak $W$ and $Z$ bosons. It features an additional unbroken global $G' = SU(2)_L'\times SU(2)_R'$ symmetry acting on the heavy spin-1 resonances.  
The effective Lagrangian thus features two global symmetries $G$ and $G'$. The former breaks to $H = SU(2)_V$ and the latter remains intact.  The breaking $G\to H$ is identified with the GSW custodial symmetry breaking pattern  $SU(2)_L\times SU(2)_R\to SU(2)_V$ while $G'$ acts only on the new heavy vector resonances and serves to protect the $S$-parameter as well as longitudinal $WW$ scattering from large contributions from the heavy resonances. 

We model the Higgs sector as in the GSW theory. By construction, our model then has the GSW theory as a well defined decoupling limit when sending the mass of the new resonances to infinity. We shall call our model the custodial vector model (CVM).

A discussion of possible strong dynamics underlying the CVM are given in \cite{Appelquist:1998xf,Appelquist:1999dq,Foadi:2007se}. The spectral symmetry of the vector resonances in the CVM was discussed in \cite{Casalbuoni:1995yb} and built into the so-called Degenerate Breaking Electroweak Symmetry Strongly (D-BESS) model \cite{Casalbuoni:1995qt} without featuring a Higgs particle. 
The CVM can also be interpreted as an extension of
the GWS theory with multiple scalars, in which the massive spin-one bosons arise from new gauge sectors, e.g. \cite{Casalbuoni:1996wa,Casalbuoni:1997rs,DeCurtis:2003in,Hernandez:2010iu}.

In this paper we introduce the CVM and investigate its LHC phenomenology. The model features a very distinct pattern of narrow spin-1 resonances in the diboson, dilepton and associated Higgs search channels allowing, in principle, to pin it down. Specifically the CVM predicts closely spaced spin-1 resonance double peaks in the dilepton invariant mass distributions, single resonance peaks in the single charged lepton channels, and suppressed peaks or no signal in the diboson channels. This is in contrast to general effective descriptions of composite dynamics leading to broad, well-spaced resonances with large branching ratios to diboson channels, \emph{e.g.} \cite{Belyaev:2008yj,Contino:2011np,Greco:2014aza,Brooijmans:2014eja}, or to specific spin-1 spectra appearing in Composite Higgs models and extra-dimensional theories, \emph{e.g.} \cite{Agashe:2007ki,Agashe:2008jb,Agashe:2009bb}.

Higgs production in association with vector bosons is also an important search channel, which depending on the parameter space can be substantially enhanced with respect to the GSW theory.

The outline of the paper is as follows. In section~\ref{sec:MI} we discuss current LHC constraints on spin-1 resonances. The CVM Lagrangian
 is discussed in section~\ref{sec:Model}. Here we also outline the qualitative phenomenology. We compare the model predictions with the electroweak precision measurements in section~\ref{sec:EWPTLimits}. The detailed phenomenological analysis is provided in section~\ref{sec:LHCpheno}. Finally in section~\ref{sec:conclusions} we summarise our findings and discuss further developments.

\section{Current Constraints on Generic Vector Resonances}
\label{sec:MI}
Several studies have been dedicated to the LHC phenomenology of heavy spin-1 particles, see for instance \cite{Pappadopulo:2014qza} for a recent discussion. Here, we focus on the latest experimental results to summarise the relevant LHC searches for vector resonances that will be used to constrain the CVM parameter space in section~\ref{sec:Model}. 

We consider a set of narrow charged and neutral spin-one resonances,  $\mathcal{R}_i^\pm$ and $\mathcal{R}_i^0$  respectively, with $i$ counting the number of independent mass eigenstates. With $H$ we denote the 125 GeV Higgs-like particle. The relevant effective interaction vertices are summarised via contact operators  in the  Lagrangian:
\bea
\mathcal{L}^\mathcal{R}=
\mathcal{L}^\mathcal{R}_{\rm kinetic}
+\mathcal{L}^\mathcal{R}_{\rm self}
+\mathcal{L}^\mathcal{R}_{\rm fermion}
+\mathcal{L}^\mathcal{R}_{\rm gauge}
+ \mathcal{L}^\mathcal{R}_H  \ .
\eea
The vertices linking the spin-one resonances with the SM fermions are 
\bea
\mathcal{L}^\mathcal{R}_{\rm fermion}&=&
\sum_i \sum_{u,d} \bar{u}\slashed{\mathcal{R}}_i^+\left(g^L_{\mathcal{R}_i u d}P_L + g^R_{\mathcal{R}_i u d}P_R\right) d + {\rm h.c.}
+\sum_i \sum_{f}  \bar{f} \slashed{\mathcal{R}}_i^0 \left(g^L_{\mathcal{R}_i f}P_L+g^R_{\mathcal{R}_i f }P_R\right) f \ \nonumber \\
&=& \sum_i \sum_{u,d} \bar{u}\slashed{\mathcal{R}}_i^+\left(g^V_{\mathcal{R}_i u d}- g^A_{\mathcal{R}_i u d}\, \gamma_5\right) d + {\rm h.c.}
+\sum_i \sum_{f}  \bar{f} \slashed{\mathcal{R}}_i^0 \left(g^V_{\mathcal{R}_i f }-g^A_{\mathcal{R}_i f}\, \gamma_5\right) f \ ,
\eea
where $u$ ($d$) runs over all up-type (down-type) quarks and leptons, $f$ runs over all quark and lepton flavors, and we have expressed the vertices both in left-right and vector-axial basis, with $P_{L/R}=(1\pm \gamma_5)/2$.

The $CP$-invariant trilinear interactions of the spin-one resonances with $H$ are
\bea
\mathcal{L}^{\mathcal{R}}_{H} &\supset& 
\sum_{i}  g_{\mathcal{R}_i Z H} \, \mathcal{R}^0_{i\mu} Z^\mu   H +  g_{\mathcal{R}_i W H} \left(\mathcal{R}^+_{i\mu} W^{-\mu} + \mathcal{R}^-_{i\mu} W^{+\mu} \right) H \nonumber \\
&+&\frac{1}{2} \sum_{i,j} g_{\mathcal{R}^0_i \mathcal{R}^0_j H} \,  \mathcal{R}^0_{i\mu} \mathcal{R}_j^{0\mu} H + g_{\mathcal{R}^+_i \mathcal{R}^-_j H} \, \mathcal{R}^+_{i\mu} \mathcal{R}_j^{-\mu}  H  \ .
\eea
Note that a vertex with one spin-one resonance and two scalars $H$ is not $CP$-invariant, and is therefore not included in $\mathcal{L}^{R}_{H}$ ~\cite{Hagiwara:1986vm}.

For single resonance production and subsequent decay, in $\mathcal{L}^{R}_{\rm gauge}$ we only need to consider the vertices with one resonance and two SM gauge bosons, as tri-boson final states are suppressed compared to the di-boson ones, due to smaller available phase-space. 
The $C$ and $P$ invariant interactions are
\bea
&& \mathcal{L}^\mathcal{R}_{\rm gauge} \supset \sum_i \Bigg[
g_{\mathcal{R}_i W W}^{(1)} [[W^+ W^- \mathcal{R}_i^0]] + g_{\mathcal{R}_i W W}^{(2)} [[\mathcal{R}_i^0 W^+ W^-]]   \nonumber \\
&& + g_{\mathcal{R}_i W Z}^{(1)}  (\mathcal{R}_{i\mu\nu}^+ W^{-\nu}-\mathcal{R}_{i\mu\nu}^- W^{+\nu}+W_{\mu\nu}^+ \mathcal{R}_i^{-\nu}-W_{\mu\nu}^- \mathcal{R}_i^{+\nu}) Z^\mu  \nonumber \\
&& + g_{\mathcal{R}_i W Z}^{(2)} (\mathcal{R}_{i\mu}^+ W^{-\nu}-\mathcal{R}_{i\mu}^- W^{+\nu} ) Z^{\mu\nu}
\Bigg] 
\eea
where
\bea
 [[V_1 V_2 V_3]]&\equiv i\partial_\mu V_{1\nu} V_2^{[\mu} V_3^{\nu]} + {\rm h.c.} \nonumber \\
 \mathcal{R}_{\mu\nu}&\equiv \partial_\mu  \mathcal{R}_\nu -  \partial_\nu  \mathcal{R}_\mu
\eea
The 2-body decay modes of $\mathcal{R}_i^\pm$ and $\mathcal{R}_i^0$ may be then summarized as

\bea
&& \Gamma_{\mathcal{R}^\pm_i}= \sum_{u,d} \Gamma_{\mathcal{R}_i^\pm}^{u d}+\sum_{\nu,e} \Gamma_{\mathcal{R}_i^\pm}^{\nu e}+\Gamma_{\mathcal{R}_i^\pm}^{WZ}  + \Gamma_{\mathcal{R}_i^\pm}^{WH}  \ ,  \nonumber \\
&& \Gamma_{\mathcal{R}^0_i}= \sum_{q} \Gamma_{\mathcal{R}_i^0}^{q \bar{q}}+\sum_{\ell } \Gamma_{\mathcal{R}_i^0}^{\ell \bar{\ell}}+\sum_{\nu}  \Gamma_{\mathcal{R}_i^0}^{\nu\bar{\nu}}
+\Gamma^{WW}_{\mathcal{R}_i^0} + \Gamma_{\mathcal{R}_i^0}^{ZH} \; ,
\eea
where the formulae for the partial widths are provided in Appendix~\ref{sec:widths}. We disregard the subdominant 3- and 4-body decay modes.

\begin{figure}[htb!] 
\begin{center}
 \includegraphics[width=.55\columnwidth]{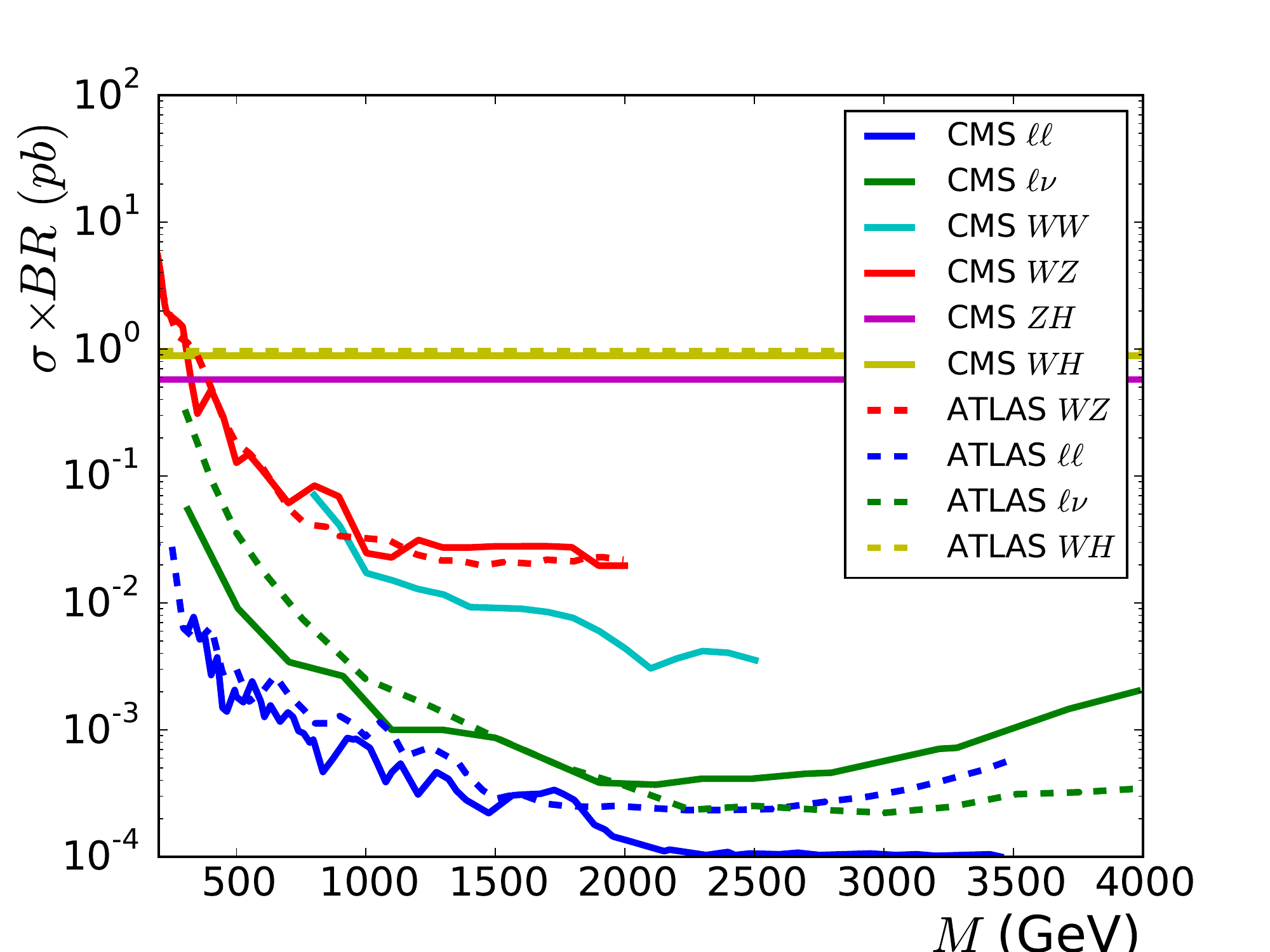}
\caption{ATLAS and CMS 95\% CL exclusion limits on production cross-section times branching ratio, $\sigma \times BR$, for a new neutral or charged vector resonance.  The charged vector final states are $\ell \nu$, $WZ$ and $WH$. Data references are given in \tab{table:R0searches}.}
\label{fig:xsbr}
\end{center}
\end{figure}

The relevant current LHC limits for a single charged or neutral vector resonance are given in Fig.~\ref{fig:xsbr} and the corresponding data listed in \tab{table:R0searches}. The dilepton limits are at least an order of magnitude stronger than any of the diboson limits at any resonance mass.

The ATLAS dilepton limit on the figure is the one relevant for a sequential standard model (SSM) $Z'$ in \cite{Aad:2014cka}.
The CMS limits on the $\ell^+\ell^-$ production \cite{CMS-PAS-EXO-12-061} are expressed in terms of 
$R_\sigma\equiv \frac{\sigma(pp\ra Z'+X \ra \ell\ell+X)}{\sigma(pp\ra Z+X \ra \ell\ell+X)}.$
We convert the bounds on $R_\sigma$ to bounds on the total inclusive cross section.  
We use the total standard model cross section for the Drell Yan Z boson production  given in \cite{CMS:2014hga}.
Similarly, for the associated Higgs production, the limits in \cite{PhysRevD.89.012003} are given in terms of the signal strength, $\mu\equiv \sigma/\sigma_{SM}$.  We convert this to a limit on the cross section, $\sigma_{BSM}=\sigma-\sigma_{SM}$. For $\sigma_{SM}$ we use the prediction at NNLO QCD and NLO electroweak accuracy\cite{Dittmaier:2011ti}.
The $WZ$ channel CMS search gives exclusion for the fully decayed 3 leptons and missing energy final state, therefore we obtain the limit on $WZ$ cross section by correcting for the $W$ and $Z$ branching ratios.

\begin{table}[!htp]
\setlength{\tabcolsep}{5pt}
\renewcommand{\arraystretch}{1.7}
\center
\begin{tabular}{||c|c|c|c||} 
\hline \hline 
Channel [Exp] &  L  $[{\rm fb}^{-1}]$ & Mass range [GeV]  & Reference
\\
\hline
$pp \rightarrow  \ell^+ \ell^-  \,\, [{\rm CMS}] $ & 20.6 (8 TeV) & $300-2500$ &\cite{CMS-PAS-EXO-12-061}
\\
$pp \rightarrow  \ell^+ \ell^-  \,\, [{\rm ATLAS}] $ & 20.3 (8 TeV) & $200-3000$  &\cite{Aad:2014cka}
\\
\hline
$pp \rightarrow  W \, Z \,^{1}\, [{\rm ATLAS}] $ & 20.3 (8 TeV) & $200-2000$  &\cite{Aad:2014pha}
\\
$pp \rightarrow  W \, Z \,^{2}\, [{\rm CMS}] $ & 19.6 (8 TeV) & $200-2000$ &\cite{CMS-PAS-EXO-12-025}
\\
\hline
$pp \rightarrow  W \, W \,^{3}\, [{\rm ATLAS}] $ & $4.7$ (7 TeV) & $200-1500$  &\cite{Aad:2012nev} 
\\
$pp \rightarrow  W \, W \,\, [{\rm CMS}] $ & $19.7$ (8 TeV) & $800-2500$  &\cite{CMS-PAS-EXO-12-021} 
\\
\hline
$pp \rightarrow  Z/W \, H \,\, [{\rm CMS}] $ & 18.9 (8 TeV) & $*$  &\cite{PhysRevD.89.012003} 
\\
$pp \rightarrow  W \, H \,^{4}\, [{\rm ATLAS}] $ & 20.3 (8 TeV) & $*$  &\cite{Aad:2014xzb} 
\\
\hline
$pp \rightarrow  \ell \, \nu \,\, [{\rm CMS}] $ & 20& $300-4000$  & \cite{CMS-PAS-EXO-12-060} 
\\
$pp \rightarrow   \ell \, \nu \,\, [{\rm ATLAS}] $ & 20.3& $300-4000$  &\cite{ATLAS-CONF-2014-017} 
\\
\hline \hline
\end{tabular}
\caption{LHC searches used to constrain the CVM. $^{1}$ Fully leptonic analysis, see \cite{Aad:2014xka} for similar limits from a semi leptonic analysis. $^2$ Semi leptonic analysis, see \cite{Khachatryan:2014gha} for a boosted semi-leptonic analysis. $^3$ Not shown in \fig{fig:xsbr} due to the low luminosity. $^4$ The $ZH$ analysis of ATLAS is not relevant as explained in \sec{sec:Hassoc}.}
\label{table:R0searches} \vspace{-0.35cm}
\end{table}

\section{The Custodial Vector Model}
\label{sec:Model}

The CVM, like the GSW theory, possesses a global $SU(2)_L\times SU(2)_R$ chiral symmetry which breaks spontaneously to the diagonal $SU(2)_V$ symmetry. It is well known that this custodial symmetry protects the $T$-parameter. The electroweak gauge symmetry group $SU(2)_L\times U(1)_Y$ is embedded in  $SU(2)_L\times SU(2)_R$ and therefore provides a small breaking of the custodial symmetry.  Upon spontaneous symmetry breaking the final intact gauge symmetry is $U(1)_{\rm QED}$.   
  
The CVM features the Higgs state $H$ and two weak isospin triplet vector resonances. The GSW custodial symmetry is encoded in the Higgs Lagrangian and 
the model includes yet another custodial symmetry acting on the vector sector. The new custodial symmetry is simply $SU(2)_L^\prime \times SU(2)_R^\prime$  and protects the $S$-parameter and $WW$ scattering from large corrections coming from the vector sector ~\cite{Casalbuoni:1995yb,Casalbuoni:1995qt,Appelquist:1999dq,Duan:2000dy,Casalbuoni:2000gn,Foadi:2007se,Foadi:2014wza}, as we shall show below.  
 
To elucidate the patterns of chiral symmetry breaking we use a linear representation of the original chiral symmetry group, both for the Higgs and vector sector. The Higgs $H$ and the electroweak Goldstone bosons $\Pi^a$ constitute a weak doublet that can be represented via  
\bea
\Sigma=\frac{1}{\sqrt{2}}\left[v+H+2\ i\ \Pi^a\ T^a\right]\ ,
\eea
where $T^a = \tau^a/2$ with $\tau^a$ the Pauli matrices. Here $v$ is the vacuum expectation value (VEV) and $\Sigma$ transforms as a bi-fundamental of the chiral symmetry group: 
\bea
\Sigma\to u_L \Sigma u_R^\dagger \ , \qquad u_{L/R}\in SU(2)_{L/R} \ .
\eea
The electroweak gauge boson interactions with $\Sigma$ are introduced via the covariant derivative \bea
D_\mu \Sigma &=& \partial_\mu \Sigma -i\ g\ {\widetilde W}_\mu^a\ T^a \Sigma + i\ g^\prime \ \Sigma\ \widetilde{B}_\mu\ T^3\ ,
\eea
where the tildes over the gauge fields indicate that these are not yet mass eigenstates. The new heavy vectors , $A_L\equiv A_L^a T^a$ and $A_R\equiv A_R^a T^a$, are formally introduced, following \cite{Appelquist:1999dq}, as gauge fields transforming under the original chiral symmetry group, {\em i.e.}:
\bea
A_{L/R}^\mu\to u_{L/R}\left(A_{L/R}^\mu+\frac{i}{\tilde{g}}\partial^\mu\right)u_{L/R}^\dagger \ , \quad u_{L/R}\in SU(2)_{L/R} \ ,
\eea
where $\tilde{g}$ is the self-coupling. Note that we have used a single coupling for both $A_L$ and $A_R$: in fact we assume that the new CVM sector is invariant under parity, {\em i.e.}
\bea
\label{eq:parity}
\quad \Sigma(t,\vec{x}) \leftrightarrow \Sigma^\dagger (t,-\vec{x}) \ , \qquad A_L(t,\vec{x})  \leftrightarrow A_R(t,-\vec{x})  \ .
\eea
The linear combinations \cite{Appelquist:1999dq}
\bea
C_{L\mu}\equiv A_{L\mu}-\frac{g}{\tilde{g}}\widetilde{W}_\mu\ , \qquad
C_{R\mu}\equiv A_{R\mu}-\frac{g^\prime}{\tilde{g}}\widetilde{B}_\mu\ ,
\eea
transform homogeneously under the electroweak subgroup and can be immediately used to build Lagrangian invariants. As shown in \cite{Appelquist:1999dq} the following Lagrangian 
\bea
{\cal L}_{\rm boson}&=&-\frac{1}{2}{\rm Tr}\left[\widetilde{W}_{\mu\nu}\widetilde{W}^{\mu\nu}\right]
-\frac{1}{4}\widetilde{B}_{\mu\nu}\widetilde{B}^{\mu\nu}
-\frac{1}{2}{\rm Tr}\left[F_{L\mu\nu} F_L^{\mu\nu}+F_{R\mu\nu} F_R^{\mu\nu}\right] \nonumber \\
&+& \frac{1}{2}{\rm Tr}\left[D_\mu \Sigma D^\mu \Sigma^\dagger\right]
+\frac{\tilde{g}^2 f^2}{4}\ {\rm Tr}\left[C_{L\mu}^2+C_{R\mu}^2\right]
+ \frac{\tilde{g}^2 s}{4} {\rm Tr}\left[C_{L\mu}^2+C_{R\mu}^2\right] {\rm Tr}\left[\Sigma \Sigma^\dagger\right]
\nonumber  \\
&+&\frac{\mu^2}{2} {\rm Tr}\left[\Sigma \Sigma^\dagger\right]-\frac{\lambda}{4}{\rm Tr}\left[\Sigma \Sigma^\dagger\right]^2\ ,
\label{eq:boson}
\eea
preserves $SU(2)_L \times SU(2)_R \times SU(2)'_L \times SU(2)'_R $ when the electroweak gauge interactions are switched off. 
It is straightforward to see that in this limit the vectors can be transformed independently as:
\bea
 A_{L/R}\to u_{L/R}^\prime A_{L/R} u_{L/R}^{\prime\dagger} \qquad u_{L/R}^\prime\in SU(2)_{L/R}^\prime
\eea
Adding an $SU(2)_L\times SU(2)_R$ invariant term like
${\rm Tr}\left[C_{L\mu} (\Sigma D^\mu \Sigma^\dagger -  D^\mu \Sigma  \Sigma^\dagger )\right] + (L\leftrightarrow R)$
would break the $SU(2)_L'\times SU(2)_R'$ symmetry and contribute to the electroweak $S$-parameter as computed in e.g. \cite{Belyaev:2008yj}.

In the Lagrangian $\widetilde{W}_{\mu\nu}$ and $\widetilde{B}_{\mu\nu}$ are the ordinary electroweak field strength tensors, whereas $F_{L\mu\nu}$ and $F_{R\mu\nu}$ are the field-strength tensors built out of the spin-one fields $A_L$ and $A_R$, respectively. The coupling $s$ is real and $f$ is a new mass scale for the heavy vectors. 

Because $\mu^2$ is positive $\Sigma$ acquires a VEV, given at tree-level by  
\bea
v=\mu/\sqrt{\lambda}\ . 
\eea
 Upon diagonalising the mass matrices we end up with the ordinary GSW gauge bosons, and two nearly mass-degenerate triplets of heavy vectors. The physical heavy vectors are denoted by $L^{\pm, 0}$ and $R^{\pm,0}$.  They are dominantly $A_L$ and $A_R$ respectively. In the appendices~\ref{appendix:diag}, \ref{appendix:couplings} and \ref{sec:widths} we diagonalise the mass matrices, evaluate the couplings and widths of the spin-one resonances.

It is useful to sketch the basic qualitative features of the CVM phenomenology before the quantitative study presented in  \sec{sec:LHCpheno}. The new $SU(2)'_L\times SU(2)'_R$ custodial symmetry over the vectors has an immediate impact on the partial decay widths of the vectors into either fermions or bosons, which scale as   \begin{align}
\label{scaling}
\Gamma_{\mathcal{R}_i}^{\bar{f}f'}\sim \Gamma_{\mathcal{R}_i}^{V V'} \sim \frac{1}{a}\Gamma_{\mathcal{R}_i}^{H V} \sim \frac{M_R}{\tilde{g}^2} ,
\end{align}
where 
\bea
\label{Eq:Massparam}
M_R^2\equiv \frac{\tilde{g}^2}{4}\left(f^2+s\, v^2\right) \ ,
\eea
 is the mass of the vectors in the absence of the subdominant electroweak corrections.  $V,V'$ denote the $W,Z$ bosons.  We also trade the parameter $s$ for the parameter 
 \begin{equation} 
a = \frac{f^2}{f^2 + s\,v^2} = \frac{\gt^2 f^2}{4\,M_R^2}  \ , 
\label{def:a}
 \end{equation} 
  because 
 it controls the ratio of the partial width $\Gamma_{\mathcal{R}_i}^{H V}$ to the other partial widths of the model, see \eqref{scaling}. 

The dominant production mode of the CVM vectors is the Drell-Yan (DY) process. From the partial widths scaling above and from the LHC limits shown in Fig.~\ref{fig:xsbr} it follows that for the CVM the strongest constraints arise from the dilepton final state provided $a$ is not much larger than unity. The LHC constraint from the associate Higgs production final states $H V$ is dominant when $a$ is large\footnote{Notice $a$ can assume negative values if a negative $f^2$ is allowed (compensated by a positive $s\,v^2$), the interpretation of which, however, is unclear though we still allow it in our analysis; while DY production depends on $a$ only through the Higgs contribution to the CVM vector width, which is proportional to $|a|^2$, associated Higgs production receives contributions proportional to $a$ and $a-1$.}. We detail this in~\sec{sec:Hassoc}.
$a$ is not expected to be too large either: this would correspond to a situation where $M_R$, because of a cancellation between $f^2$ and $s\,v^2$, becomes significantly smaller than $f$ (for reasonable values of $\gt$).
For instance if $M_R\sim f$ and $\gt\sim10$, then $a\sim 25$;
we will not consider values of $a$ larger than this.
Typically, $a$ should be of $\mathcal{O}(1)$ while $a=0$ corresponds to $f=0$ and the mass scale of new physics provided by $s\,v^2$ alone.

It is useful to define $\delta$, the fractional difference between the $ZZ$-Higgs coupling in the CVM with respect to the GSW Higgs, as an alternative to $a$;
$a$ is indeed the parameter directly controlling $\delta$.
 \bea
\label{eq:deltadef}
1-\delta\equiv\frac{g_{HZZ}}{g^{SM}_{HZZ}}
\qquad
\text{with}
\qquad g^{SM}_{HZZ} \equiv \frac{2}{v}\, M_Z^2,
\eea
where $M_Z$ is the Z mass, and the sign of $\delta$ is chosen such that it coincides with the one of $a$.
From the explicit expression of $g_{HZZ}$ derived in~\eq{Eq:HiggscouplingsZ} we get the approximate expression: 
\bea
\label{eq:deltaZ}
\delta\simeq a\ \frac{v^2 \left(g'^4+g^4\right)}{4\, \gt^2 M_R^2}\ \simeq a\ \frac{1}{\tilde{g}^2}\left(\frac{55\GeV}{M_R}\right)^2 \ ,
\eea
obtained assuming $M_R \gg M_W$ and  $\tilde{g} \gg 1$;
note that $\delta$ is exactly proportional to $a$, not only in the limit of large $M_R$ and $\gt$.
On the other hand, the deviation from the GSW relation for the $WW$-Higgs coupling, $\delta_W$, does not vanish when $a=0$ because the tree-level $W$-boson mass is modified in the CVM.
The relation is given in \eq{Eq:Higgscouplings}, and is approximately
\bea
\label{eq:deltaW}
\delta_W \simeq 0.9\ \delta - \frac{1}{\tilde{g}^2}\left(\frac{40\GeV}{M_R}\right)^2 \simeq (a-0.6)\ \frac{1}{\tilde{g}^2}\left(\frac{50\GeV}{M_R}\right)^2.
\eea

The Yukawa sector of the CVM is modelled after the GSW theory to include minimal flavour violation and consequently minimise tension with experimental results from flavour physics. 

Intriguingly the CVM is challenging to uncover at the LHC even for vector masses in the TeV region and not too large values of $\gt$. The reason being that, for order unity values of all the couplings, the vectors  are very narrow and therefore their line shapes are hard to reconstruct with current experimental resolution. Furthermore for $\tilde{g} \gtrsim 2$ even the spacing in mass of the two resonances is less than the current experimental resolution in the dilepton invariant masses making it impossible to resolve them. 

{}For sufficiently large values of $a$ the partial width of $R\rightarrow HV$ grows and of course the overall width grows too. In this case one can reconstruct the overall line shape but cannot resolve the two closely spaced resonances because they significantly overlap. 

\begin{figure}[t!] 
\begin{center}
 \includegraphics[width=.55\columnwidth]{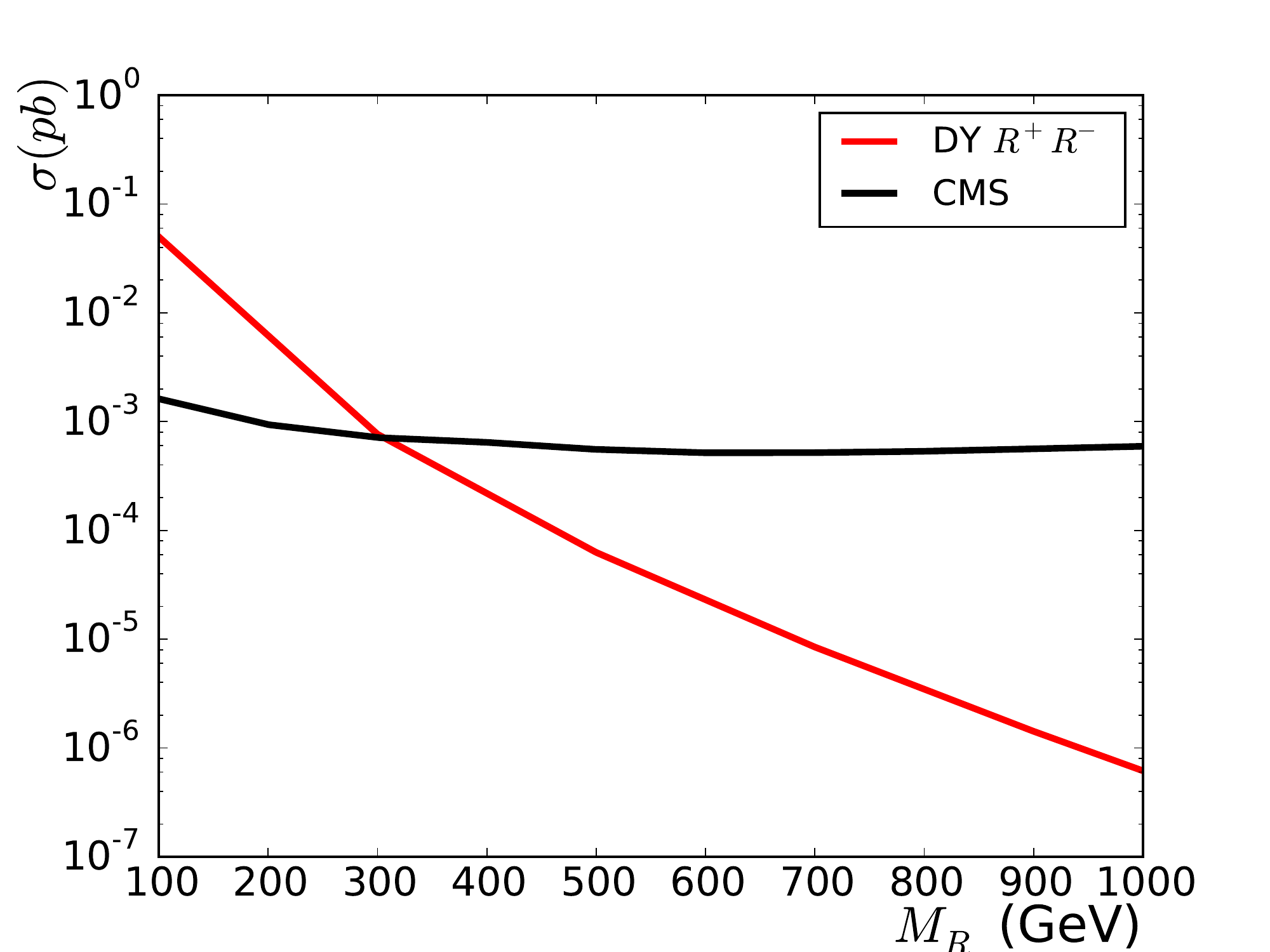}
\caption{ Full signal cross section for $R^+R^-$ pair production at the LHC with $\sqrt{s}=8\TeV$. The 95\% exclusion limit on long-lived charged particles provided by CMS is given in the black curve \cite{Chatrchyan:2013oca}.}
\label{fig:exclDYlongliv}
\end{center}
\end{figure}

We also note that due to the enhanced symmetry over the vectors the charged right resonances $R^{\pm}$ are stable. However we expect the CVM symmetry to be only approximate in the full theory. If the breaking is very small the $R^\pm$ are long lived. They are pair produced via a Drell-Yan process and will leave tracks in the CMS tracker and muon system \cite{Chatrchyan:2013oca}. The exclusion limit shown in \fig{fig:exclDYlongliv} is independent of $\gt$ and $a$ to leading order and
rules out values of $M_R$ below $\simeq 300\GeV$ so this constraint is currently weak.

\section{Electroweak Precision Tests}
\label{sec:EWPTLimits}
Contributions from the CVM to the electroweak observables are suppressed relative to generic models with vector resonances and scalars for two reasons:
the global symmetry acting on the vectors, and the presence of a very SM-like Higgs state. We will now discuss in turn these contributions.


\subsection{Vector sector}
The CVM contributions to the electroweak parameters $S$ and $T$ from the heavy vector bosons vanish because of the two custodial symmetries of the model.
The vector contribution to the $T$ parameter is zero at tree-level because the model respects ordinary custodial symmetry. 
The contribution to the $S$ parameter stemming from the heavy spin-1 resonances vanishes because the $SU(2)_L'\times SU(2)_R'$ insures parity doubling of the vector spectrum and decay constants. To elucidate this point  we observe that the S-parameter contribution from a genetic vector and axial resonance contribution reads:  
\begin{equation}
S=4\pi \left[  \frac{F_V^2}{M_V^2} - \frac{F_A^2}{M_A^2}\right]
\end{equation}
with the expressions for the decay constants and masses given in \cite{Belyaev:2008yj}. The $SU(2)_L'\times SU(2)_R'$ symmetry implies that $F_V=F_A$ and $M_V=M_A$ thus the vector contribution to the $S$ parameter vanishes as discussed further in \cite{Casalbuoni:1988xm,Appelquist:1999dq,Foadi:2007se}. More generally, the electroweak $S$ parameter, after integrating out all heavy vector and scalar states, can be described in the effective Lagrangian by the operator ${\rm Tr}[W^{\mu\nu} \Sigma B^{\mu\nu} \Sigma^{\dagger}]$. Any new contribution, before integrating out heavy vectors, would involve insertion of $A_L$ or $A_R$ but that is not allowed by the $SU(2)_L'\times SU(2)_R'$ symmetry.
Note that the discrete $Z_{2L}'\times Z_{2R}'$ symmetry acting as $A_{L/R}\to z_{L/R} A_{L/R}$, with $z_{L/R}=\pm 1$ is sufficient to ensure the vanishing of the vector contribution to $S$ at tree-level. 
 
According to the parameterization of electroweak observables \cite{Barbieri:2004qk}, only the custodial and isospin preserving parameters $W$ and $Y$ are now non-vanishing \cite{Foadi:2007ue,Casalbuoni:2007dk,Foadi:2007se}:
\bea
\label{eq:WY}
W = \frac{4\cos^4\theta\, M_Z^4}{\gt^2\, v^2\, M_R^2} \ ,\qquad 
Y = \frac{\sin^2(2\theta)\, M_Z^4}{\gt^2\, v^2\, M_R^2} \ .
\eea

\subsection{Higgs sector}

The presence of a light SM-like scalar --- now experimentally established --- provides important corrections to electroweak observables for a good agreement with data, such that effects from new physics need only be small rather than having to mimic a Higgs.
While the contributions from the spin-1 resonances in our model are under control thanks to the custodial symmetry, we do allow a small misalignment between the vector-boson-mass matrix and the scalar-coupling matrix, parametrized by $a$ (or equivalently $\delta$ \eqref{eq:deltaZ}).
This implies a deviation in the couplings of the Higgs boson to the electroweak bosons in the CVM and thus contributions to $S$ and $T$.
These have previously been determined in full in \cite{Foadi:2012ga}.
We will however conclude that this additional effect is even less important than the spin-1 contributions.

Approximate expressions of the contributions to the electroweak parameters read:
\bea
\label{eq:STdelta}
\hat{S} \approx \delta\,\frac{\alpha}{12\pi \sin^2\theta}\, \ln{\frac{\Lambda}{M_h}}, \qquad
\hat{T} \approx -\delta\,\frac{3\,\alpha}{4\pi \cos^2\theta}\, \ln{\frac{\Lambda}{M_h}}, \qquad
\hat{U} \approx 0,
\eea
where $M_h$ is the Higgs mass, $\alpha$ is the electromagnetic coupling at the $Z$ pole and $\theta$ the Weinberg angle defined as
\bea
\label{eq:thetadef}
\sin^2{2\theta} &\equiv& \frac{4 \pi\, \alpha}{\sqrt{2}\, G_F M_Z^2} = \frac{e^2\, v^2}{M_Z^2}. 
\eea
To provide simple constraints on $\delta$ we approximate here the renormalisation procedure  \cite{Foadi:2012ga} by the presence of a physical cutoff $\Lambda$ which is expected to be around the new resonances mass scale, i.e. $4\pi v$. Ignoring for an instant the contributions from the vector resonances, we deduce the following approximate bounds on $\delta$ at $95\%$ CL --- adapting the analysis in \cite{Ciuchini:2013pca,Ciuchini:2014dea}:
\bea
-0.09<\delta<0.03\,.
\eea
The limits are comparable to the ones from the direct Higgs couplings measurements \cite{ATLAS-CONF-2014-009} that at two-sigmas yield,
\bea
-0.31<\delta<0.01\,.
\label{eq:HVVdirectlim}
\eea


\subsection{Limits}

The effect on electroweak observables is best expressed in terms of the $\epsilon$ parameters \cite{Altarelli:1990zd}, and one has (with \eqref{eq:WY}-\eqref{eq:STdelta} and $V=X=0$) \cite{Barbieri:2004qk}:
\bea
\delta\epsilon_1 = \hat{T} - W - \tan^2\theta\, Y, \qquad
\delta\epsilon_2 = \hat{U} - W, \qquad
\delta\epsilon_3 = \hat{S} - W - Y.
\eea

\begin{figure}[t!] 
\begin{center}
  \includegraphics[width=.55\columnwidth]{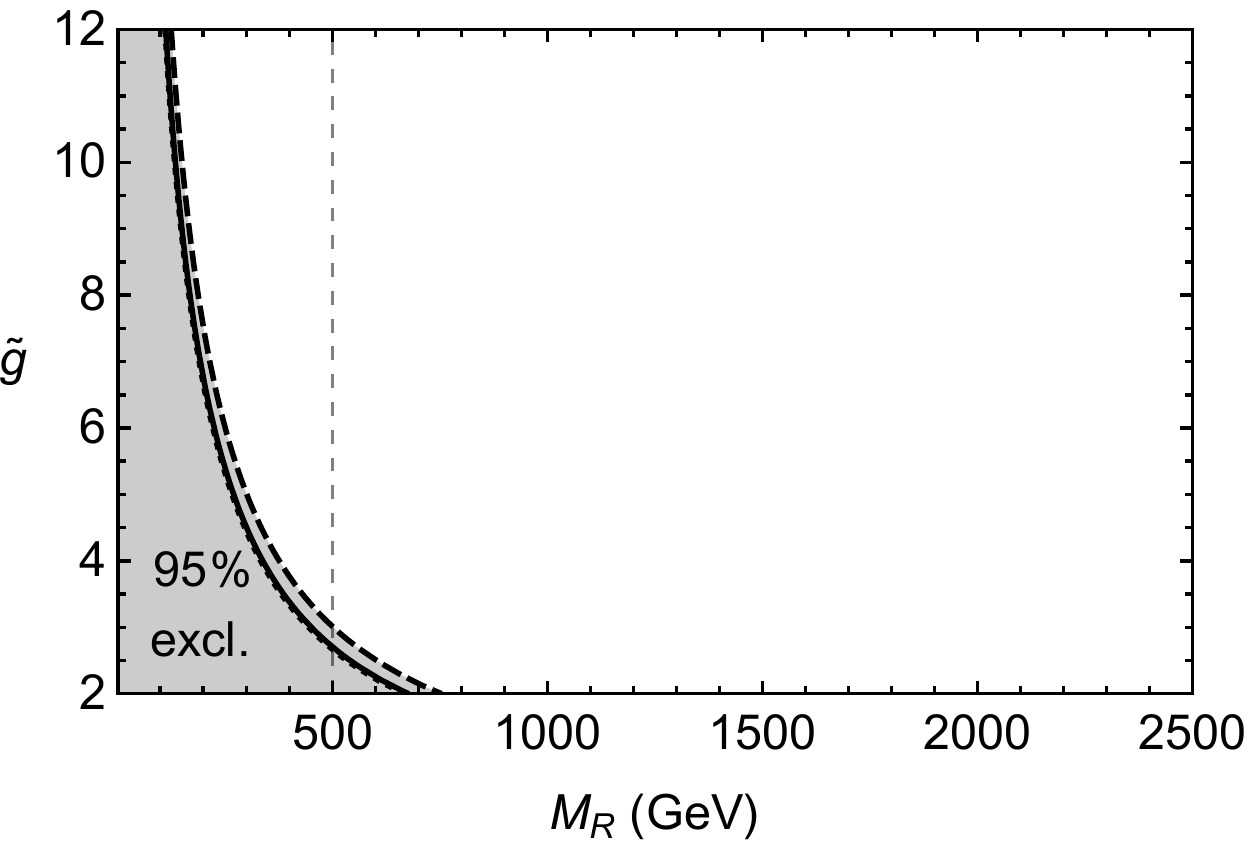}
\caption{Exclusion limits at 95\% CL in the $(M_R, \gt)$ plane from electroweak precision measurements on the CVM vector resonances. Dotted, continuous and dashed lines for $a=25$, $0$ and $-25$, respectively.
The vertical separation line is meant to guide the eye: plots at the end of the paper only start at $M_R=500\GeV$.}
\label{fig:ewlimit}
\end{center}
\end{figure}

A recent fit from \cite{Ciuchini:2014dea} gives 
\bea
\begin{matrix}
10^3\, \delta\epsilon_1 =&   &0.7\pm 1.0 \\
10^3\, \delta\epsilon_2 =& -&0.1\pm 0.9 \\
10^3\, \delta\epsilon_3 =&   &0.6\pm 0.9
\end{matrix}
\quad
\text{with correlation matrix}
\quad
\rho =
\begin{pmatrix}
1 & 0.80 & 0.86 \\
0.80 & 1 & 0.51 \\
0.86 & 0.51 & 1
\end{pmatrix},
\eea
and performing a simple $\chi^2$ test, we obtain the exclusion limits on ($M_R$, $\gt$), shown in \fig{fig:ewlimit}, adding the Higgs contributions above for fixed values of $a$ and $\delta$ given by \eqref{eq:deltaZ}.
Even for the extreme values $|a|=25$, at the edge of the parameter space we will be considering, the dominant effect is the one from the vector resonances.

However, due to the the double suppression --- in $\gt^{-2}$ and $M_R^{-2}$ --- of the new physics contributions to $W$ and $Y$, electroweak constraints are overall very weak and direct searches for the vector resonances are much more important.

In the low $(\gt,\,M_R)$ region of parameter space shown in \fig{fig:ewlimit}, direct measurements of the Higgs boson couplings can be competitive with electroweak precision test for extreme values of $a$. The $\delta$-parameter measuring the Higgs coupling deviations from their SM values can reach the percent level here. In this region the difference between $W$ and $Z$ boson couplings to the Higgs boson could also be experimentally accessed. In practice, however, direct searches for the vector resonances are much more constraining and rule out this parameter region, as we are going to see in the next section.



\section{LHC Phenomenology}
\label{sec:LHCpheno}

In this section we present the detailed LHC phenomenology of the CVM, previously sketched in sec. \ref{sec:Model}. To aide numerical computations, the model is implemented in {\tt MadGraph 5} \cite{Alwall:2011uj} using the {\tt FeynRules} package \cite{Alloul:2013bka}. In our computations we use the following electroweak parameters:
\begin{flalign}
&M_Z=91.2 \GeV\\ \nonumber
&G_F=1.16637\times 10^{-5} \GeV^{-2}\\  \nonumber
&\alpha^{-1}(M_Z)=127.9 \\  \nonumber
&M_t=172\GeV\, \ . 
\end{flalign}
In addition to these, the CVM is parameterized by the three parameters, characterizing the new spin one resonances
\bea
M_R \ ,   \quad \gt \ ,  \quad a \ . 
\eea
where $M_R$ is the mass scale of the heavy resonances, $\gt$ is their self-coupling and $a$ was defined in~\eq{def:a}. Instead of $a$ we will sometimes use $\delta$ defined in~\eq{eq:deltadef}. Values of $\delta/a$  range from $\delta/a \approx 0.000003 - 0.003$ for ($\tilde{g}$, $M_R$) between $(12,2500\GeV)$ and $(2,500\GeV)$. So unless $a$ is large, the $HZZ$ and $HWW$ couplings are very SM-like in the model.

\begin{figure}[t!] 
\begin{center}
 \includegraphics[width=.49\columnwidth]{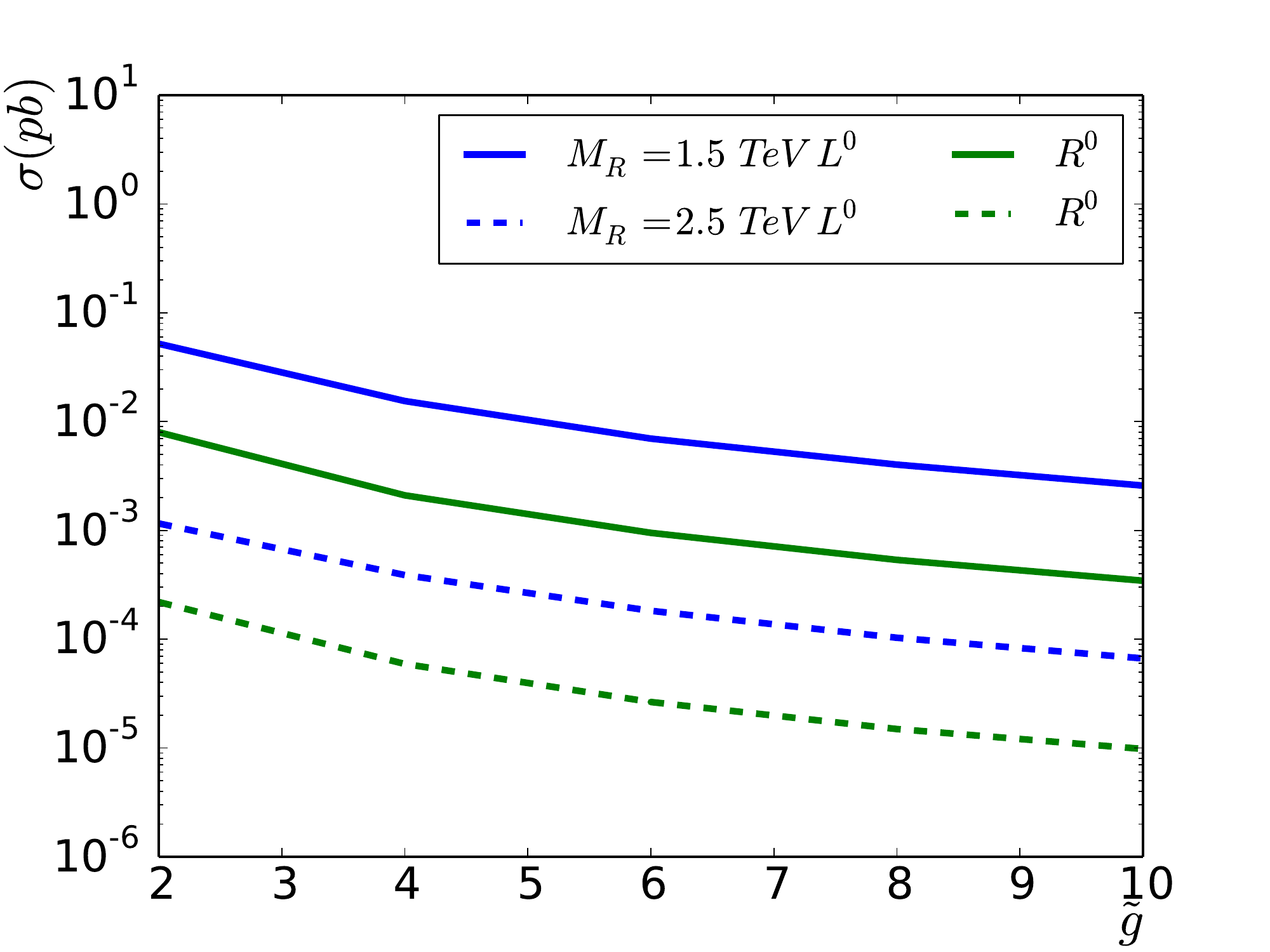}
  \includegraphics[width=.49\columnwidth]{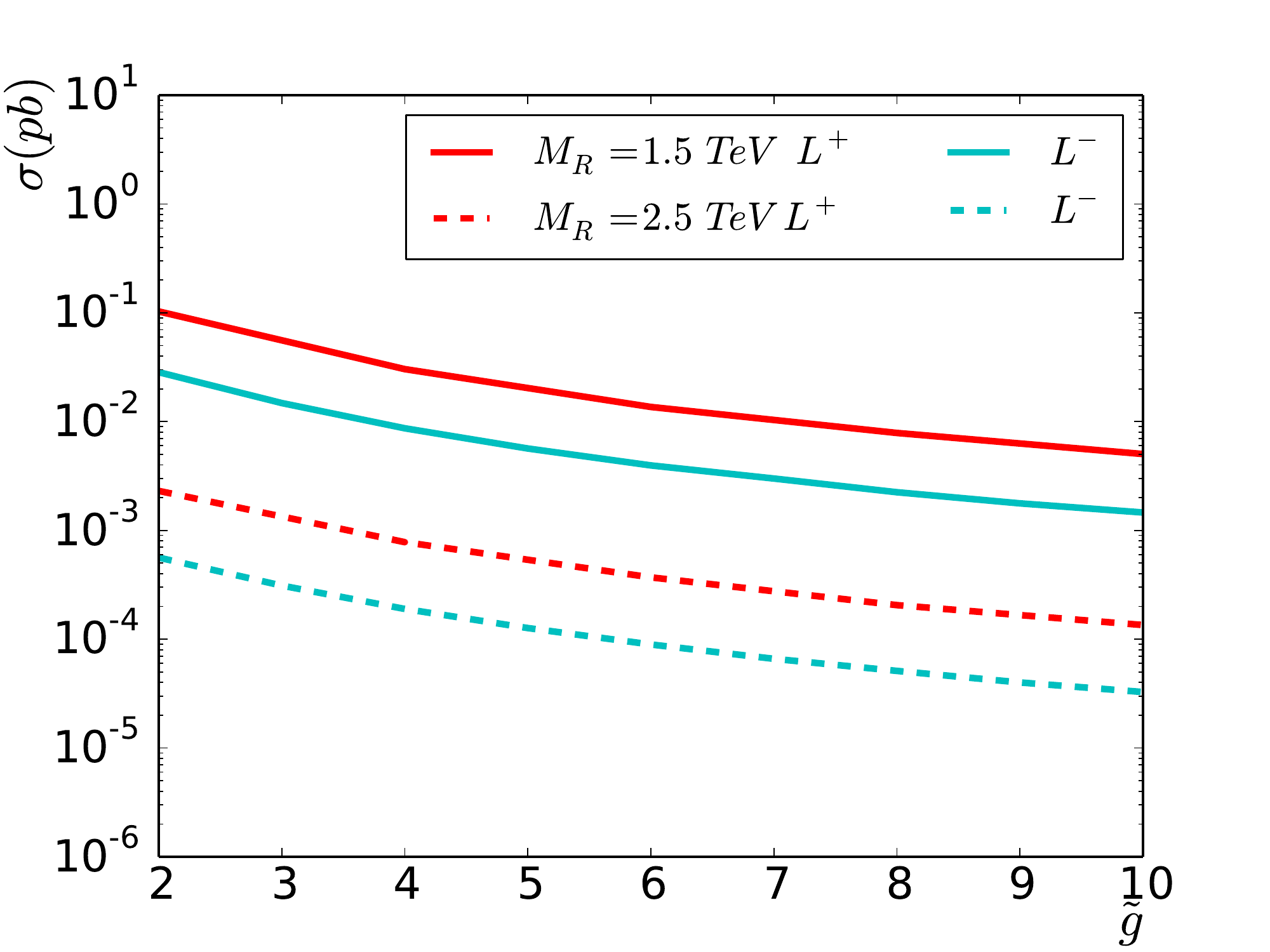}
\caption{LHC production cross section of the heavy CVM vector resonances at $\sqrt{s}=8$ TeV as a function of $\tilde{g}$ for $M_R=1500 \GeV$ (solid lines) and  $M_R=2500\GeV$ (dashed lines). On the \emph{left} we show $L^0$ (blue) and $R^0$ (green) . On the \emph{right} we show $L^+$ (red) and $L^-$ (cyan).}
\label{fig:xs}
\end{center}
\end{figure}

The LHC production cross sections of the new vector resonances, at $\sqrt{s}=8\TeV$ center of mass energy, are shown in \fig{fig:xs} as a function of $\tilde{g}$ for different values of $M_R$. Due to the factorisable nature of the QCD corrections for the Drell-Yan production, the inclusive cross section at NNLO accuracy in QCD is given by 
\bea
\sigma_{NNLO}=\sigma_{LO}\times K, 
\eea
where $\sigma_{LO}$ is the leading order prediction and the $K$ factor depends only on the mass of the resonance. 
We use $K=1.16$ for the neutral vector resonance production and $1.2$ for the charged. These choices of $K$ factors mean that our exclusion limits are slightly  conservative\footnote{These choices correspond to the smallest $K$ factors used by ATLAS and CMS in the resonance mass range from 1 to 3 TeV --- the variation of $K$ factors in this mass range for the neutral resonances are  $K=1.16-1.22$ \cite{Aad:2014cka}  and for the charged resonances $K=1.2-1.3$ \cite{Khachatryan:2014tva}}.

\begin{figure}[]
\begin{center}
 \includegraphics[width=.49\columnwidth]{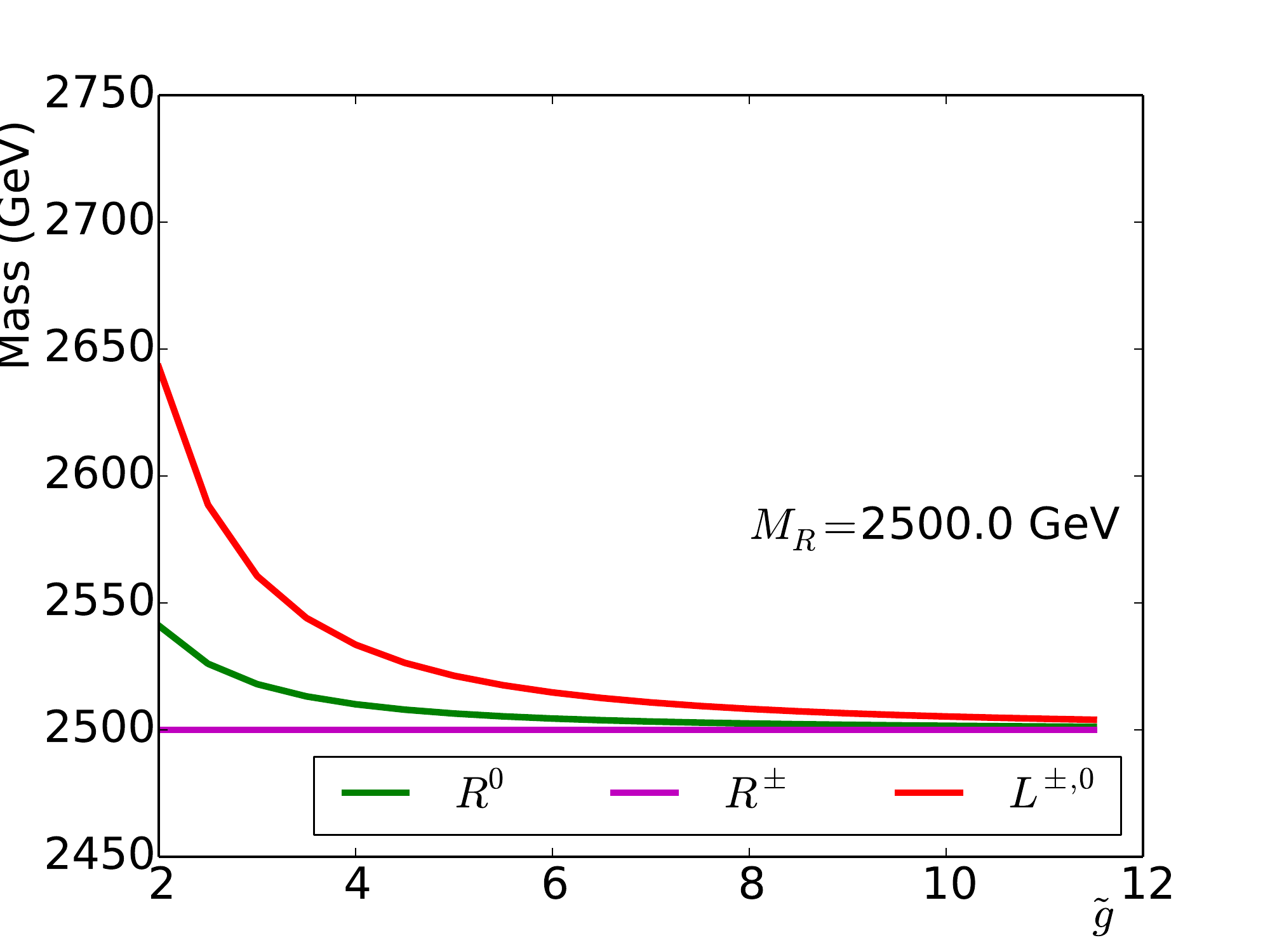}
\caption{Masses of the heavy CVM vector resonances for $M_R=2500\GeV$. The difference in mass between the charged and neutral $L$ states is negligible and cannot be seen on the plot.}
\label{fig:masses}
\end{center}
\end{figure}

\begin{figure}[t!] 
\begin{center}
 \includegraphics[width=.49\columnwidth]{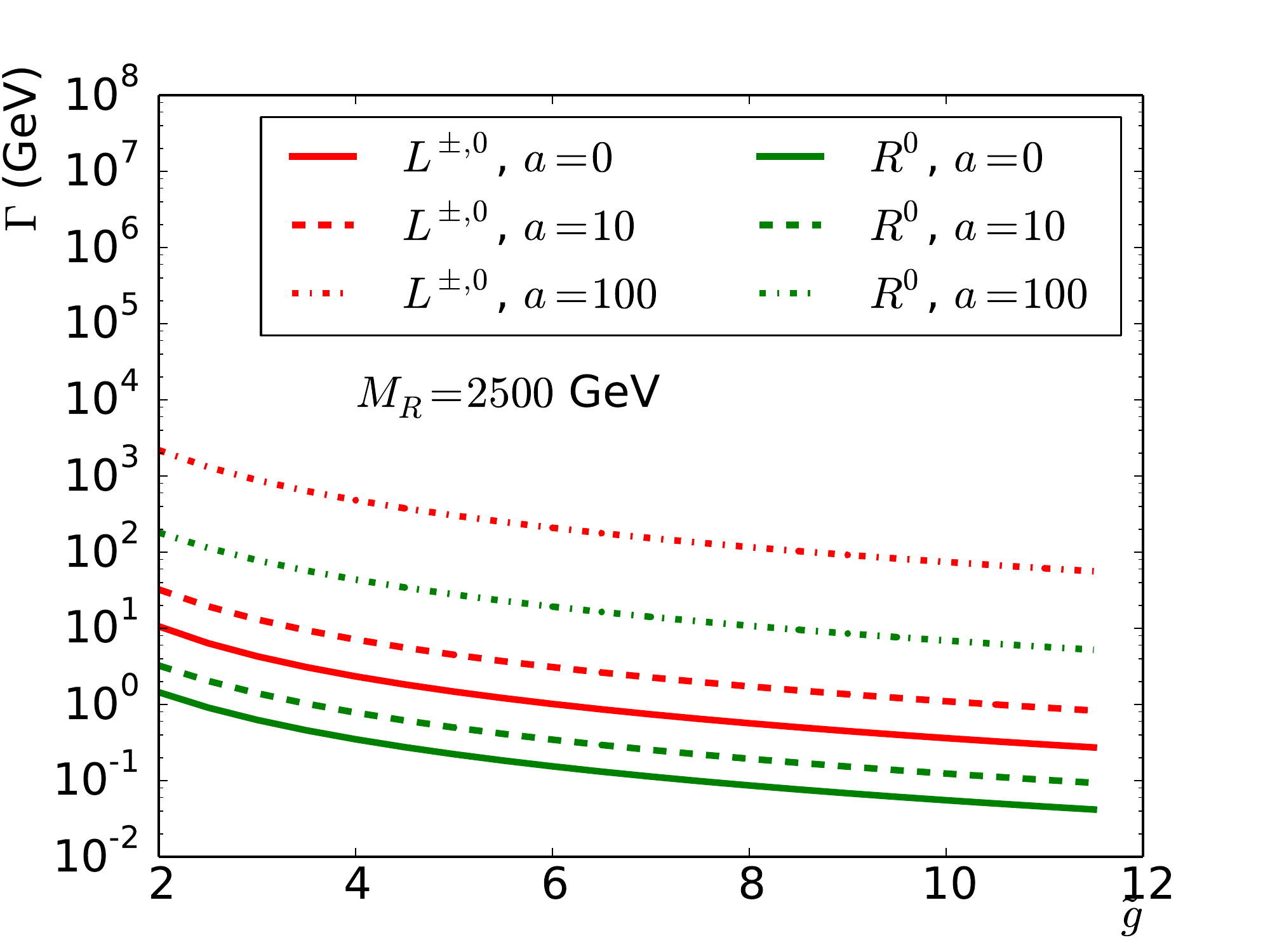}
 \includegraphics[width=.49\columnwidth]{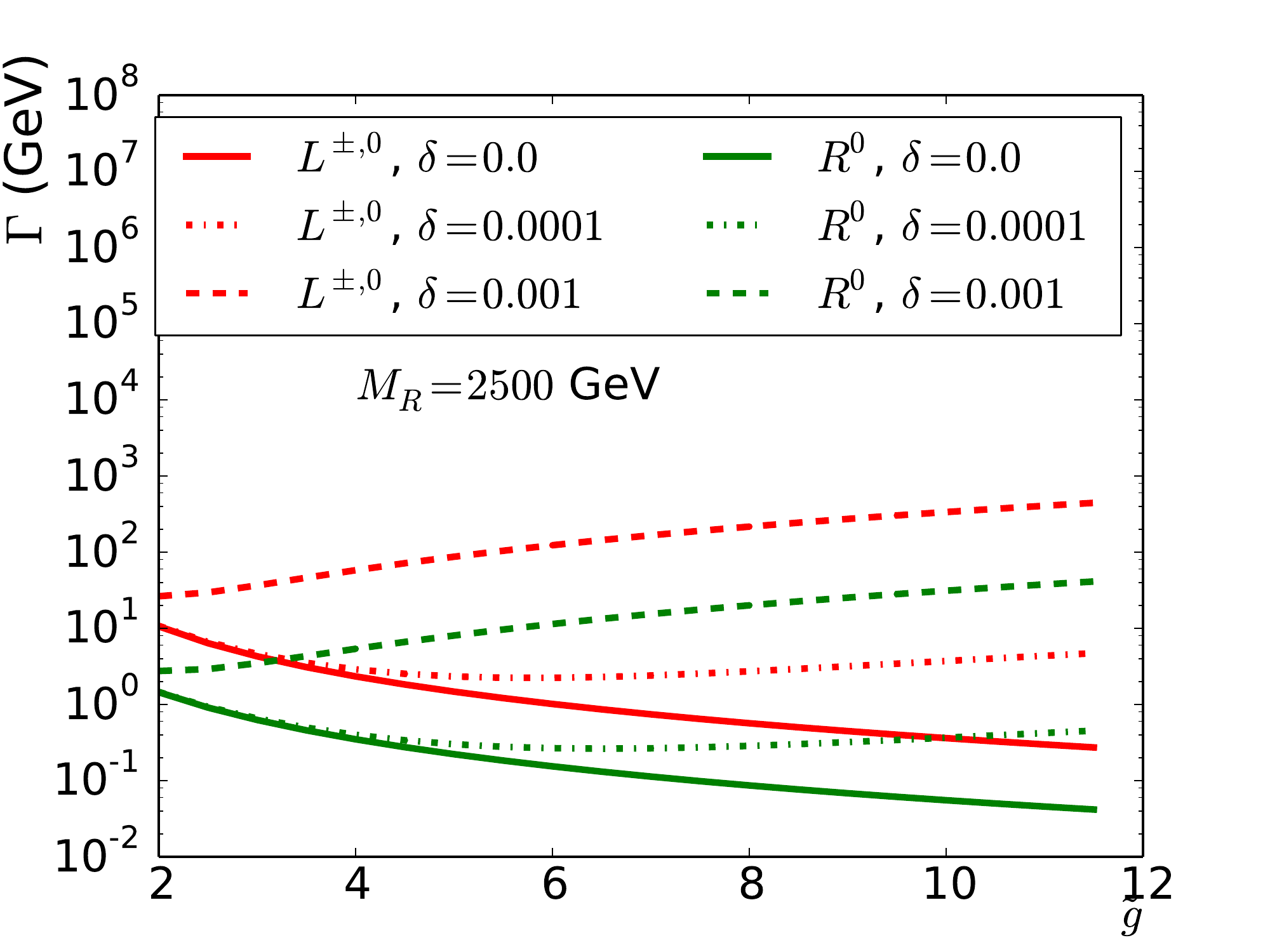}
\caption{Widths of the heavy CVM vector resonances as a function of $\tilde{g}$ for different values of $a$ (\emph{left}) and $\delta$ (\emph{right}). We keep $M_R=2.5$ TeV fixed. }
\label{fig:widths}
\end{center}
\end{figure}

As explained in the previous section, the masses of the heavy resonances are degenerate for large $\tilde{g}$ and only become appreciably different 
when $\tilde{g}\lesssim 1 $. In fact, the left triplet $L^{0,\pm}$ states remain highly degenerate for all parameter values. The vector spectrum as a function of $\tilde{g}$ can be seen in \fig{fig:masses}. 
The corresponding widths are shown in \fig{fig:widths} as a function of $\gt$ for different values of $\delta$ (left panel) and $a$ (right panel) for $M_R=2.5$ TeV.  For fixed $a$ the widths simply scale as $\Gamma\sim \frac{1}{\tilde{g}^2}$ as noted in eq. \eqref{scaling}.

For small values of 
$a$ the partial widths $\Gamma_{\mathcal{R}_i}^{HW/HZ}$ are small compared to the other decay channels. In this case, the heavy resonances are very narrow and the separation in masses between the two neutral resonances is always larger than their widths. 
Furthermore the branching ratios are nearly constant as a function of $\tilde{g}$ and $M_R$, apart from corrections due to the mass differences of the final states.

\begin{figure}[t!] 
\begin{center}
 \includegraphics[width=.49\columnwidth]{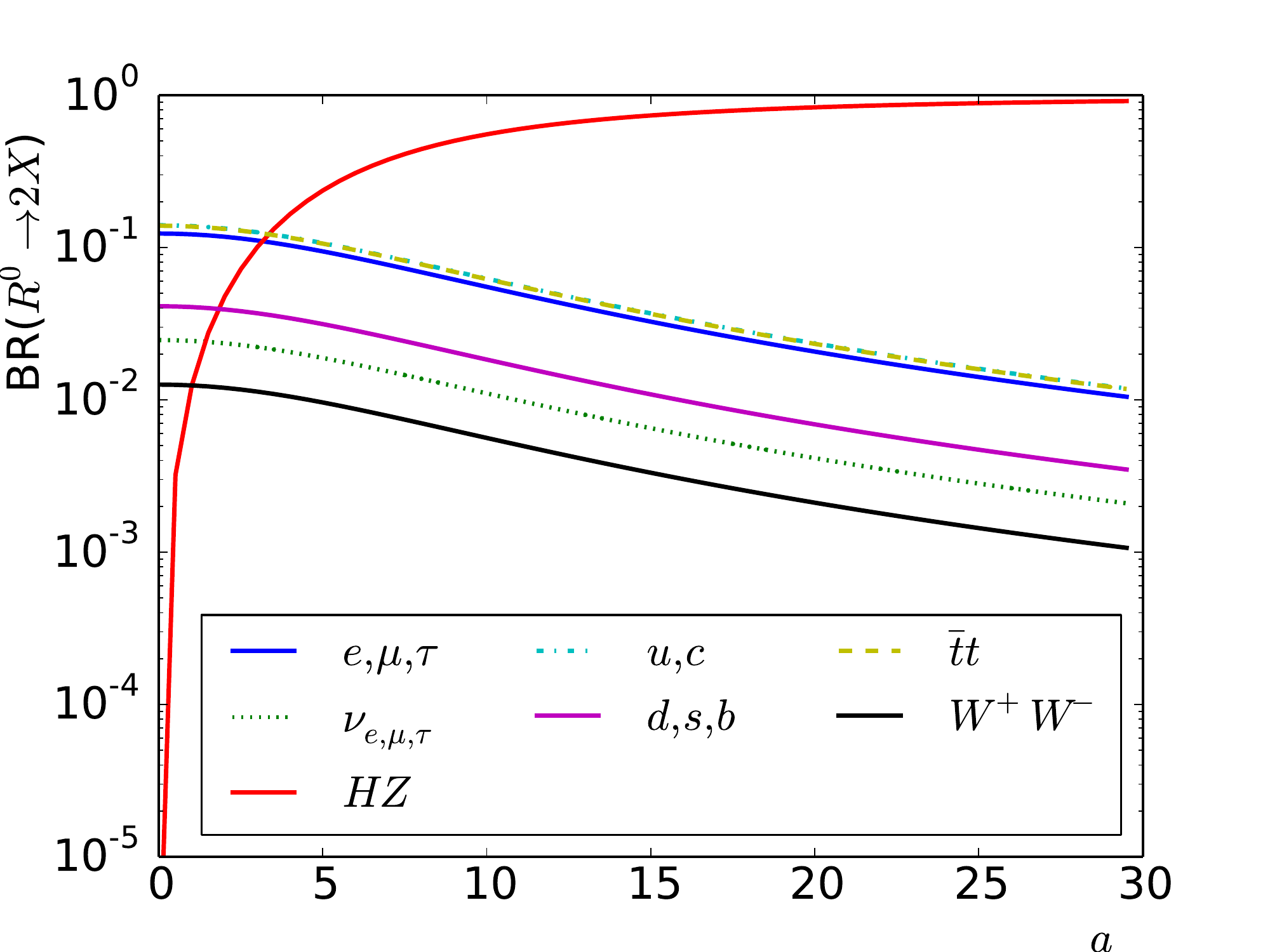}
 \includegraphics[width=.49\columnwidth]{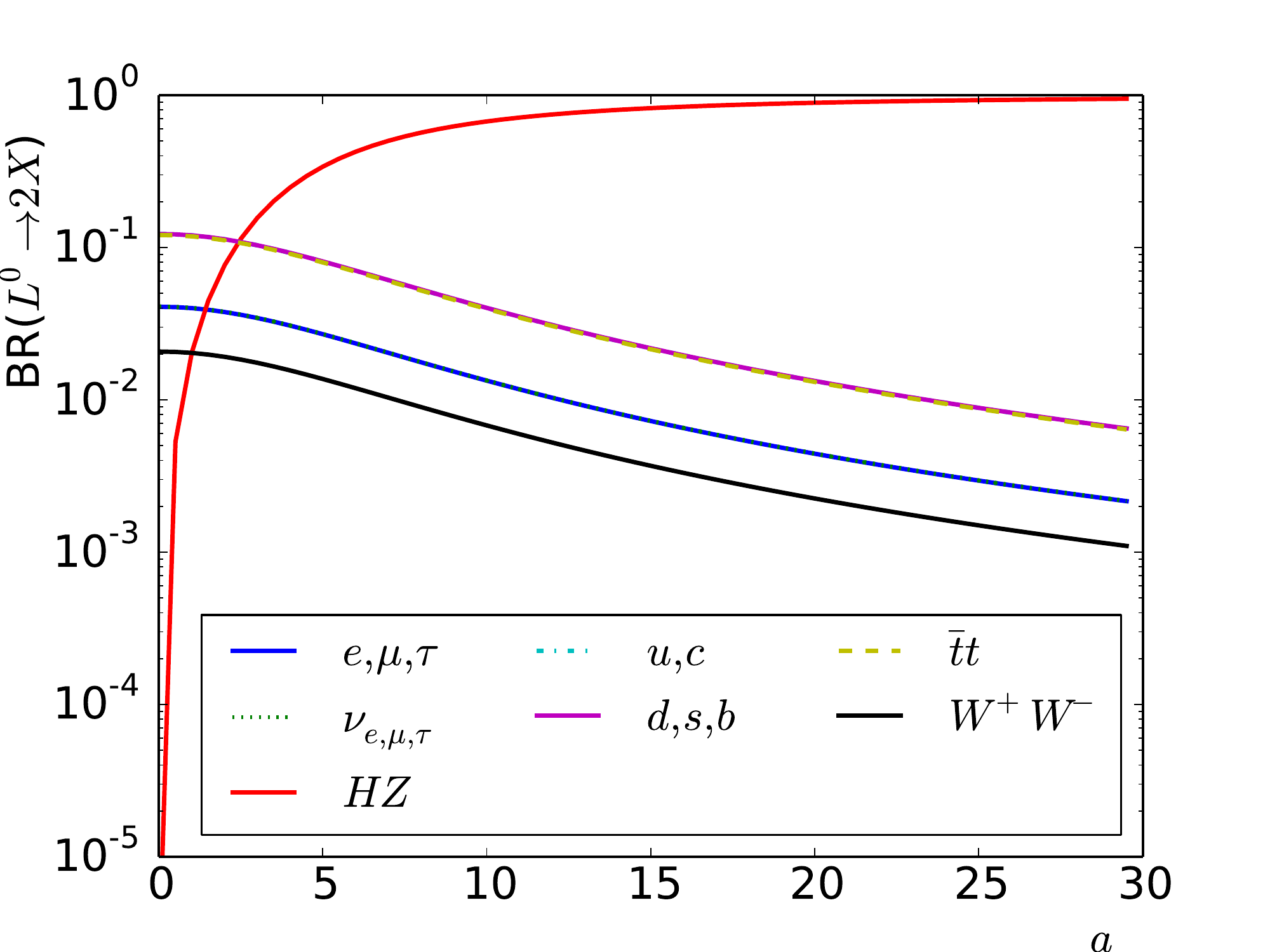}
  \includegraphics[width=.49\columnwidth]{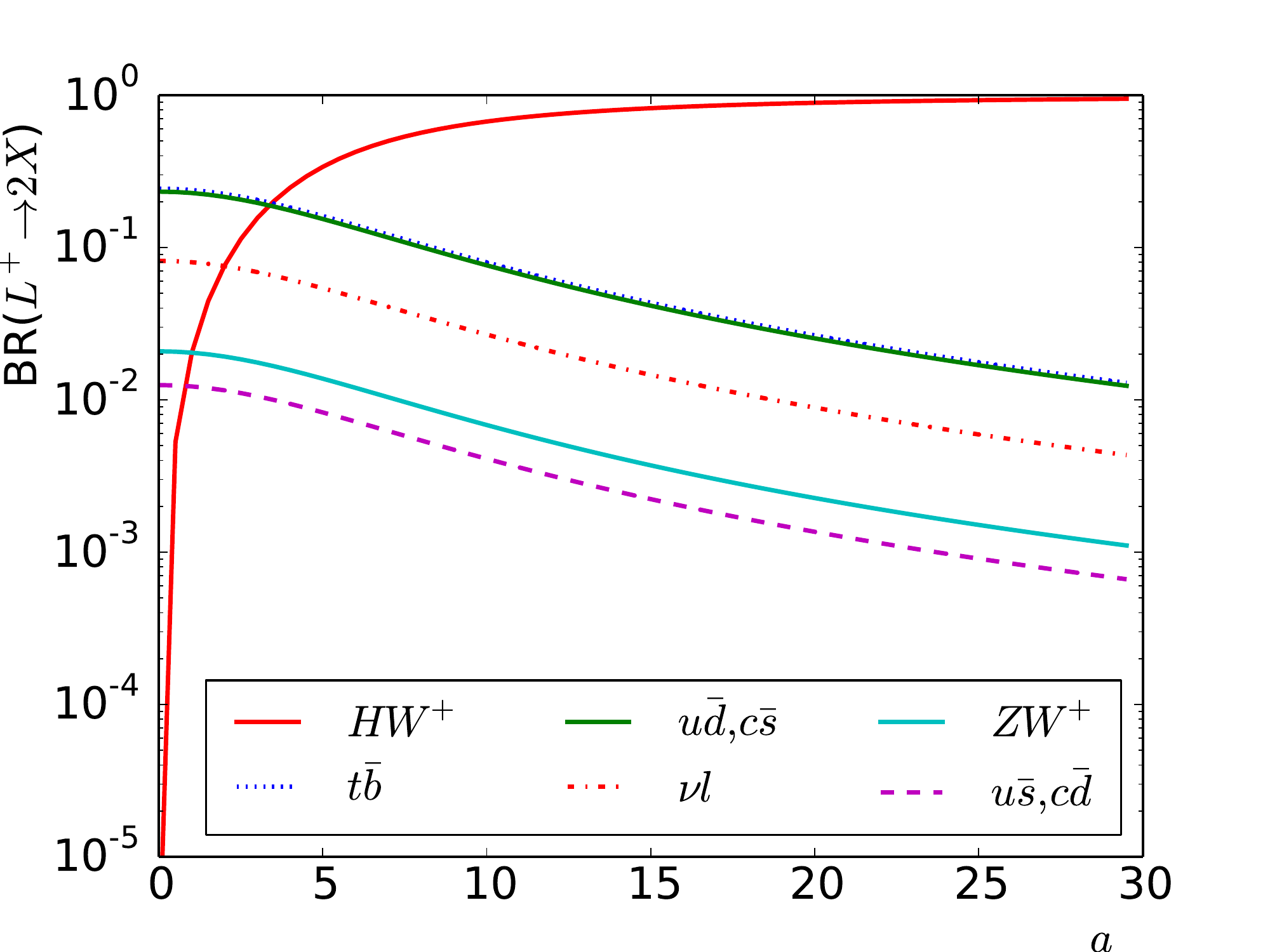} 
 \caption{Branching ratios of the heavy CVM vector resonances $R^0$, $L^0$ and $L^+$.}
\label{fig:brgt}
\end{center}
\end{figure}

Once $a$  grows, 
the $HW^\pm/HZ$ channels become important and eventually dominate the widths of the heavy resonances. This phenomenon is shown at the branching ratios level as a function of $a$ in  \fig{fig:brgt}. For fixed $a$ the branching ratios are constant to leading order in $M_R$ and $\gt$.

\subsection{Dilepton searches}
\label{sec:dilepton}

The current ATLAS  \cite{Aad:2014cka} and CMS  \cite{CMS-PAS-EXO-12-061} exclusion limits on neutral vector resonances in the dilepton channels are based on modelling the signal as a single resonance. In the CVM, the two resonances are nearly degenerate as shown in  \fig{fig:masses}. Two questions then arise: Is it possible to resolve a two peak structure? And, is it possible to resolve the line-shape of each peak? 

The fractional dimuon mass resolution at CMS is $\sigma(\mu\mu)/m_{\mu\mu}\simeq 6.5\%$ at masses around $1\TeV$. It further depletes at higher energies due to the difficulty in measuring the curvature of the track in the muon chambers. The dieletron mass resolution, $\Delta(m_{ee})/m_{ee}$ \footnote{The different symbols $\sigma$ and $\Delta$ indicate that the muon uncertainty follows a Gaussian while the electron uncertainty does not.}, on the other hand, is approximately constant above $500\GeV$ \cite{Chatrchyan:2012it} \footnote{When both electrons are detected in the barrel, this mass resolution is $1.1\%$, and when one of the electrons is in the barrel and the other is in the endcaps it is $2.3\%$ \cite{Chatrchyan:2012it}. }. Summarising, for heavy resonances whose widths are lower than $5\%$ of their masses, the search is currently dominated by the resolution of the detector and therefore the line shapes of the peaks cannot be measured \cite{Aad:2014cka}.

\begin{figure}[t!] 
\begin{center}
\includegraphics[width=.49\columnwidth]{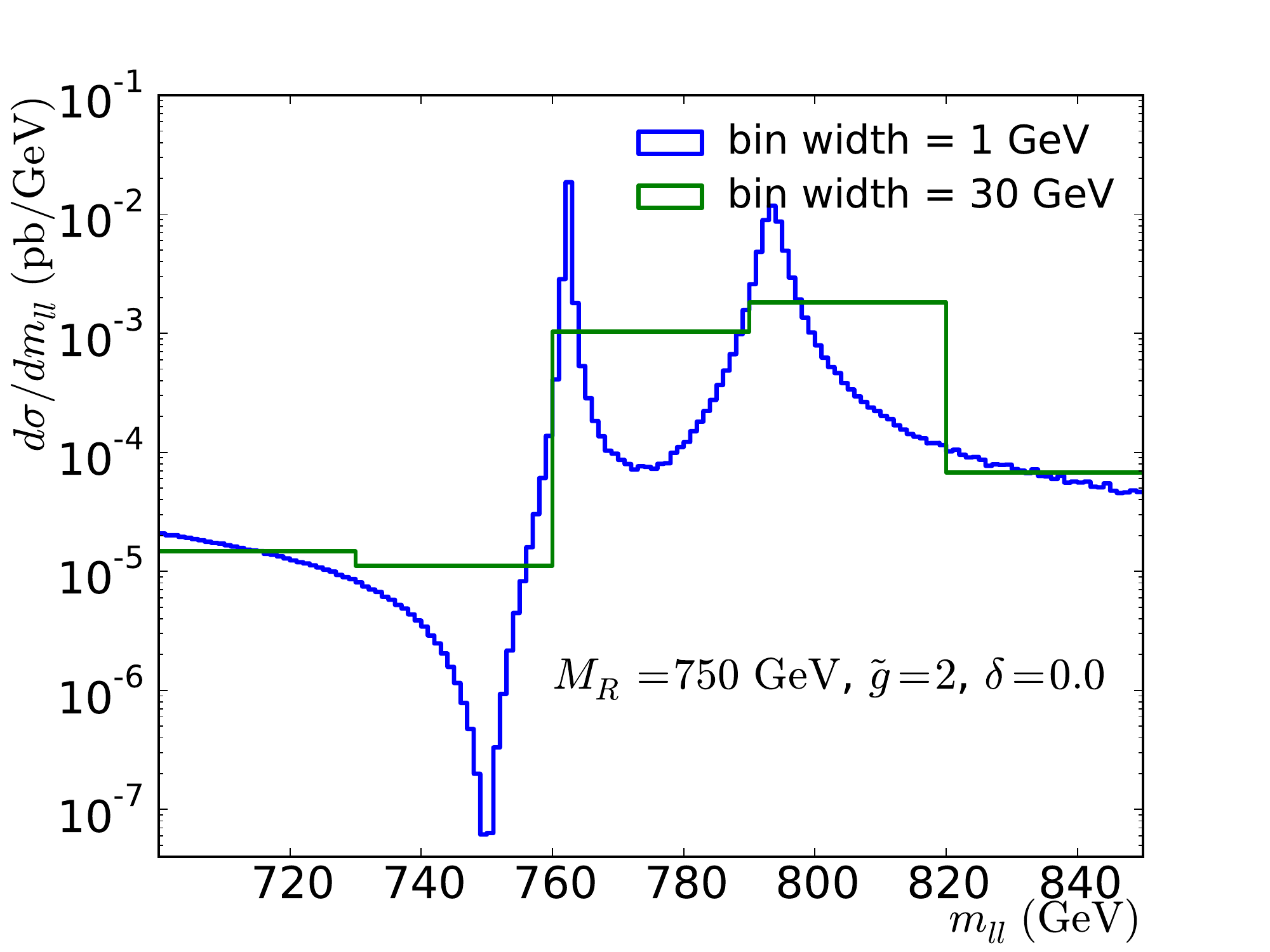}
\includegraphics[width=.49\columnwidth]{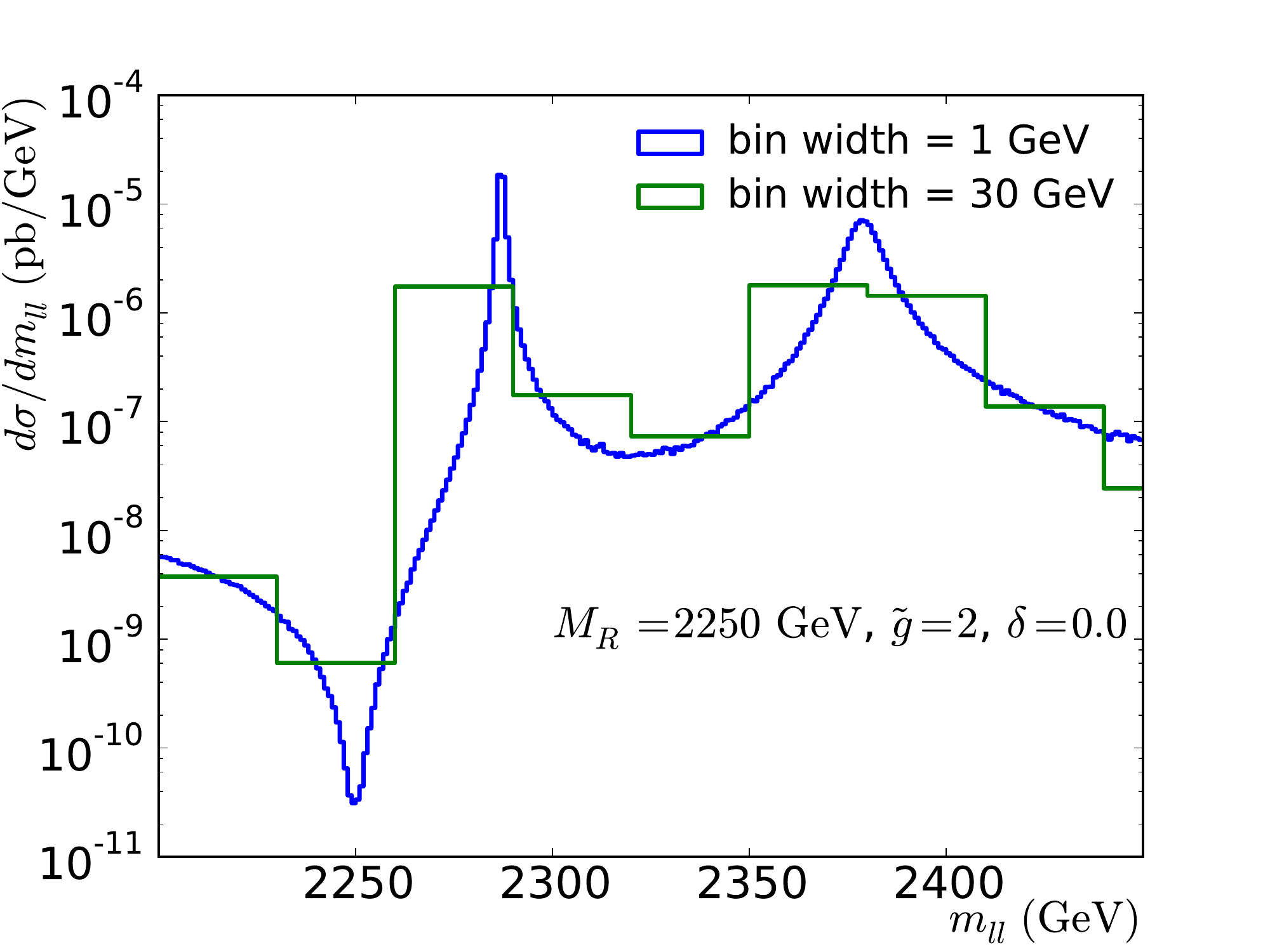}
\caption{Dilepton invariant mass distributions, $M(\ell\ell)$, in the CVM, with $1\GeV$ (blue) and $30\GeV$ (green) binning. On the (\emph{left}) $M_R=750\GeV$ and  on the (\emph{right}) $M_R=2250\GeV$.}
\label{fig:mlldet}
\end{center}
\end{figure}

For values of $a\lesssim 1$ ( or $\delta \lesssim 10^{-3}$ ) the ratio of the total width of the vector resonances to their mass satisfies $\Gamma_\mathcal{R}/m_\mathcal{R}\simeq 0.01 - 0.1\%$ which is well below the current sensitivity. This is illustrated in \fig{fig:mlldet}, showing the resonance pattern of the CVM in dilepton invariant mass distributions with two different bin widths --- the largest bin width of 30 GeV is representative of current experimental sensitivity and insufficient to reconstruct the line shapes. 
We next consider the ability to resolve the two peak structure. The relative mass splitting of our resonances is approximately
\bea
\frac{\Delta M}{M_R}\simeq  \frac{0.16}{\gt^2} \ , \qquad \Delta M=|M_{L^0}-M_{R^0}| \ .
\eea
This shows that the resolution of the detector would allow probing the presence of two peaks if $\gt \lesssim 2$ \footnote{It may eventually be possible to probe them in the electron channel up to $\gt=4$.}. This  would also allow a measurement of $\gt$ directly from the separation of the two peaks. 

\begin{figure}[t!] 
\begin{center}
  \includegraphics[width=.55\columnwidth]{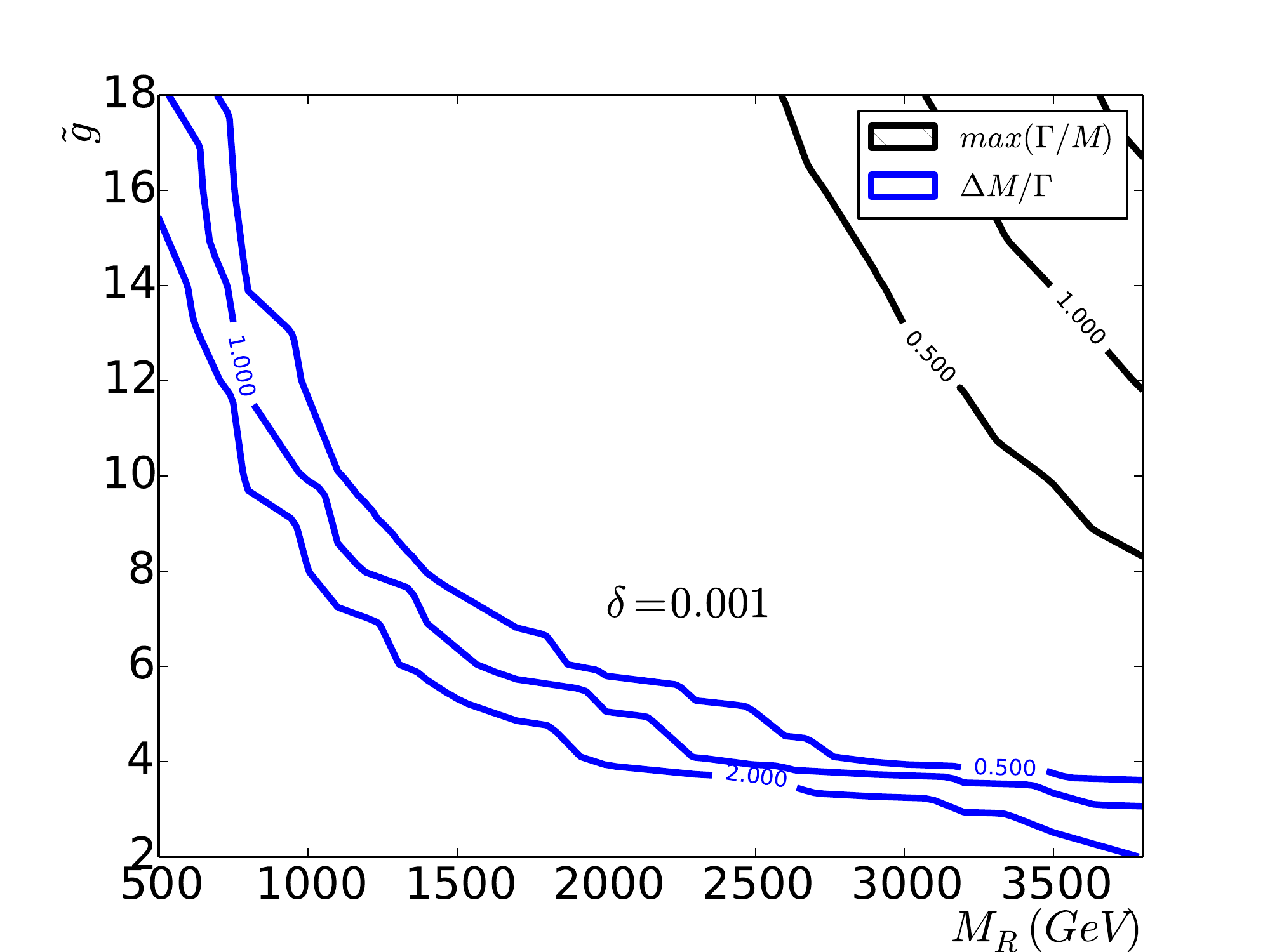}
\caption{Lines of constant mass splitting relative to width $\frac{\Delta M}{\bar{\Gamma}}$ (blue lines) of the CVM resonances relative to their width as well as lines of constant maximal resonance width relative to mass $\textrm{max}[\Gamma/M]$ (black lines), in the plane $(M_R, \gt)$.}
\label{fig:masssplit}
\end{center}
\end{figure}

\begin{figure}[t!] 
\begin{center}
\includegraphics[width=.49\columnwidth]{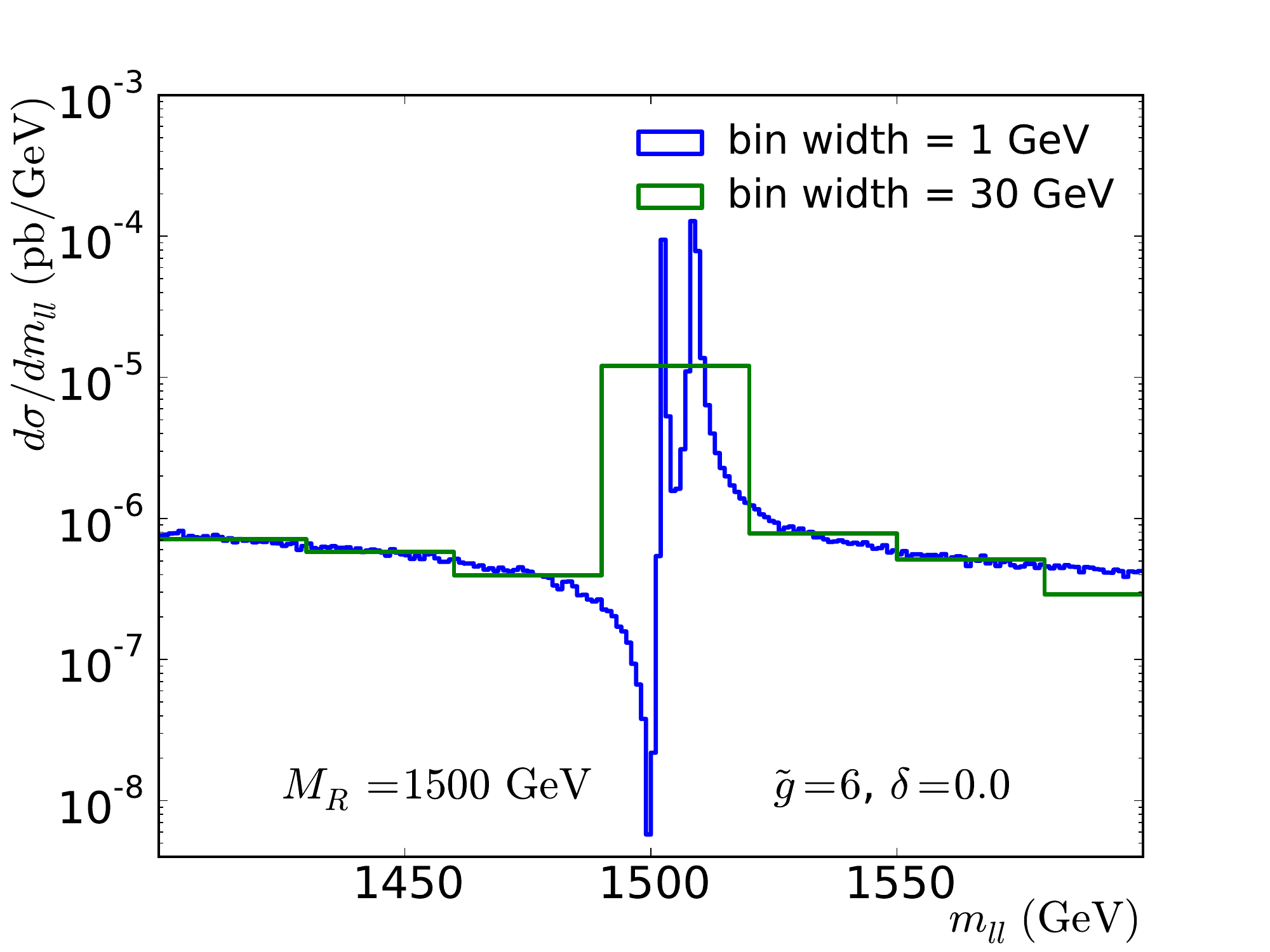}
\includegraphics[width=.49\columnwidth]{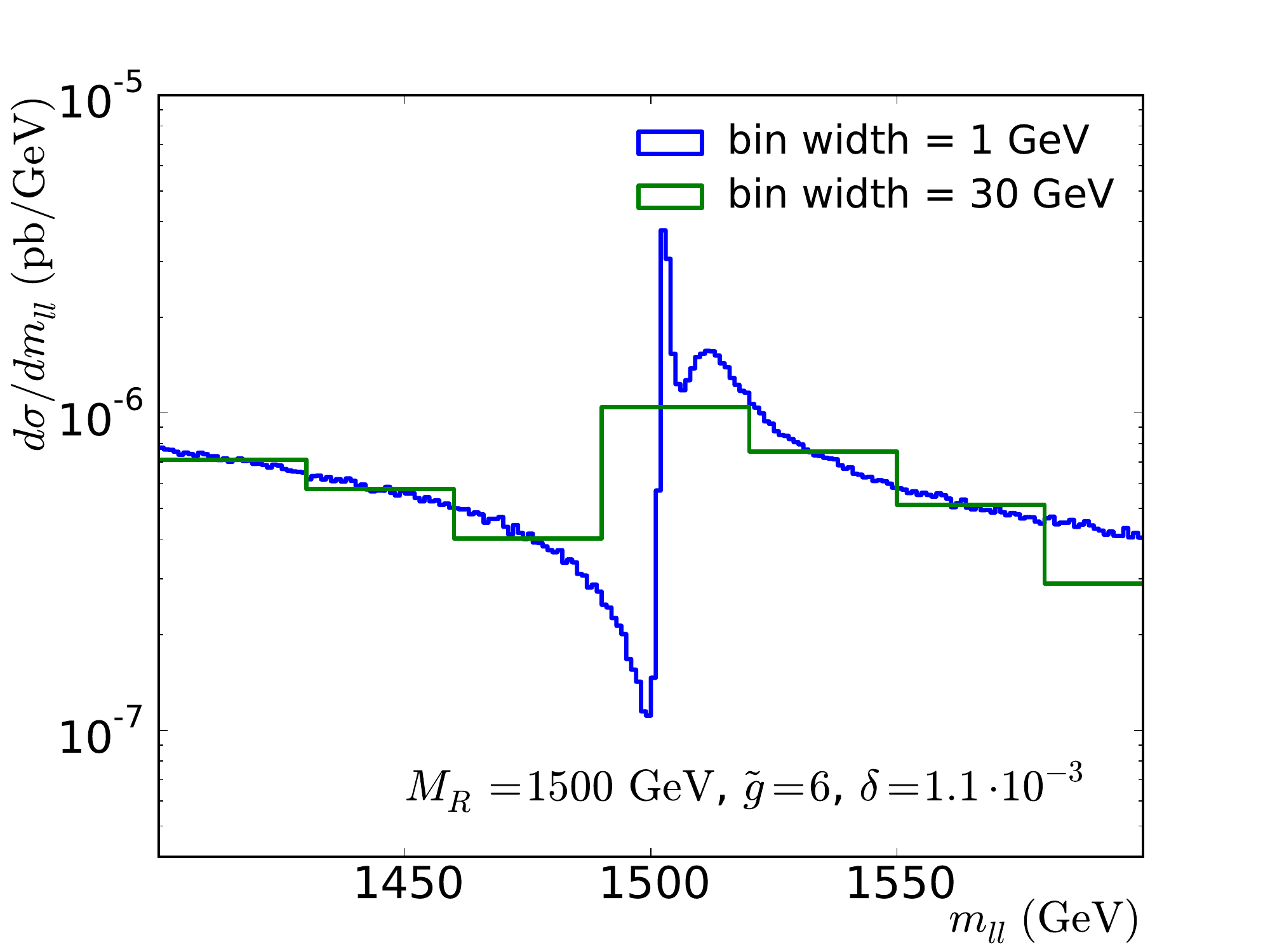}
\caption{Dilepton invariant mass distributions, $M(\ell\ell)$, in the CVM, for fixed $M_R=1500\GeV$ with $1\GeV$ (blue) and $30\GeV$ (green) binning. On the (\emph{left}) $\delta=0$ and  on the (\emph{right}) $\delta=10^{-3}$.}
\label{fig:mllphys}
\end{center}
\end{figure}

For values of $a\gtrsim 1$ ( or $\delta \gtrsim 10^{-3}$ ) the resonances can overlap.
Values of $\Delta M/\bar{\Gamma}$, where $\bar{\Gamma}=(\Gamma_{L^0}+\Gamma_{R^0})/2$, as well as the largest $\Gamma/M$ ratio are shown in \fig{fig:masssplit}. When $\Delta M/\bar{\Gamma}$ approaches unity, the resonances will overlap in the dilepton invariant mass distributions. This is shown in the right panel of \fig{fig:mllphys}. Furthermore the width over mass ratio exceeds unity for large $a$  at which point the effective description breaks down.     

Finally the interference between the signal and the SM background can be relevant. As seen in \fig{fig:mlldet} the CVM features a destructive interference between the resonances and the SM background yielding a dip just before the peaks. If the dip and the resonance peak are summed into one bin obviously this can reduce the observed cross section at the peak. The effect of interference in dilepton resonant searches has been extensively studied in \cite{Accomando:2013sfa}. 

Given the caveats above a sound strategy to set relevant constraints is to consider first the case $\gt\gtrsim 2$ and $a$ not too large. Here the peaks cannot be resolved and an overall cross section constraint can be set. Specifically, we compare the predicted cross section corresponding to the total number of events in the mass range  $M(\ell\ell)>M_R-30\GeV$ to the experimentally observed cross section limit.

In \fig{fig:exclee} we present the CVM dilepton cross section as a function of $M_R$ for different values of $\gt$ and $a$ together with the ATLAS and CMS 95\% exclusion limits with center of mass energy $\sqrt{s}=8\TeV$ and $L\approx 20\ifb$ of integrated luminosity  \cite{CMS-PAS-EXO-12-061,Aad:2014cka}. 
 
\begin{figure}[t!] 
\begin{center}
 \includegraphics[width=.49\columnwidth]{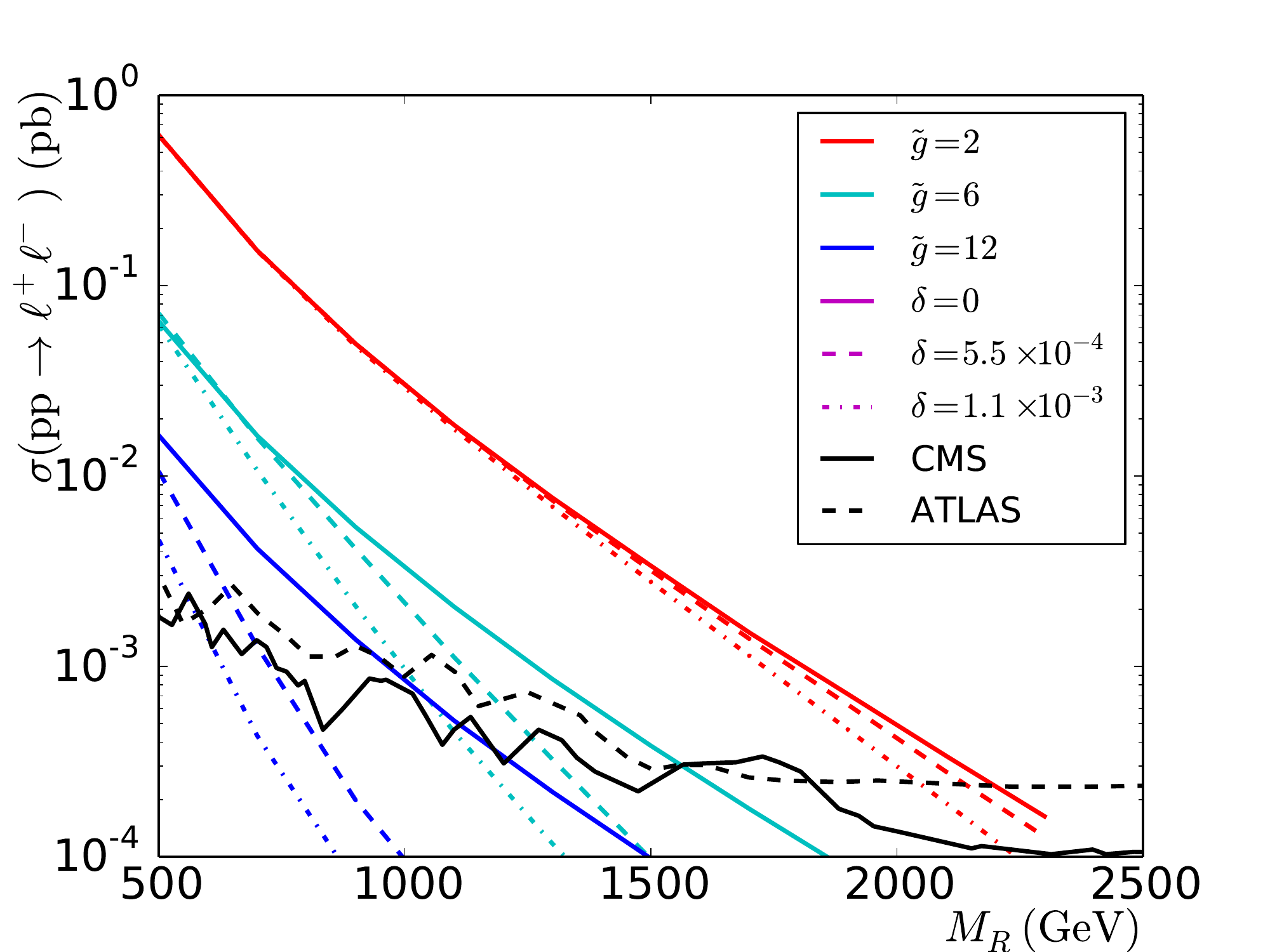}
\caption{Full LHC signal cross section for dilepton $\ell^+\ell^-$ production in the CVM with $\sqrt{s}=8\TeV$ at parton level as a function of $M_R$ for different values of $\gt$ and $\delta$ as given in the figure. Also shown in black are the 95\% exclusion limits provided by ATLAS and CMS \cite{CMS-PAS-EXO-12-061,Aad:2014cka}.}
\label{fig:exclee}
\end{center}
\end{figure}

\subsubsection*{Off-diagonal Widths}
\label{sec:offdiagonal}

In the parameter range where the resonances overlap, their off-diagonal widths can also become important --- i.e. the imaginary and real parts of the vector resonance self-energies cannot be diagonalized simultaneously --- and contribute to the amplitude.  
The basic formalism was recently discussed in \cite{Cacciapaglia:2009ic}, and we review it in appendix~\ref{sec:offdiagonalappendix}.

The contributions of fermion and vector loops  to the  imaginary parts  of the vector self-energies are reported in \cite{Cacciapaglia:2009ic}, while the Higgs contribution from the diagrams in \fig{fig:Hloop} 
\begin{figure}[t!] 
\begin{center}
 \includegraphics[width=.9\columnwidth]{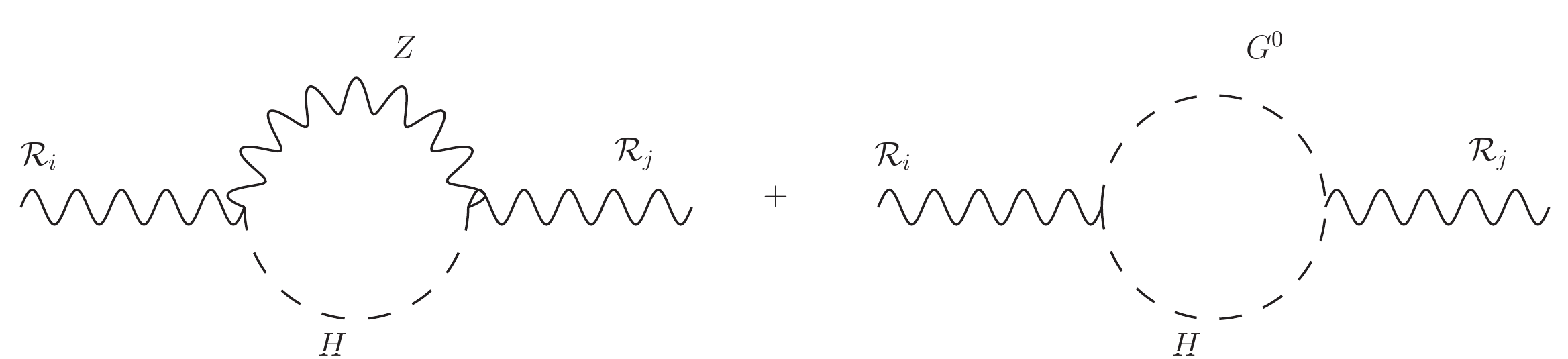}
\caption{One-loop heavy vector self-energy diagrams in the CVM with internal Higgs bosons contributing to the off-diagonal widths.}
\label{fig:Hloop}
\end{center}
\end{figure} 
are given by:
\bea
\Sigma^\mathcal{R}_H(p^2)=(g_{\mathcal{R}_i ZH}) \, \, (g_{\mathcal{R}_j ZH})\frac{\sqrt{\lambda(p^2,M_H^2,M_Z^2)}}{16\pi\,p^2}\left[1+\frac{1}{12\, M_Z^2\,p^2}\lambda(p^2,M_H^2,M_Z^2)\right]\,,
\eea
where $\lambda(x,y,z)=x^2+y^2+z^2-2xy-2yz-2zx$. Notice that these diagrams contribute only to the transverse part of the self energy, $\Pi_T$ of \eq{eq:selfen}.
In the CVM the Higgs contribution dominates the off diagonal widths. 

In \fig{fig:offdiagonal} we illustrate the effect of the off diagonal widths. We show the amplitude squared, summed and averaged over color and spin for the process $u\bar{u}\ra L^0/R^0\ra e^+e^-$ (i.e excluding purely SM contributions from $Z$ and $\gamma$) in three different schemes: In the \emph{Naive} computation the propagators are added with a fixed width; In the \emph{Running W.} computation each propagator is included with the energy dependent width and in the \emph{Full} computation, the complete amplitude including off-diagonal widths is used. The ratio between each scheme to the \emph{Full} amplitude is shown on the bottom inserts. When $\delta$ is large, the difference between the \emph{Naive} and the \emph{Full} computation can be of the order of 50\% close to the resonance peaks.

\begin{figure}[t!] 
\begin{center}
 \includegraphics[width=.49\columnwidth]{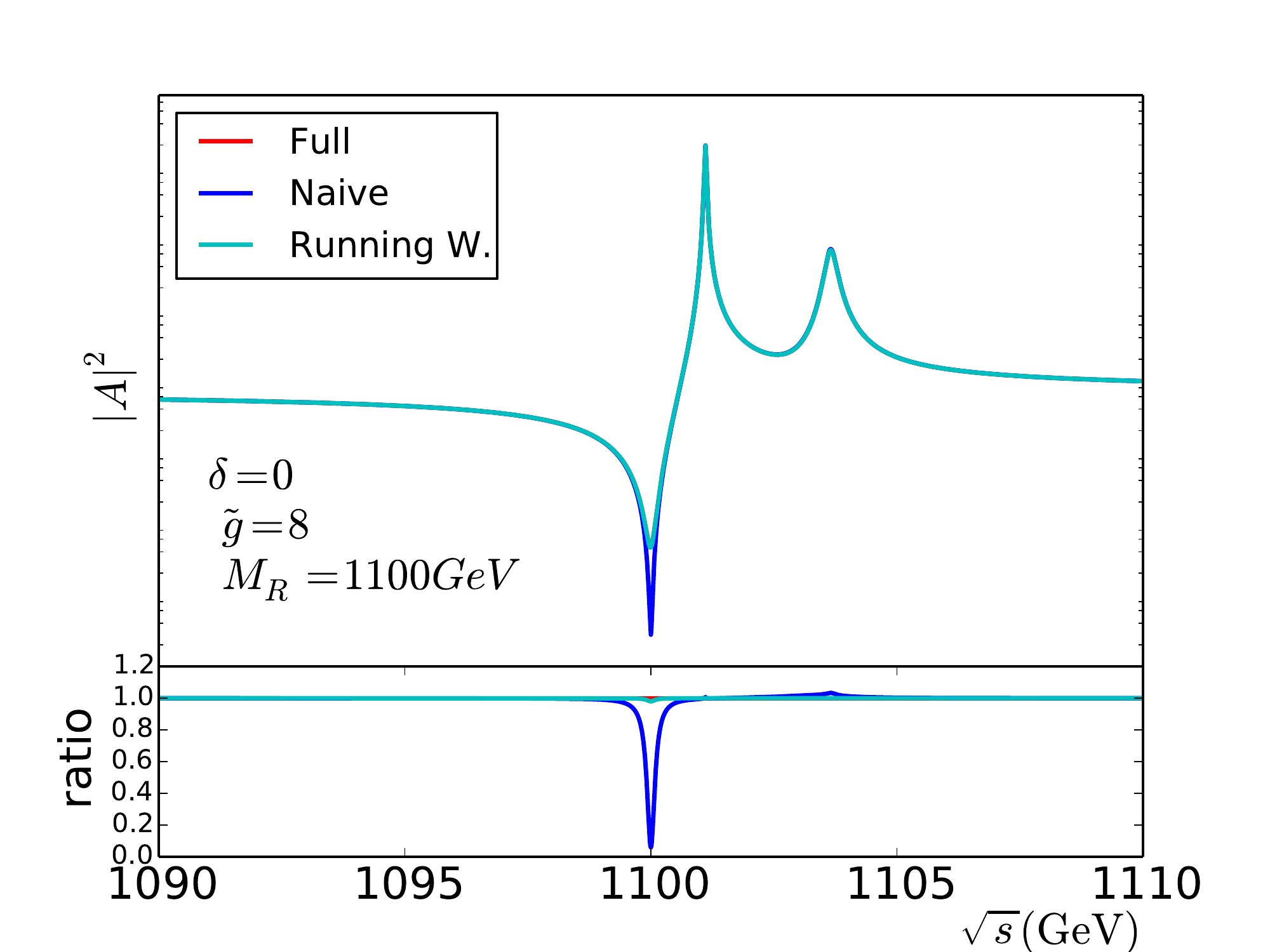}
 \includegraphics[width=.49\columnwidth]{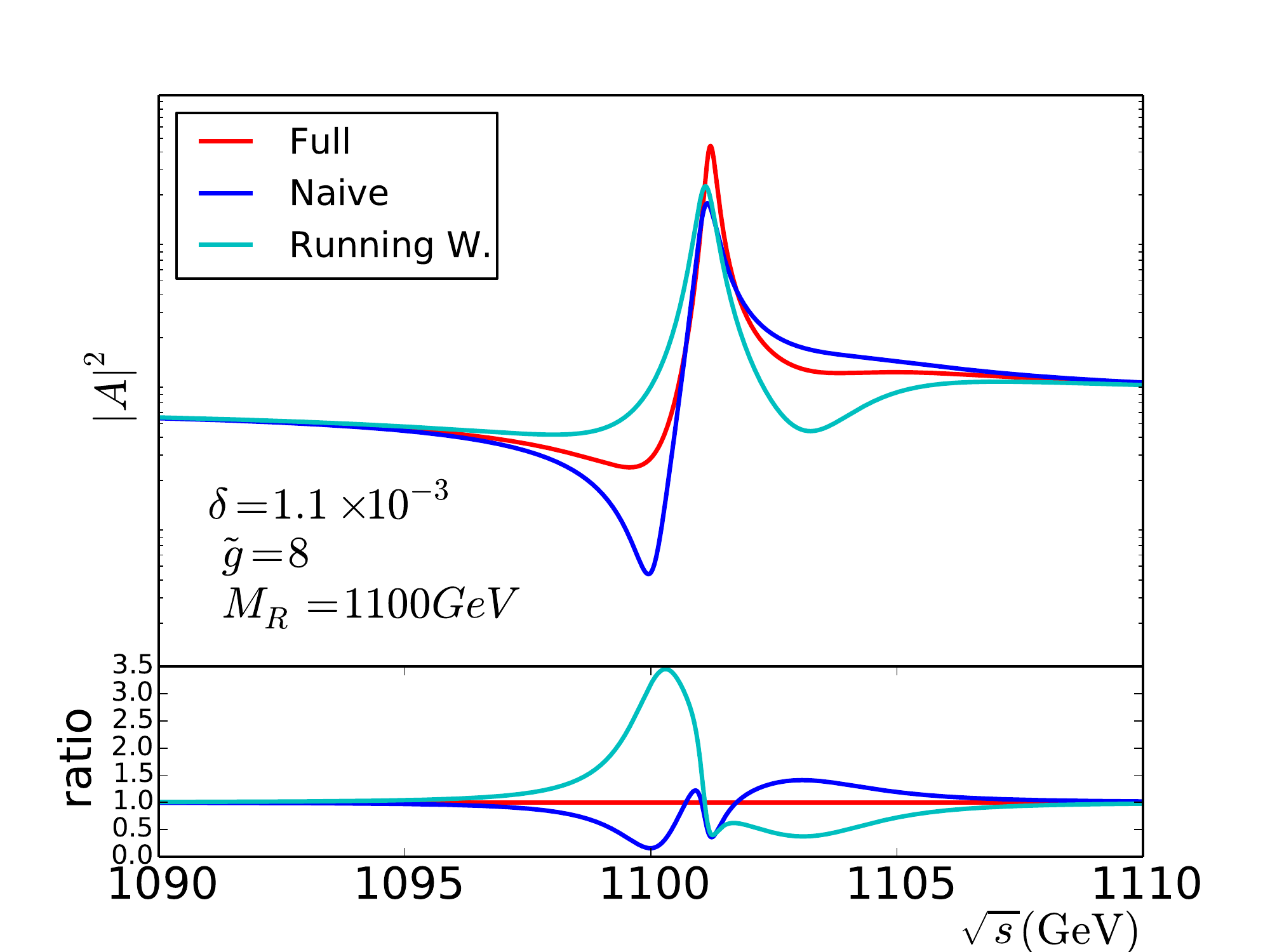}
\caption{The amplitude squared, summed and averaged over color and spin of the process $u\bar{u}\ra e^+e^-$ with the contribution of the two heavy particles and interference (background from $Z$ and photon contribution is subtracted) for three different computational schemes (see text for more details). On the \emph{left} (\emph{right}) panel $\delta=0$ ($\delta=10^{-3}$). The ratio between each scheme to the \emph{Full} amplitude is shown on the bottom inserts. }
\label{fig:offdiagonal}
\end{center}
\end{figure} 

Nevertheless, the corresponding exclusion limits derived with the full scheme are only a bit stronger as can be seen by comparing  \fig{fig:exclee} and \fig{fig:excloffdiag}.
\begin{figure}[t!] 
\begin{center}
 \includegraphics[width=.55\columnwidth]{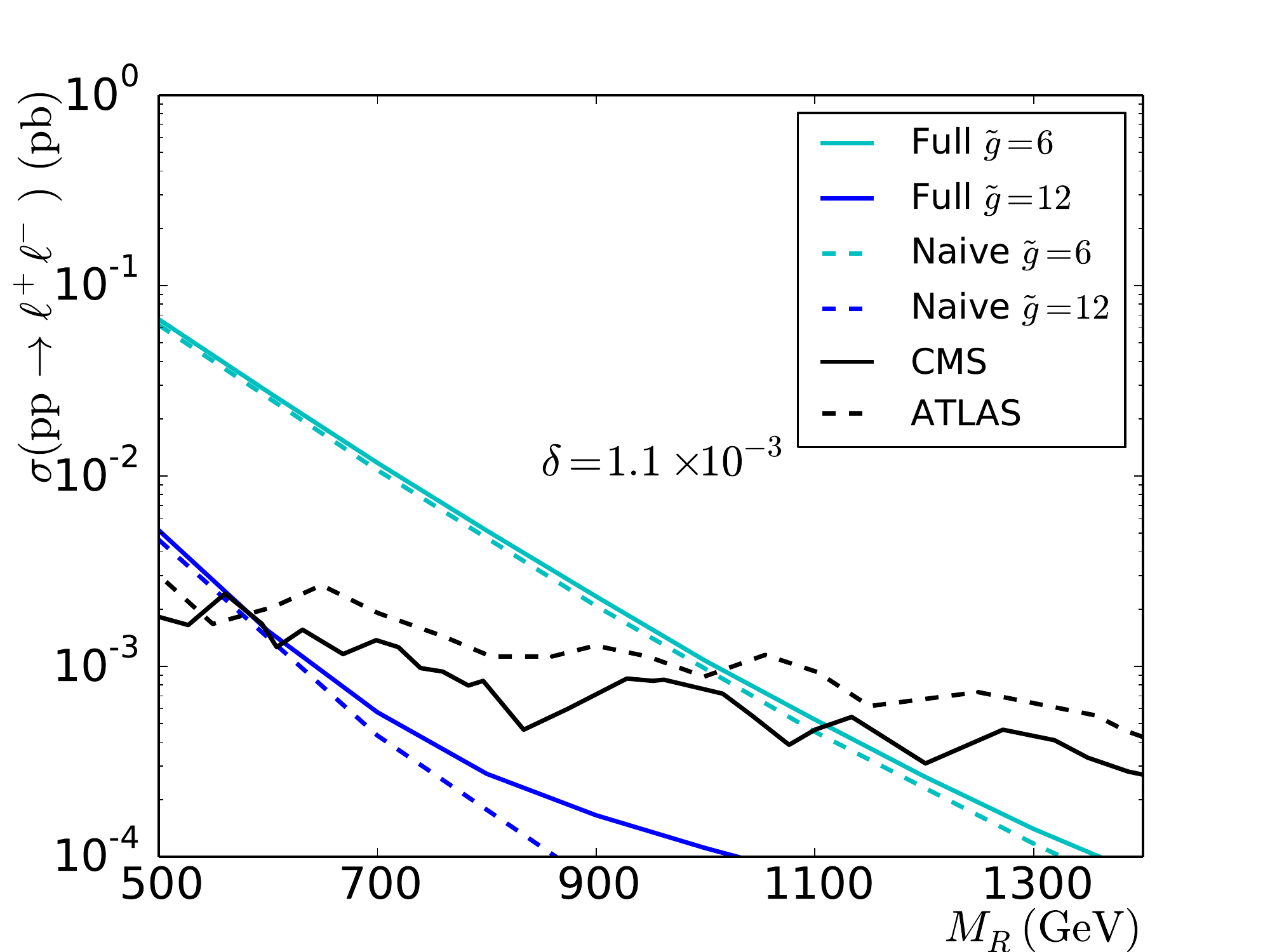}
\caption{ Full LHC signal cross section for dilepton $\ell^+\ell^-$ production in the CVM with $\sqrt{s}=8\TeV$ at parton level, taking into account off-diagonal widths, as a function of $M_R$ for different values of $\gt$ and $\delta$ as given in the figure. Also shown in black are the 95\% exclusion limits provided by ATLAS and CMS \cite{CMS-PAS-EXO-12-061,Aad:2014cka}.
}
\label{fig:excloffdiag}
\end{center}
\end{figure}

\subsection{Single Charged Lepton Searches}
\label{sec:lnu}
In the CVM, only the $L^{\pm}$ resonances contribute to the single charged lepton final states $\ell\nu$. However,
properly accounting for interference effects with the SM states in these channels  is delicate, see e.g.  \cite{Accomando:2011eu}. Due to the final state neutrino, one has to rely on the smeared transverse mass distribution to infer the presence of the new resonance, as opposed to the narrower peaks in the dilepton invariant mass distribution. 
The interference in the low energy part of the transverse mass distribution can be significant. For this reason the CMS collaboration also provides exclusion limits as a function of the minimum transverse mass cut $M_T^{\rm min}$ \cite{Khachatryan:2014tva} \footnote{Notice that the problem is not only how to define the signal region; the importance of the low energy interference will also affect the control regions since they need to be signal free.}.

The corresponding ATLAS analysis \cite{ATLAS:2014wra} does not show exclusion limits as a function of the transverse mass cut. Therefore we only use the CMS limits. The CMS exclusion limit we present is obtained with the 2012 data set of $L = 19.7\ifb$ at $\sqrt{s}=8\TeV$. On the left hand side of \fig{fig:lnuexclusion} the 95\% confidence level exclusion limit as a function of the minimum transverse mass cut is presented together with model predictions for different values of $\gt$ and $\delta$ with $M_R=1$ TeV. On the right hand side the exclusion is translated into the ($M_R$, $\tilde{g}$) plane for fixed values of $\delta$ --- for each value of $M_R$ we choose the value of $M_T^{\rm min}$ that yields the strongest limit. 

\begin{figure}[t!] 
\begin{center}
 \includegraphics[width=.49\columnwidth]{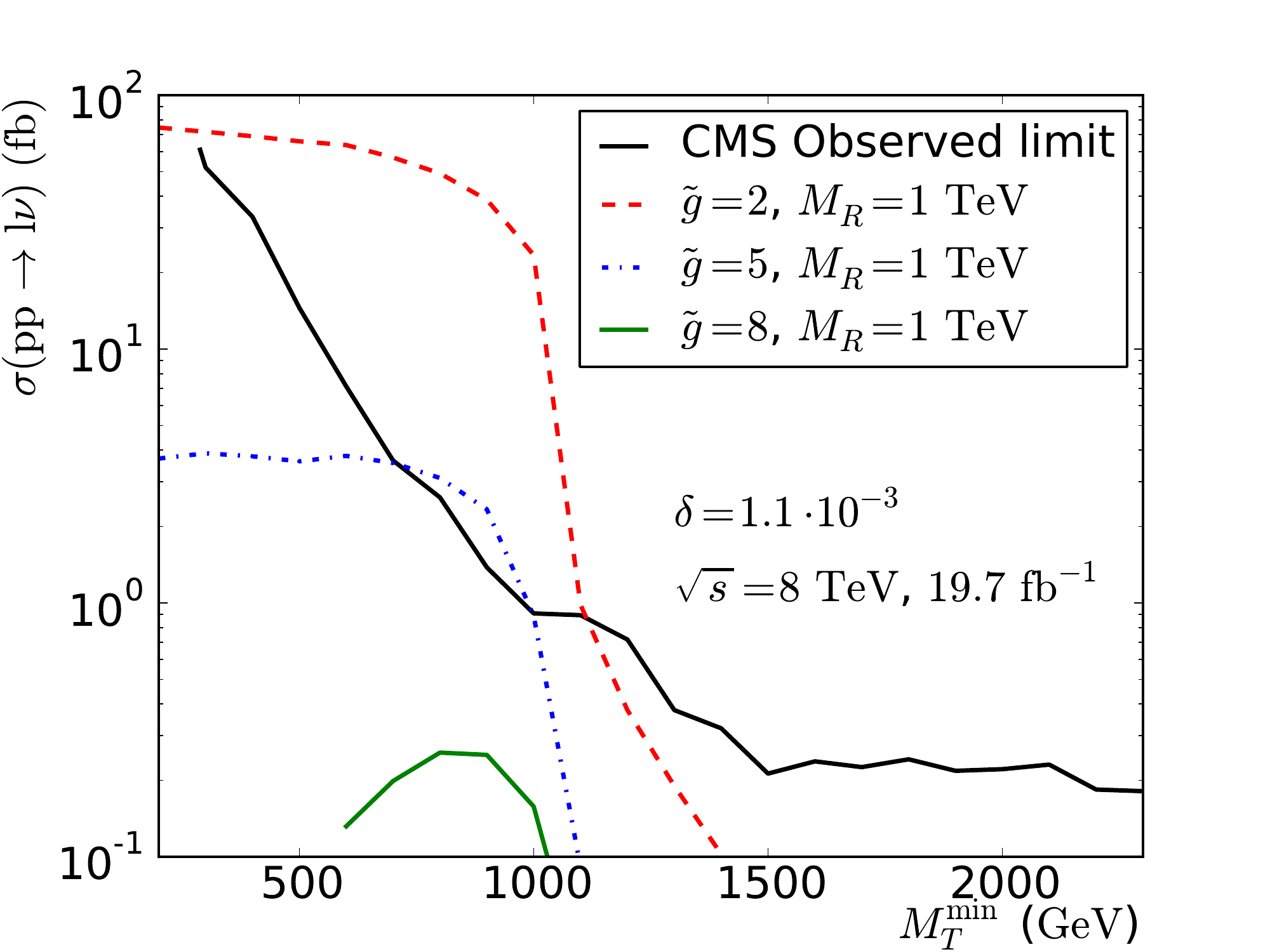}
 \includegraphics[width=.49\columnwidth]{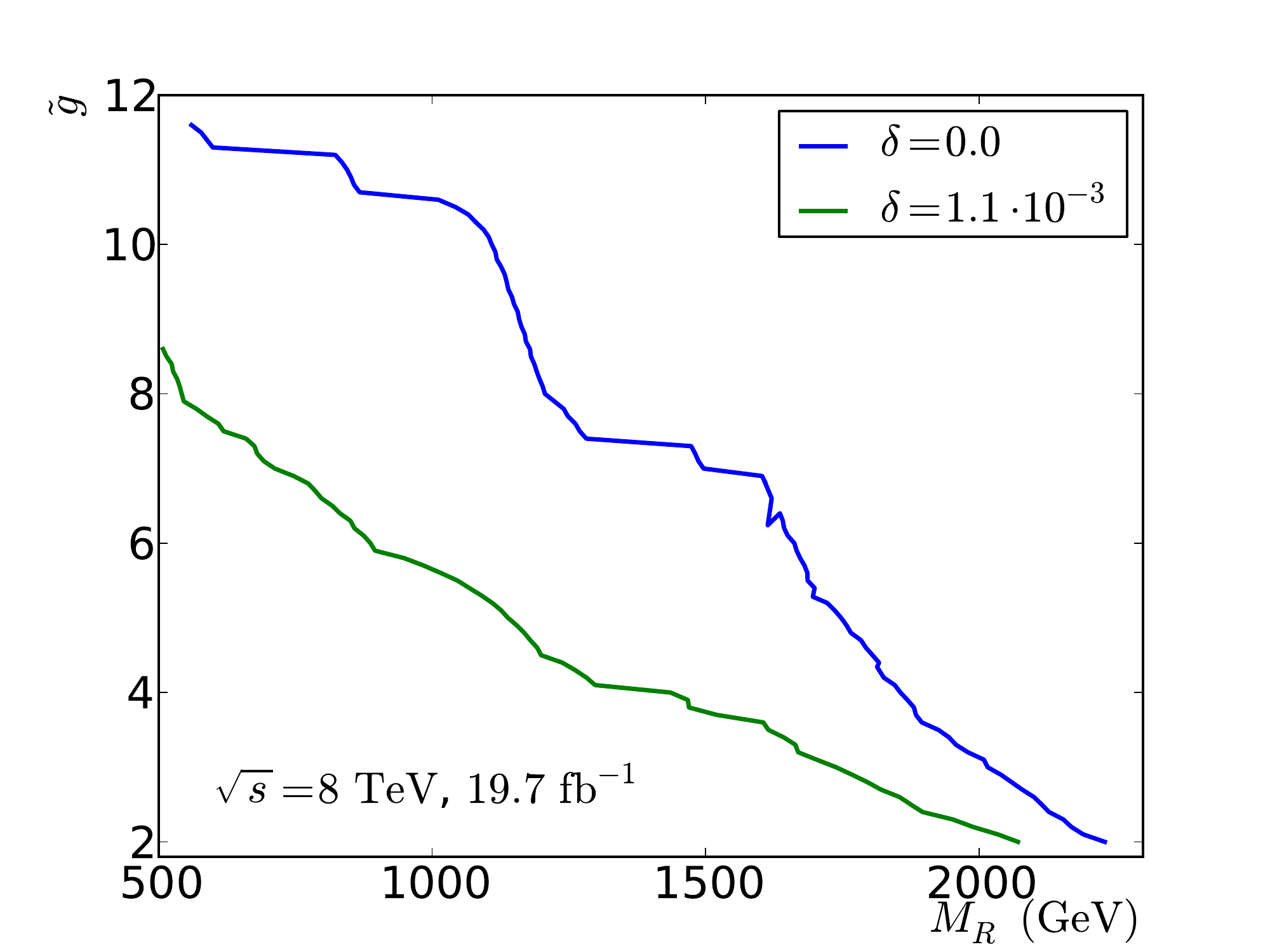}
\caption{Left panel: Full LHC signal cross section for single charged lepton $\ell \nu_\ell$ production in the CVM with $\sqrt{s}=8\TeV$ at parton level for fixed $M_R=1\TeV$ as a function of  the transverse mass cut $M_T^{\rm{min}}$.
Also shown is the 95\% confidence level exclusion limit observed by CMS \cite{Khachatryan:2014tva}. 
Right: The corresponding exclusion limit on the CVM in the $(M_R,\gt)$ plane for two different values of $\delta$. }
\label{fig:lnuexclusion}
\end{center}
\end{figure}

\subsection{Associated Higgs Searches}
\label{sec:Hassoc}

Current searches for the production of the Higgs state in association with a SM vector boson also yield relevant bounds on the CVM parameter space. The ATLAS and CMS collaborations provide upper bounds on the signal strenght $\mu=\sigma/\sigma_{SM}$, for the processes $pp\ra H(b\bar{b})Z(\ell^+\ell^-)$, $pp\ra H(b\bar{b})Z(\nu\nu)$ and $pp\ra H(b\bar{b})W(\ell\nu,\tau\nu)$ \cite{PhysRevD.89.012003,Aad:2014xzb}. 

In the CVM the final state vector resonance can be any of the states $V=Z,W^\pm$ or  $\mathcal{R}_i=L^{0,\pm}, R^0$ and the relevant diagrams are shown in \fig{fig:diagHW} and \fig{fig:diagHZ}. 
The largest contribution to the CMS analysis of the $pp\ra H(b\bar{b})W(\ell\nu,\tau\nu)$ channel typically comes from $H L^{\pm}$ production even though it is phase space suppressed with respect to $H W ^{\pm}$. This is due to the large $H L^{\pm} L^{\mp}$ coupling (\eq{Eq:HiggscouplingsLL}). Moreover, the kinematical cuts employed in the analysis tend to enhance the high energy region and consequently the new physics contribution. The CMS best-fit signal strength with 1-sigma errors is
\begin{equation}
\mu(W(\ell\nu,\tau\nu) H)=1.1\pm 0.9 \,.
\end{equation}
In the corresponding ATLAS analysis of $pp\ra H(b\bar{b})W(\ell\nu,\tau\nu)$ the transverse mass system associated with the $W$ boson (lepton and missing energy) is required to be small, $m_T^W<120\GeV$, which strongly reduce contributions from the CVM vector resonances. Therefore, we use the CMS result to set limits on the CVM parameter space. 

For the $pp\ra H(b\bar{b})Z(\nu\nu)$ search the $H L^0$ channel gives the largest contribution of new physics in both the CMS and ATLAS analysis. 
Again we choose to use the CMS result
\begin{equation}
\mu(Z(\nu\nu) H)=1.0\pm0.8 \ ,
\end{equation}
to impose limits on the CVM\footnote{The ATLAS result in this search channel gives the unphysical result $\mu(Z(\nu\nu) H)=-0.3\pm 0.5$ which essentially exludes both the SM and CVM at the 95\% of confidence level. However since a 2-sigma level deficit is also observed in the control sample $\mu(Z(\ell \ell) Z(b\bar{b}))$ and no deficit is observed in the search channel $\mu(Z(\ell \ell) H)$ we disregard the result. }.

In the search for $pp\ra H(b\bar{b})Z(\ell^+\ell^-)$ the mass of the dilepton system is required to be near $M_Z$ and therefore the $H L^0/H R^0$ channels are highly off-shell and suppressed. It is thus neglected here. 
It would be very interesting to consider a dedicated analysis looking for resonances in the dilepton mass system in this search channel as proposed in \cite{Hoffmann:2014aha,Zerwekh:2005wh,Belyaev:2008yj,Barger:2009xg,Hernandez:2010iu}. 

To set our limits we use the total CVM cross-section in the associated Higgs channels. We again believe this yields a conservative limit since the cuts employed in \cite{PhysRevD.89.012003} select high energy events and enhance the new CVM contribution with respect to the SM. 

\begin{figure}[t!] 
\begin{center}
 \includegraphics[width=.32\columnwidth]{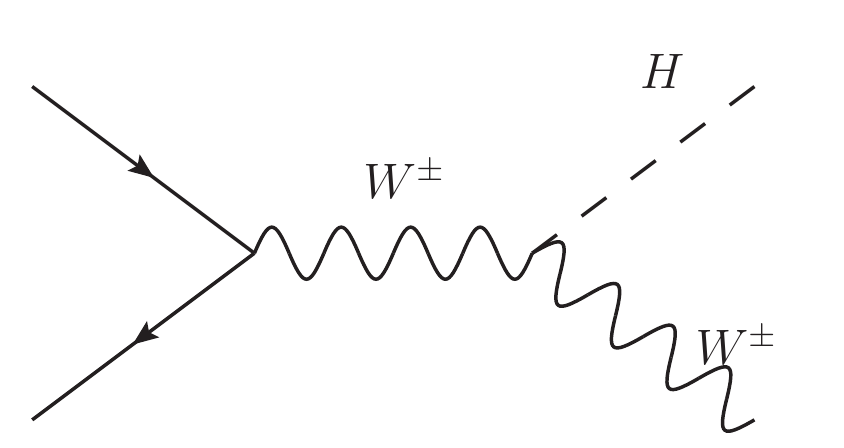}
 \includegraphics[width=.32\columnwidth]{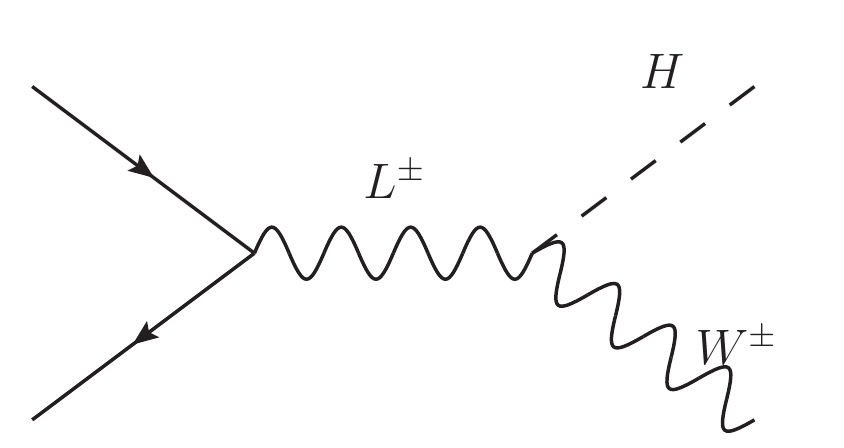}
  \includegraphics[width=.32\columnwidth]{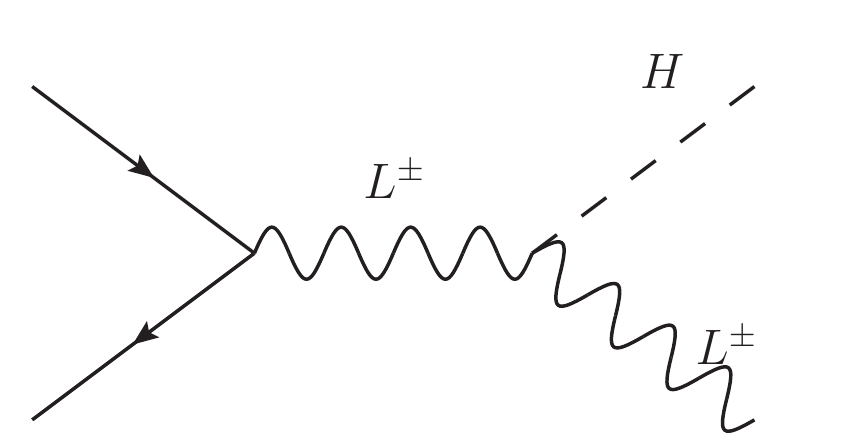}
\caption{Feynman diagrams contributing to Higgs production in association with charged vectors in the CVM.}
\label{fig:diagHW}
\end{center}
\end{figure}

\begin{figure}[t!] 
\begin{center}
 \includegraphics[width=.32\columnwidth]{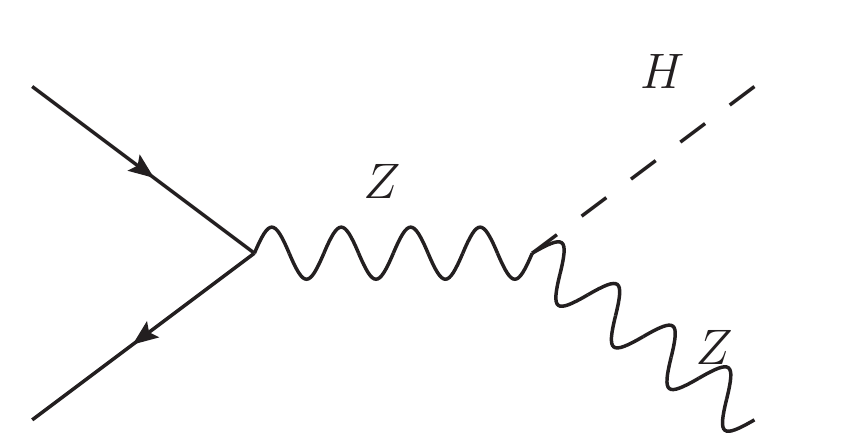}
 \includegraphics[width=.32\columnwidth]{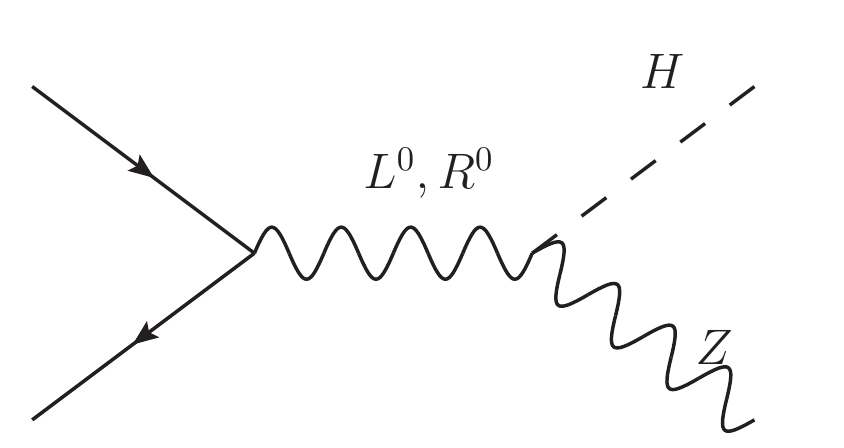}
 \includegraphics[width=.32\columnwidth]{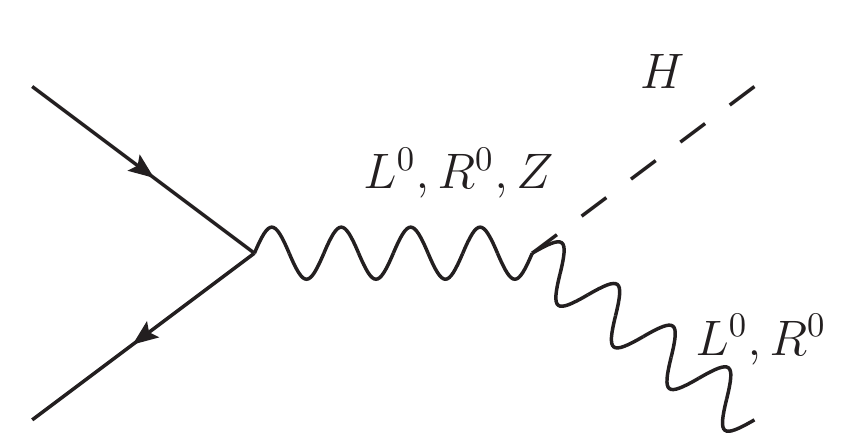}
\caption{Feynman diagrams contributing to Higgs production in association with a neutral vectors in the CVM.}
\label{fig:diagHZ}
\end{center}
\end{figure}

On the left-hand side of \fig{fig:VHexclusion} we show the predicted signal strength, $\mu=\sigma/\sigma_{SM}$ in the $pp\ra H \ell\nu$ channel for different CVM parameters. The exclusion limit on $\mu$ shown in the figure comes from the measurement of $H(b\bar{b})W(\tau\nu,\,\ell\nu)$ at CMS. Analogously, the signal strength of the $pp\ra H\nu\bar{\nu}$ process in CVM is shown on the right hand side of \fig{fig:VHexclusion} with the corresponding exclusion limits derived from the $H(b\bar{b})Z(\nu\bar{\nu})$ channel. 

\begin{figure}[t!] 
\begin{center}
\includegraphics[width=.49\columnwidth]{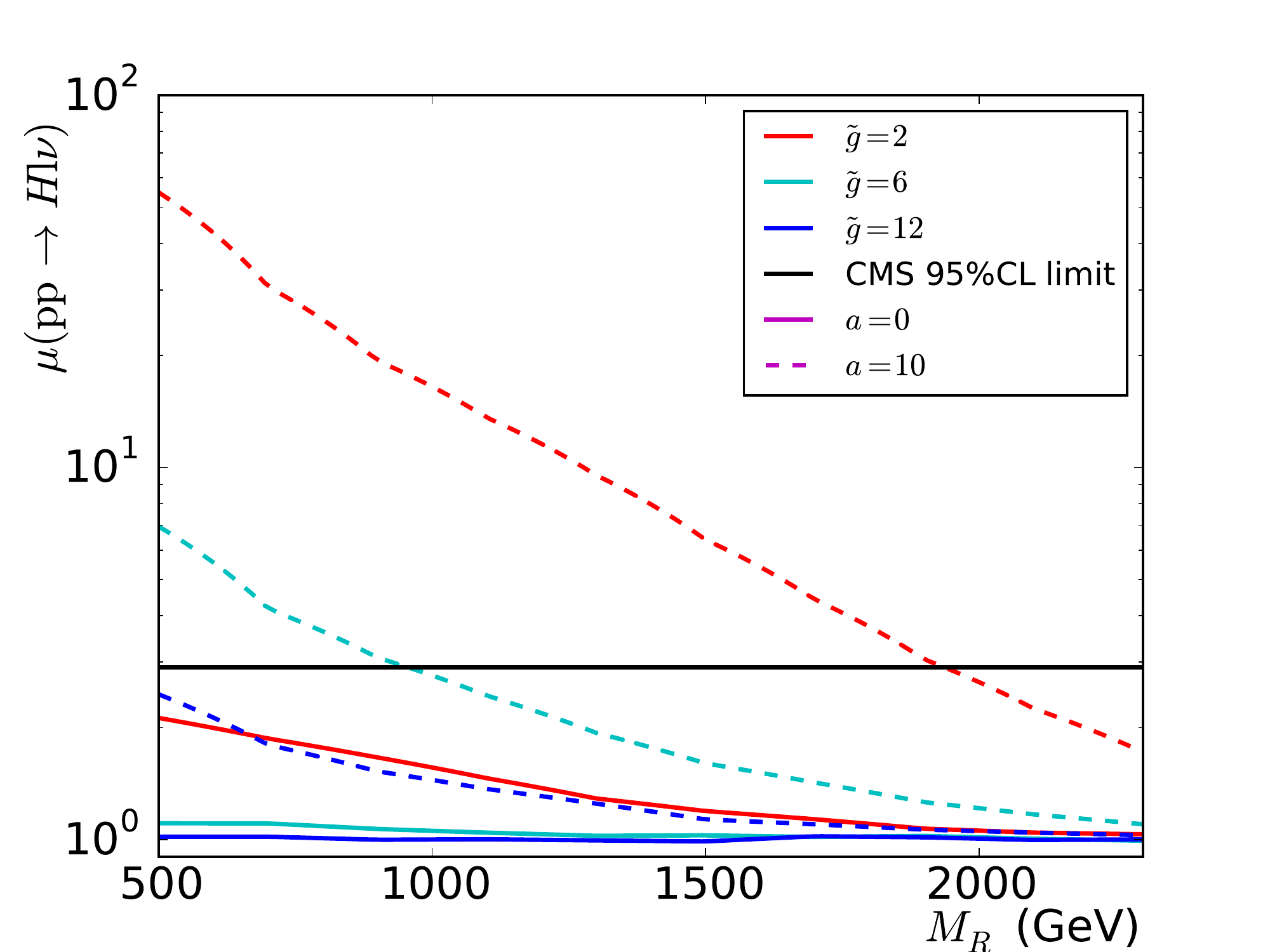}
\includegraphics[width=.49\columnwidth]{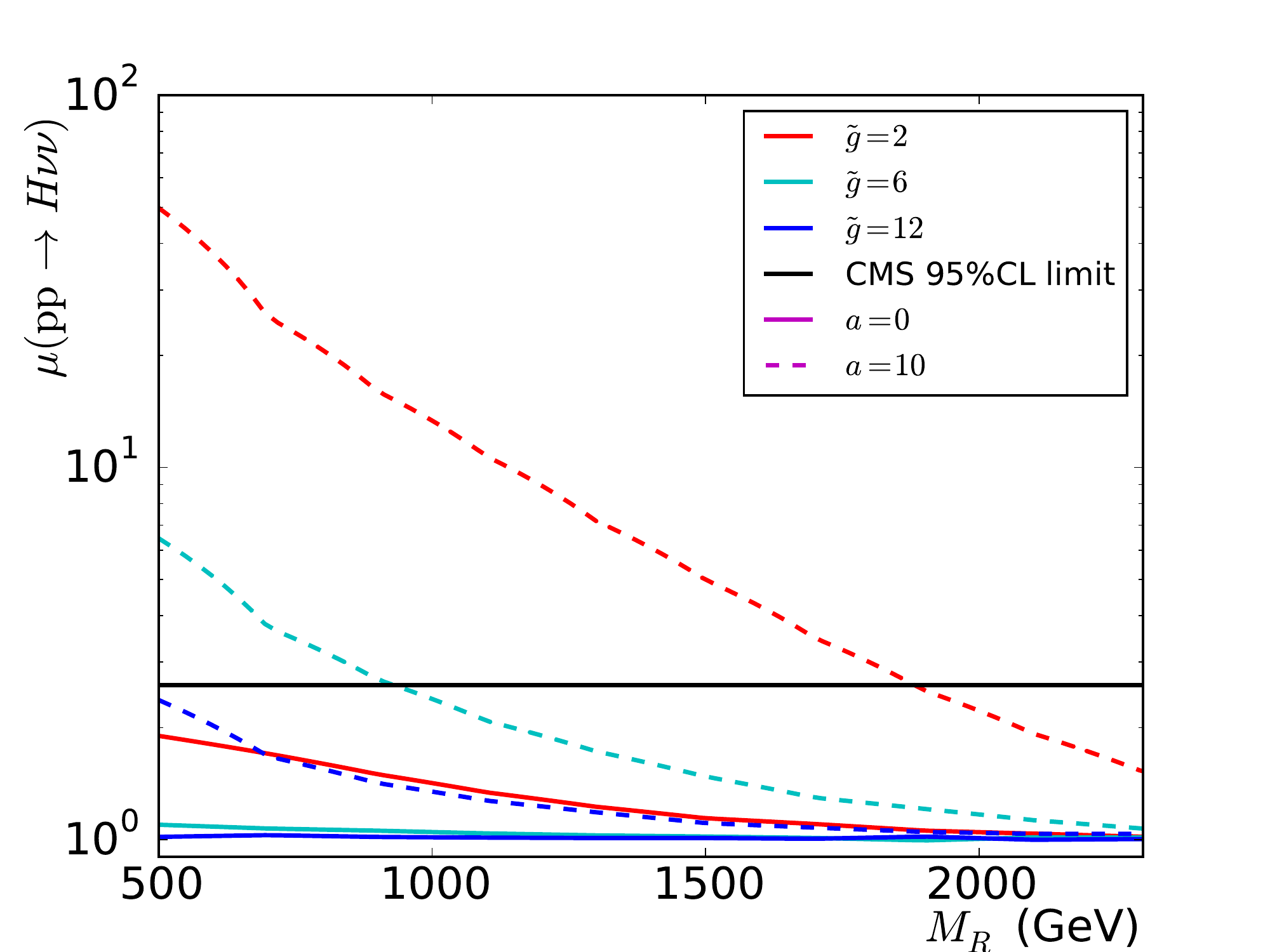}
\caption{The value of the signal strength $\mu=\sigma_{\textrm{CVM}}/\sigma_{\textrm{SM}}$ at LHC with $\sqrt{s}=8\TeV$ for the processes $pp\ra H\ell\nu$ (left) and $pp\ra H\nu\nu$ (left) as a function of $M_R$ for different values of the CVM parameters $\gt,a$. Also shown are exclusion limits from CMS on 
 $\mu$ in the $W(\ell\nu,\tau\nu)H(b\bar{b})$ channels (black).}
\label{fig:VHexclusion}
\end{center}
\end{figure}
As expected the limits are stronger than the ones from dilepton searches for large values of $a$. Moreover, a dedicated resonance search in these channels could provide more stringent limits on the parameter space. Or better, the chance to discover the interplay of multiple resonances with the Higgs.

\subsection{Parameter space}
\label{sec:combination}
We end this section by studying the allowed regions in the $M_R,\tilde{g}$ and $a$ parameter space given the constraints from dilepton $\ell^+ \ell^-$ (blue curves), single-charged lepton $\ell^\pm +\slashed{E}_T$ (magenta curves) and associated Higgs searches\footnote{We use a simple $\chi^2$-analysis to combine the $H\ell \nu$ and $H\nu\nu$ channels.} (red curves). In some of the plots we show the parameter $\delta$ instead of $a$. 

The allowed and excluded regions at 95\% CL are shown as the white and striped regions respectively in $(M_R,\tilde{g})$ planes in \fig{fig:Excl95delta} for fixed values of $a$ or $\delta$. For $a=0$ only the dilepton and (sub dominantly) the single charged lepton searches significantly constrain the parameter space as shown in the upper left panel of \fig{fig:Excl95delta}. However, a dedicated study may put further constraints via the non-zero $HL^\pm L^\pm$ interaction giving rise to diagram 3 in \fig{fig:diagHW}. As $a$ is dialed up, associated Higgs production starts to compete with the dilepton searches as shown in the 3 remaining panels. In particular for $a\gtrsim 20$ ($ | \delta | \gtrsim 10^{-3}$) the associated Higgs production provide the strongest constraint over most of the parameter space shown. 

\begin{figure}[t!] 
\begin{center}
\includegraphics[width=.49\columnwidth]{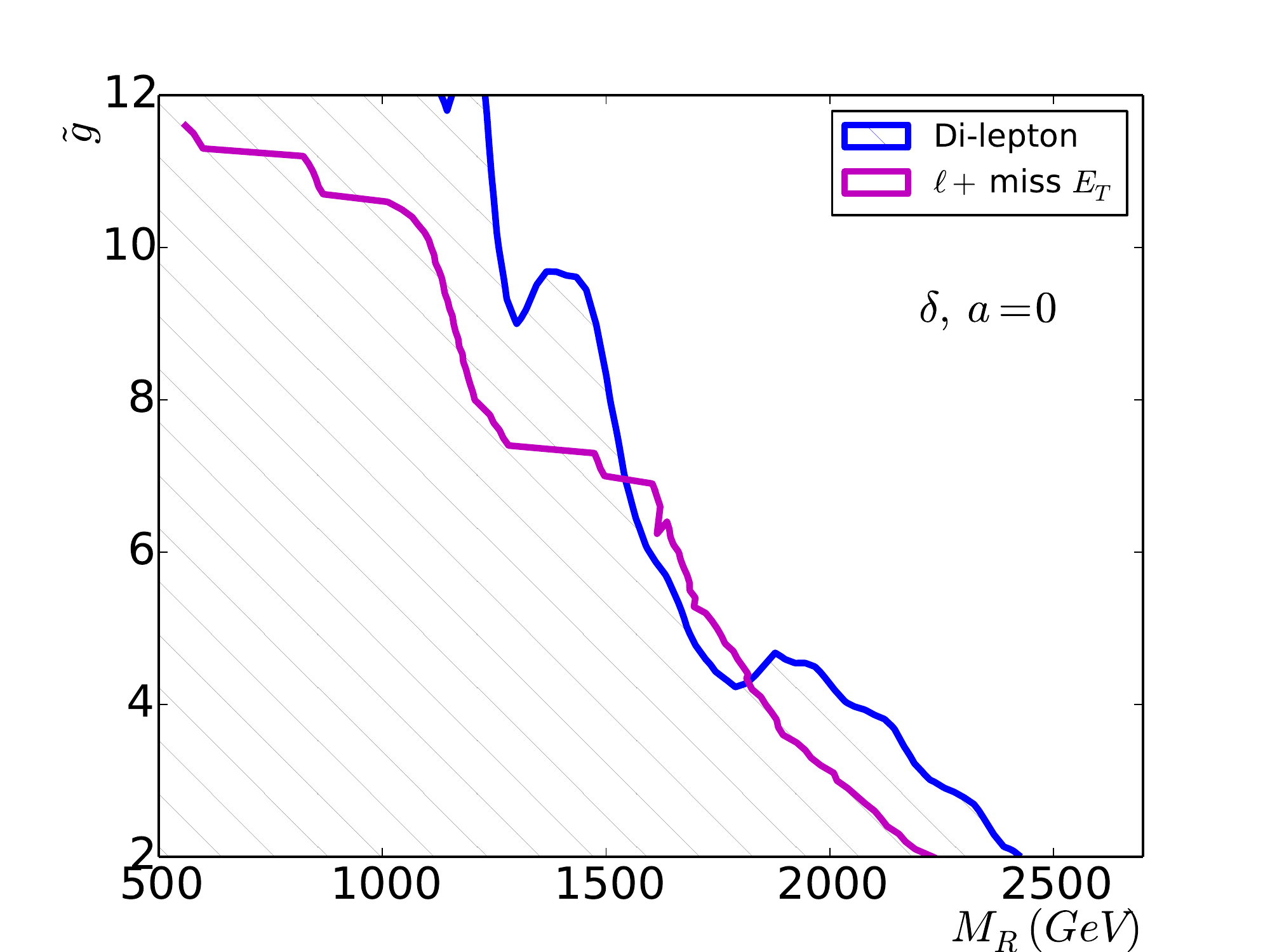}
\includegraphics[width=.49\columnwidth]{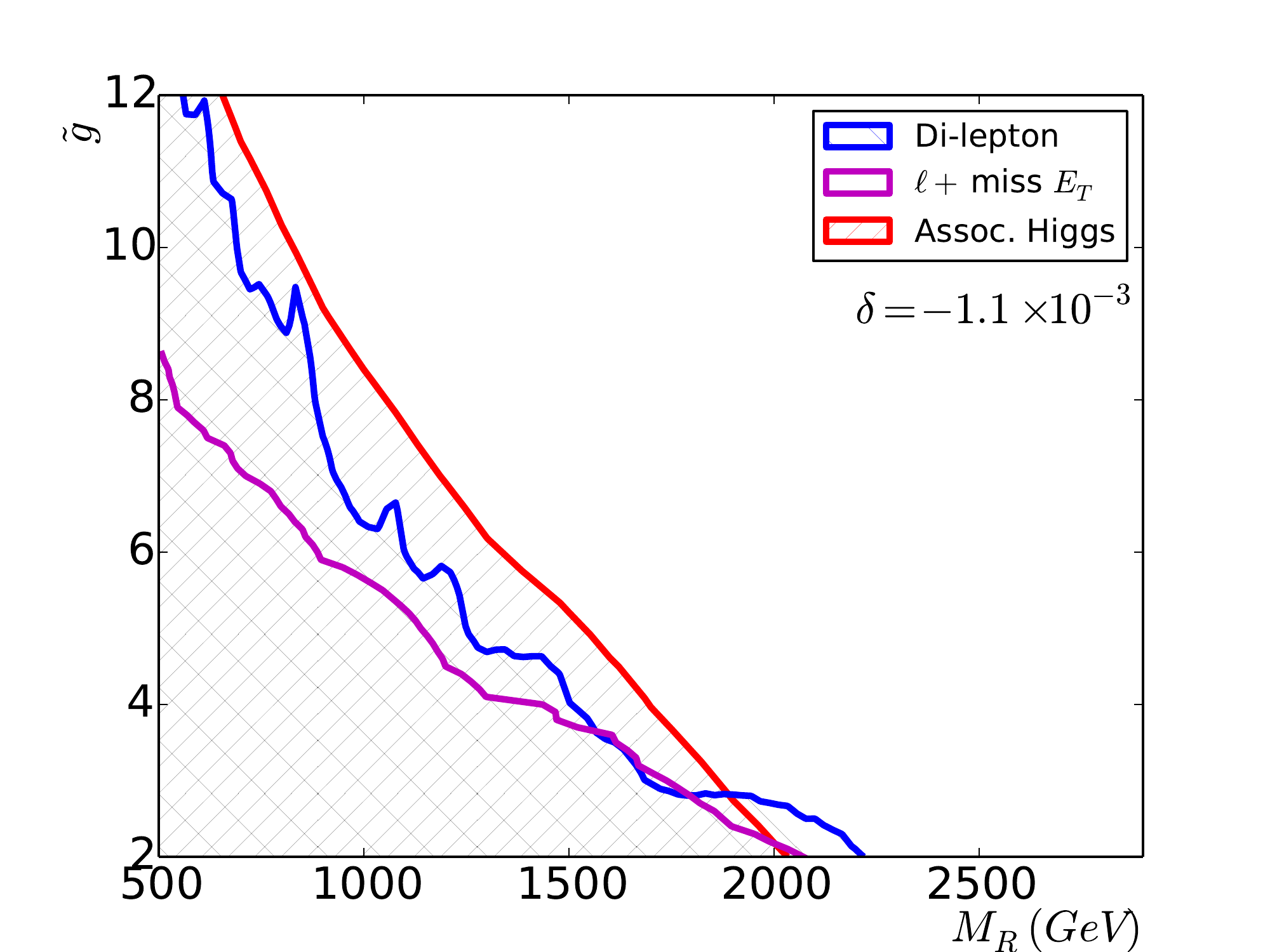}
\includegraphics[width=.49\columnwidth]{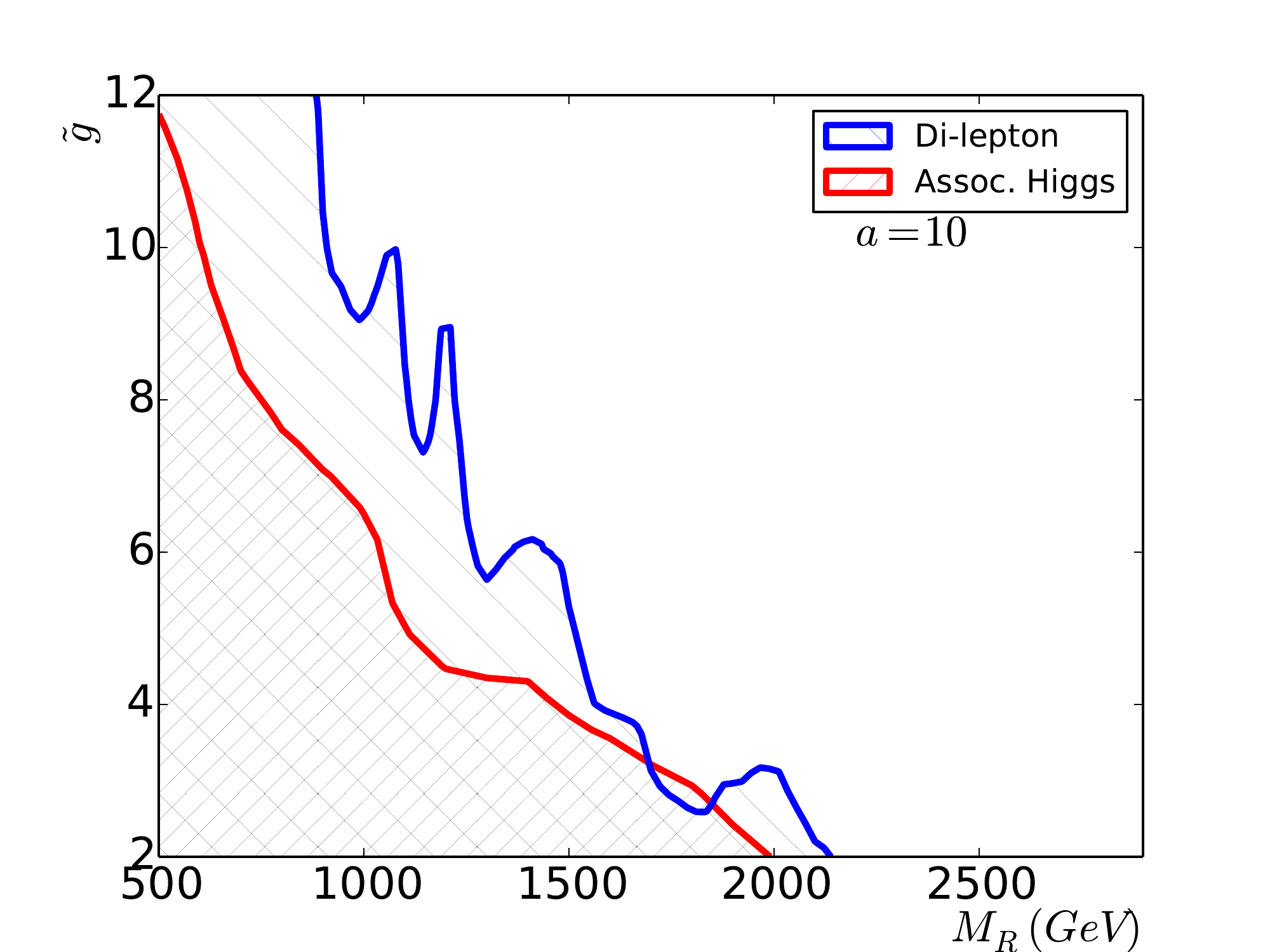}
\includegraphics[width=.49\columnwidth]{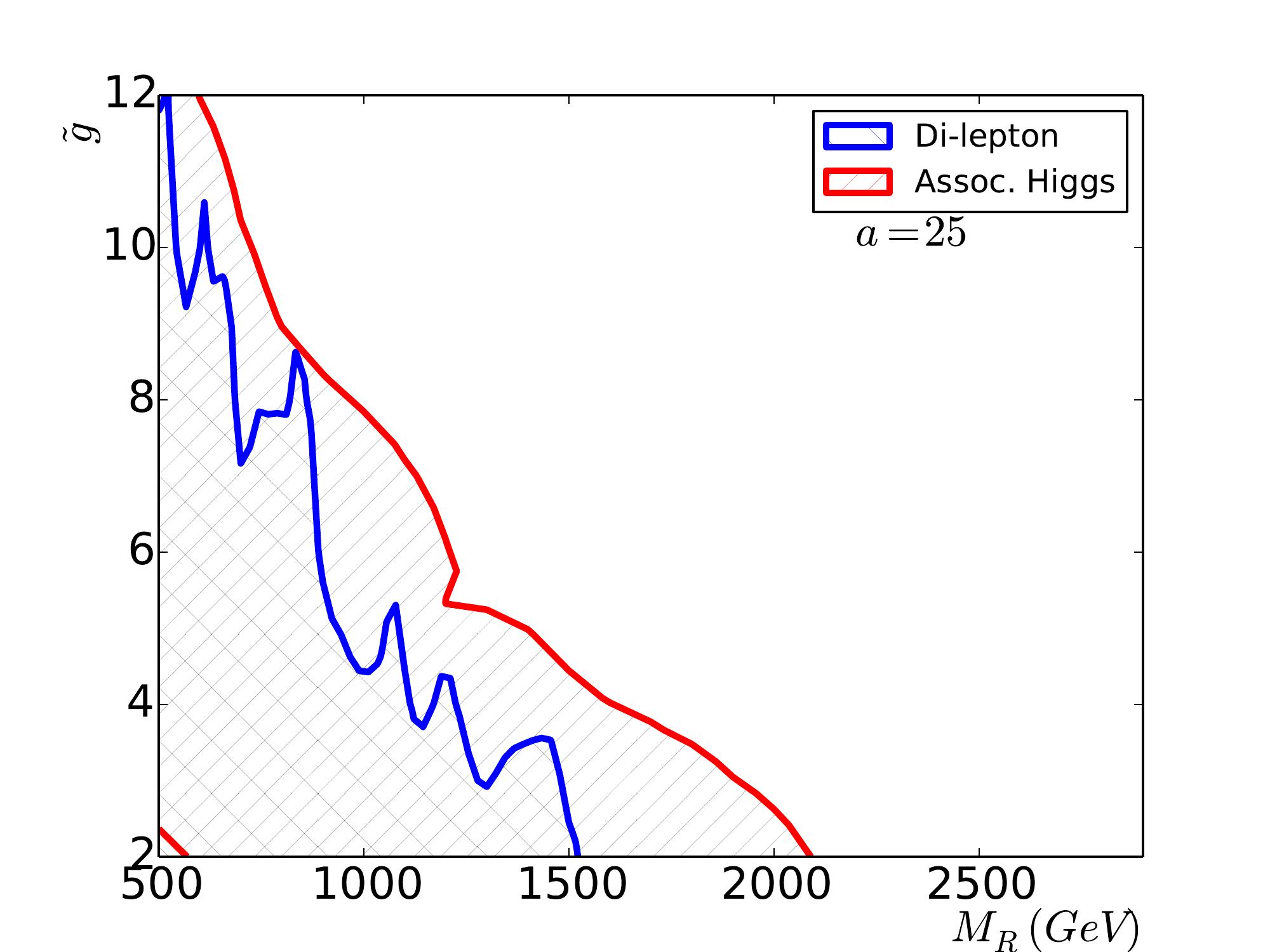}
\caption{$95\%$ exclusion limits on the CVM from LHC shown in $(M_R,\,\tilde{g})$ planes for fixed $\delta$  (\emph{upper}) and fixed $a$  (\emph{lower}) values. Shown are the limits from dilepton searches (blue), limits from single charged lepton searches (purple) and limits for associated Higgs production (red). The striped and cross striped regions are excluded. 
}
\label{fig:Excl95delta}
\end{center}
\end{figure}

The same can be seen from the exclusion limits in the $(\delta,\,\tilde{g})$ and $(a,\,\tilde{g})$ planes for different values of $M_R$ shown in \fig{fig:Excl95MR}. 
\begin{figure}[t!] 
\begin{center}
\includegraphics[width=.49\columnwidth]{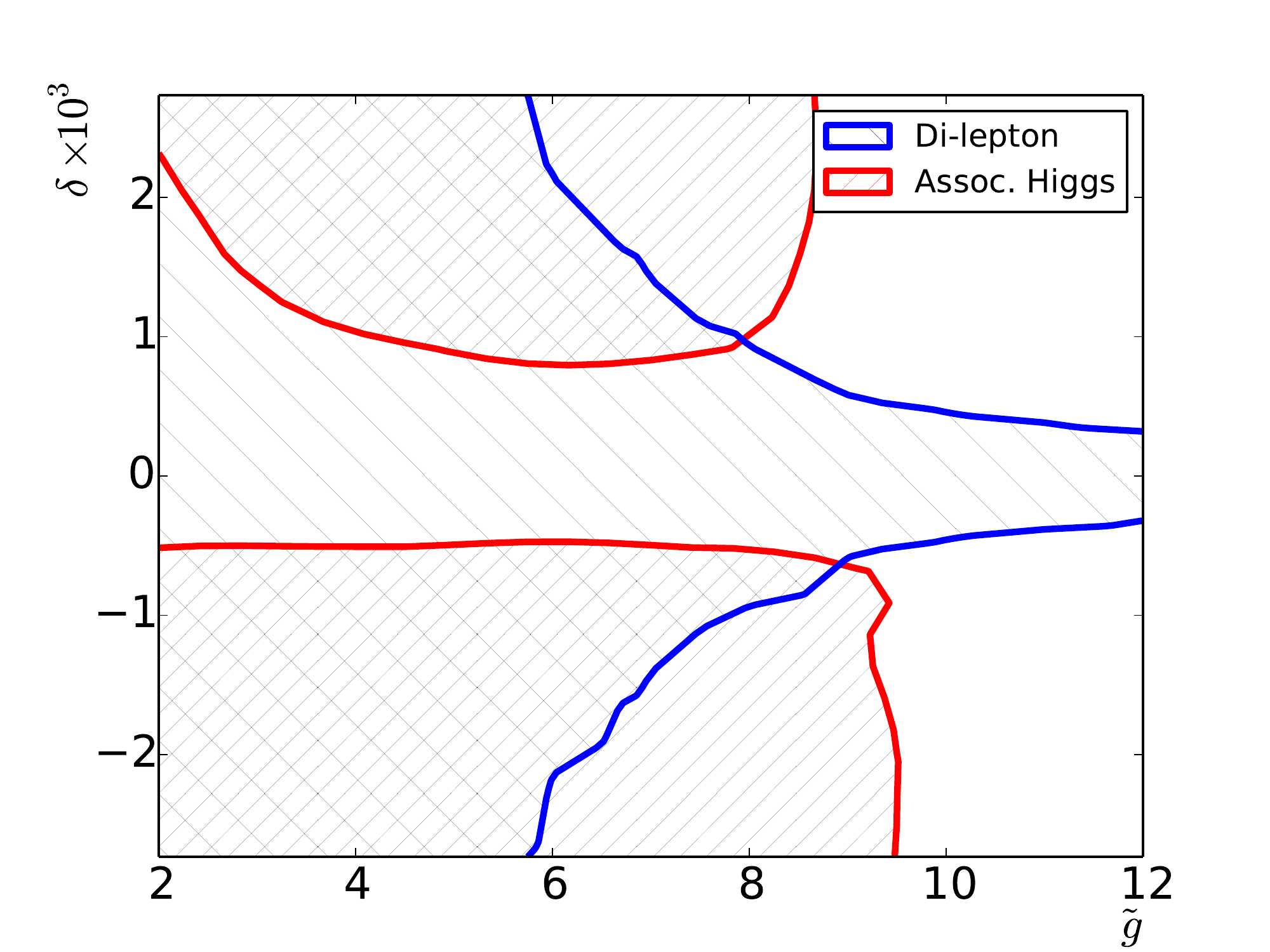}
\includegraphics[width=.49\columnwidth]{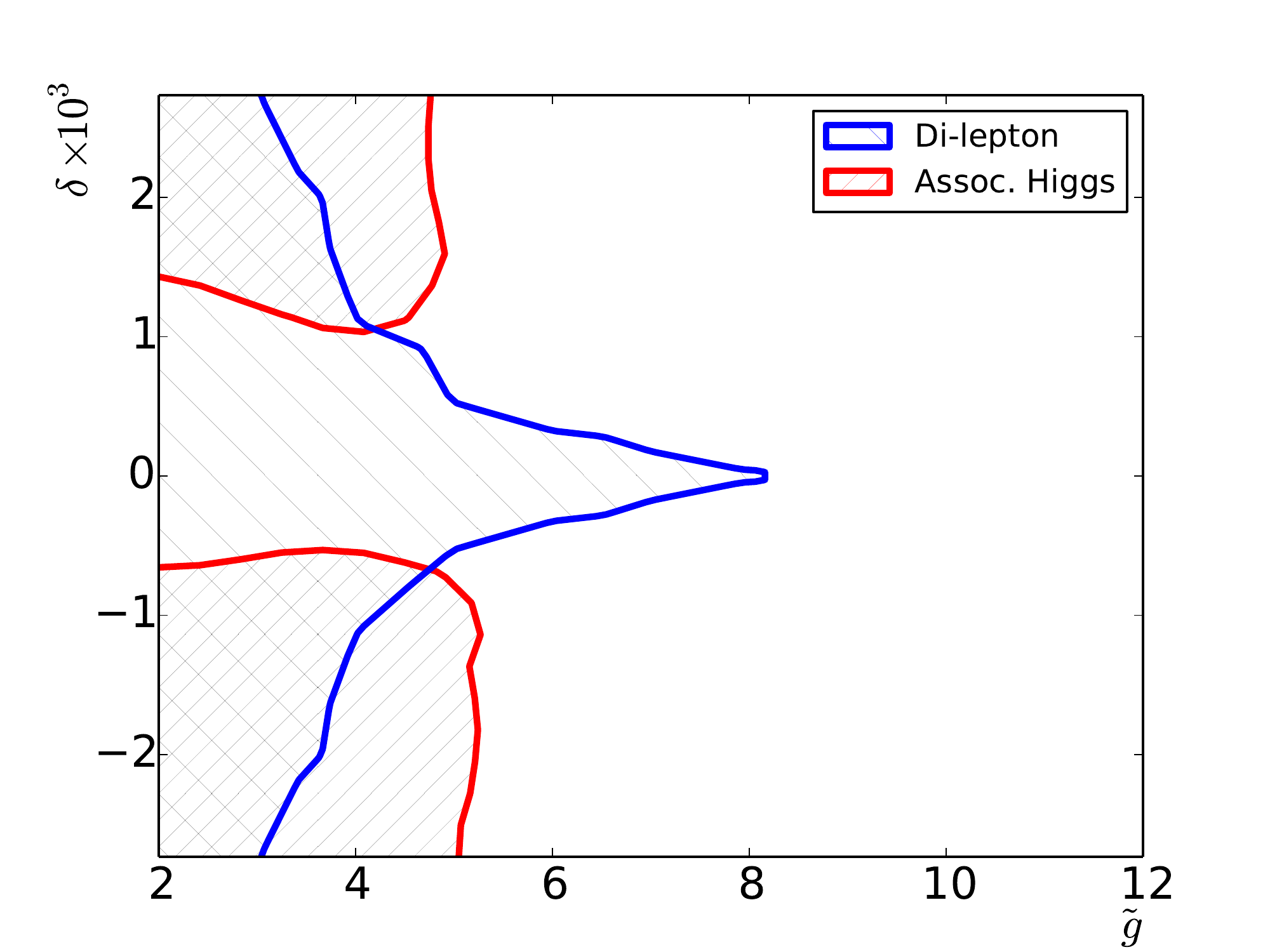}
\includegraphics[width=.49\columnwidth]{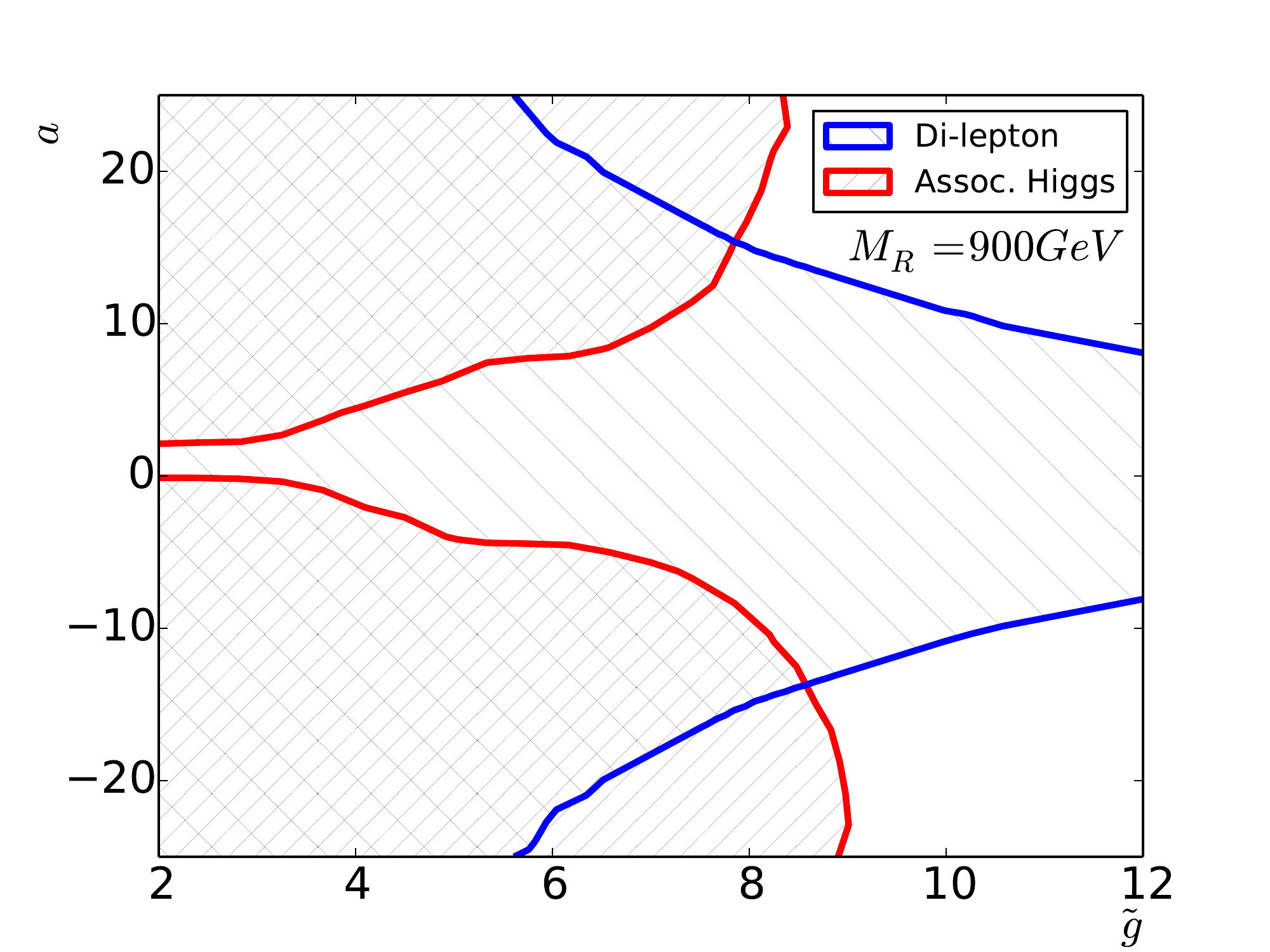}
\includegraphics[width=.49\columnwidth]{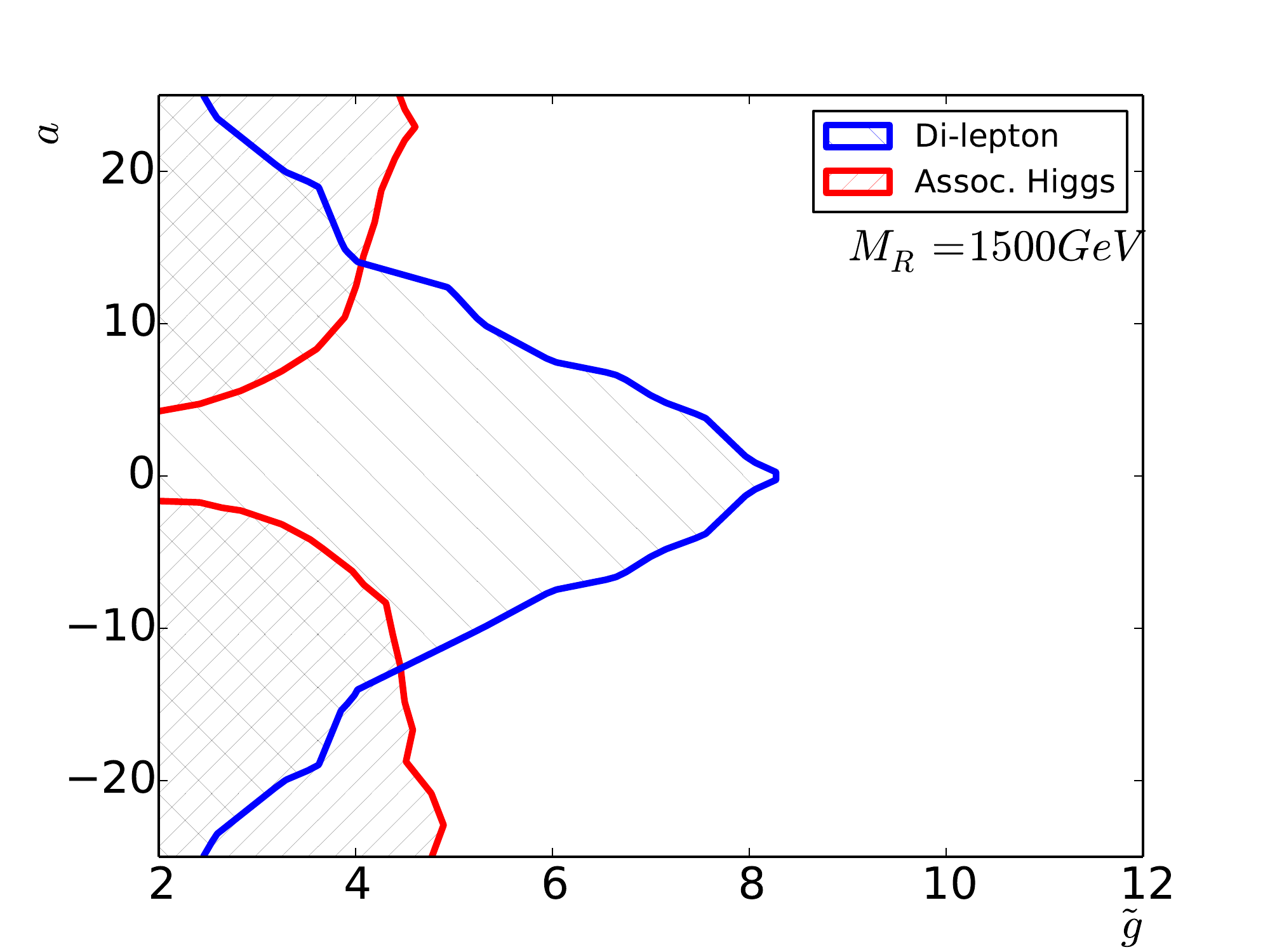}
\caption{$95\%$ exclusion limits as above, but in $(\delta,\,\tilde{g})$ (\emph{left}) and $(a,\,\tilde{g})$ (\emph{right}) planes for fixed $M_R$ values. 
}
\label{fig:Excl95MR}
\end{center}
\end{figure}
Finally, in \fig{fig:Excl95gt} we show the regions in $(M_R, a)$ and $(M_R,\delta)$ planes for different values of $\gt$.
\begin{figure}[t!] 
\begin{center}
\includegraphics[width=.49\columnwidth]{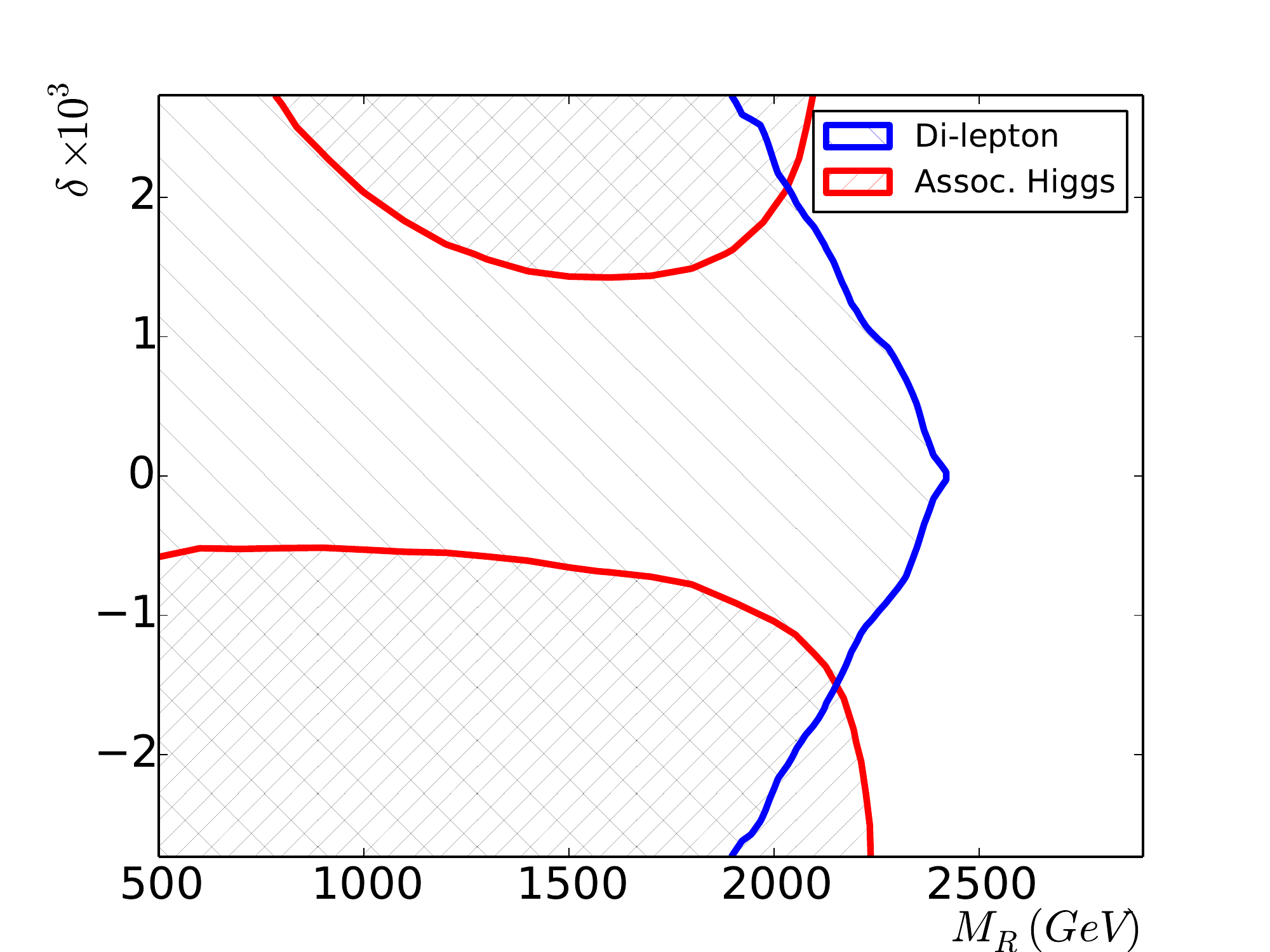}
\includegraphics[width=.49\columnwidth]{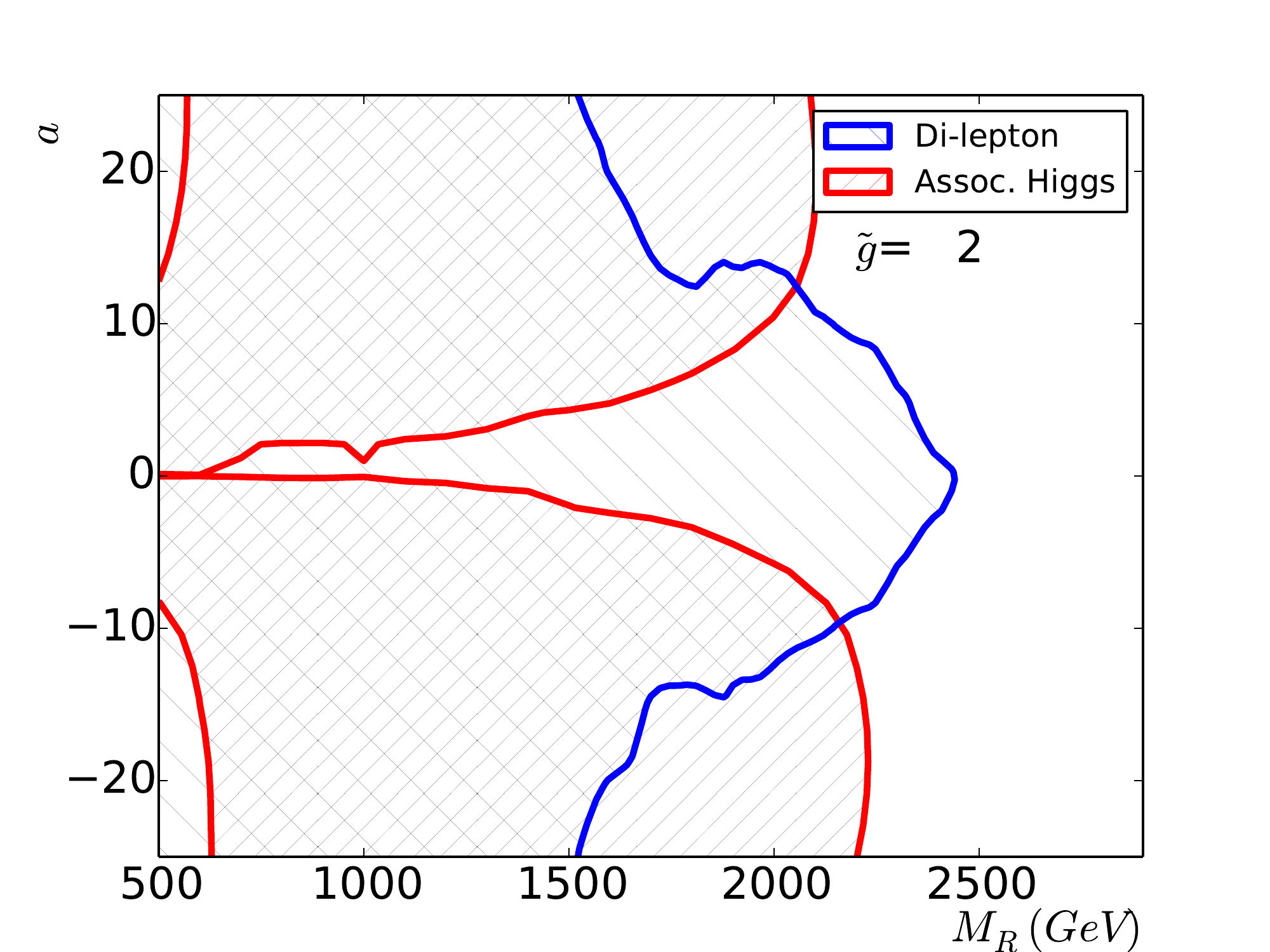}
\includegraphics[width=.49\columnwidth]{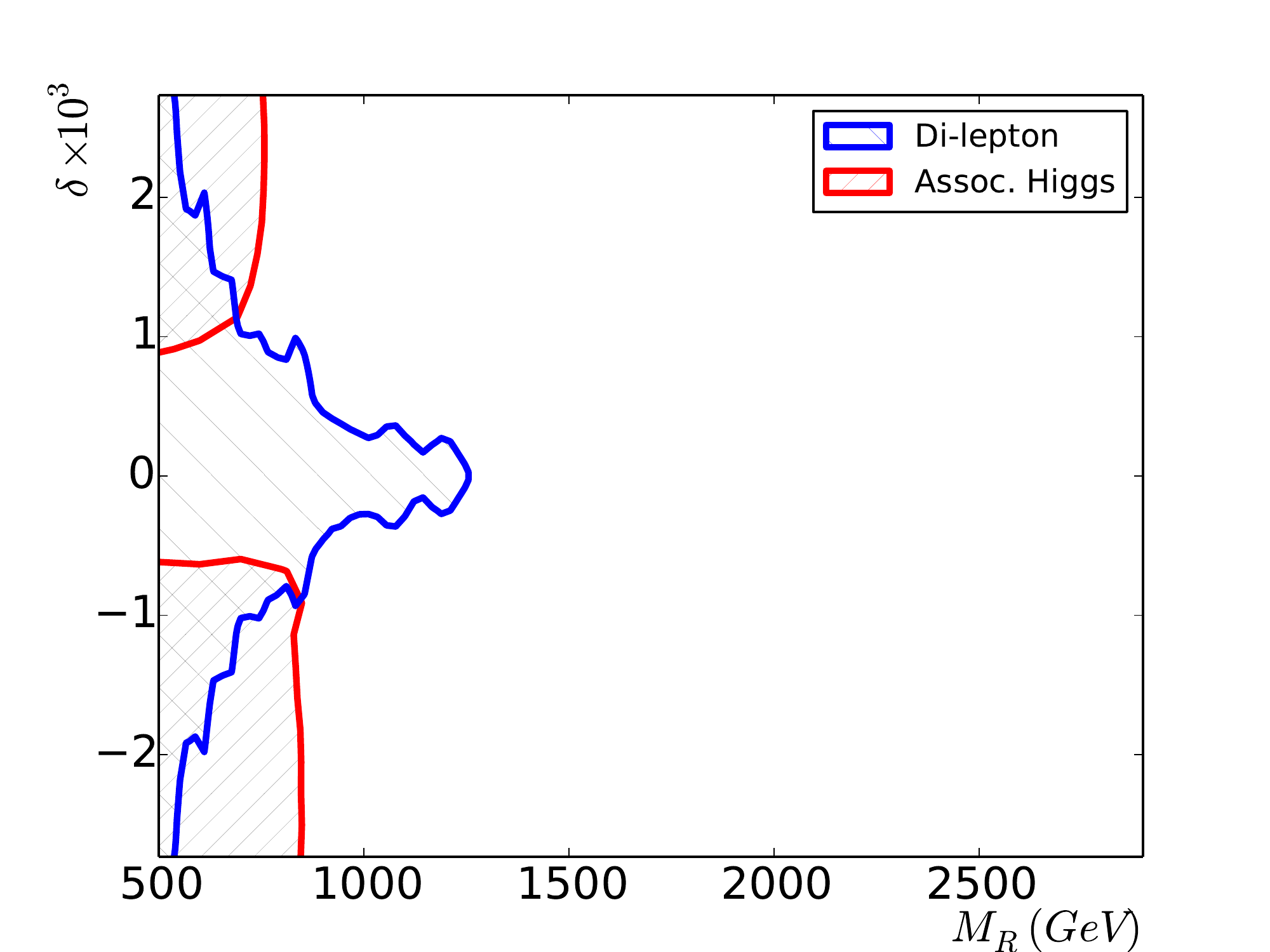}
\includegraphics[width=.49\columnwidth]{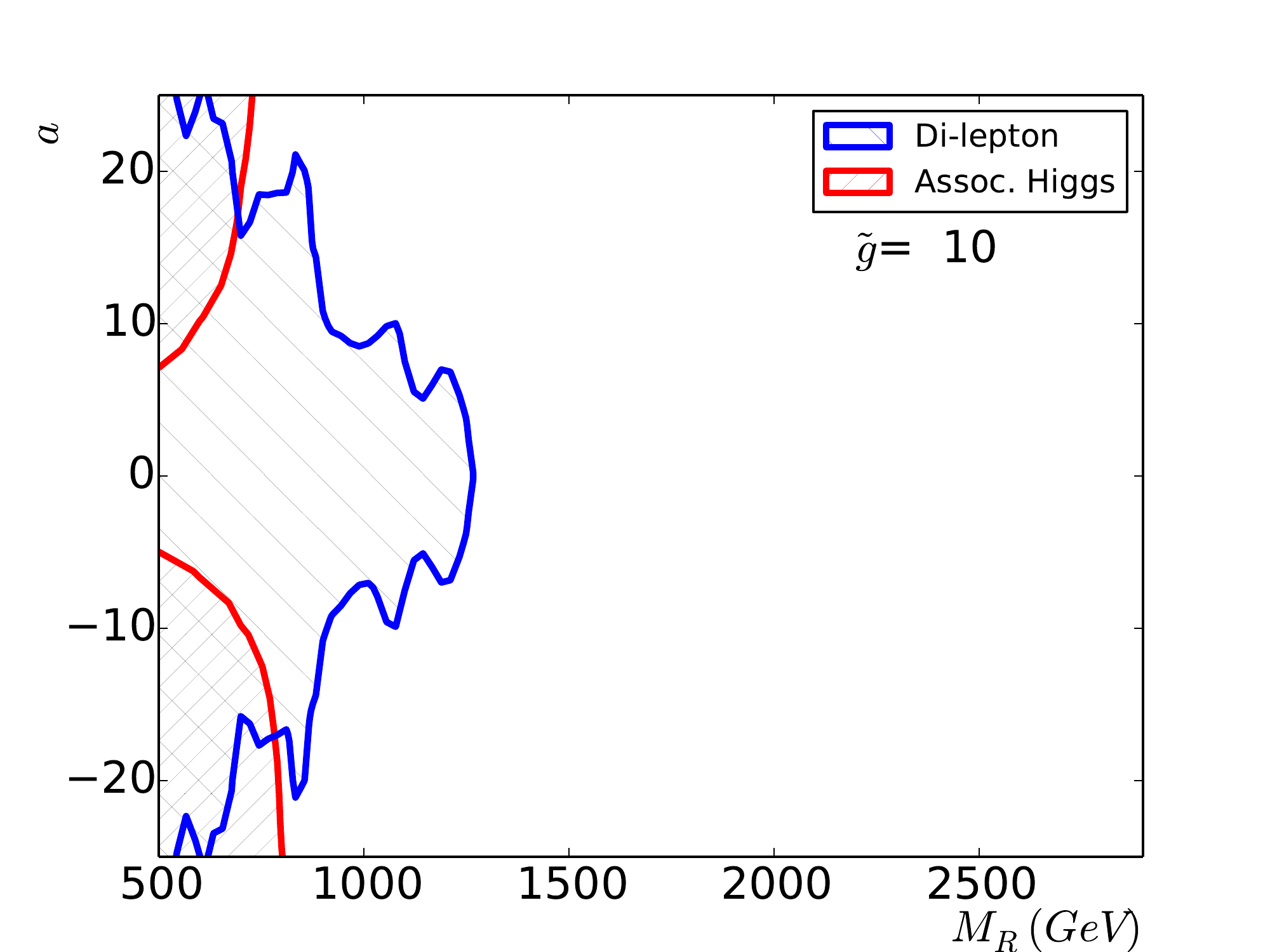}
\caption{$95\%$ exclusion limits as above but in $(\delta, M_R)$ (\emph{left}) and $(a,\,M_R)$ (\emph{right}) planes for fixed $\tilde{g}$ values.
}
\label{fig:Excl95gt}
\end{center}
\end{figure}

In summary, the LHC currently excludes roughly between a third or half of the parameter space satisfying $\gt<4\pi$ and $M_R\lesssim \gt v\simeq 3\TeV$ and $|a|\lesssim 25$. The constraints from electroweak precision measurements are negligible in comparison, due to the enhanced global $SU(2)_L'\times SU(2)_R'$ symmetry over the vector spectrum. 

\subsubsection*{Future Reach}
We show an estimate of the CMS reach in the dilepton channel at the high energy Run II of LHC at $\sqrt{s}=13\TeV$ in \fig{fig:Excl95Exp}. 
\begin{figure}[t!] 
\begin{center}
\includegraphics[width=.55\columnwidth]{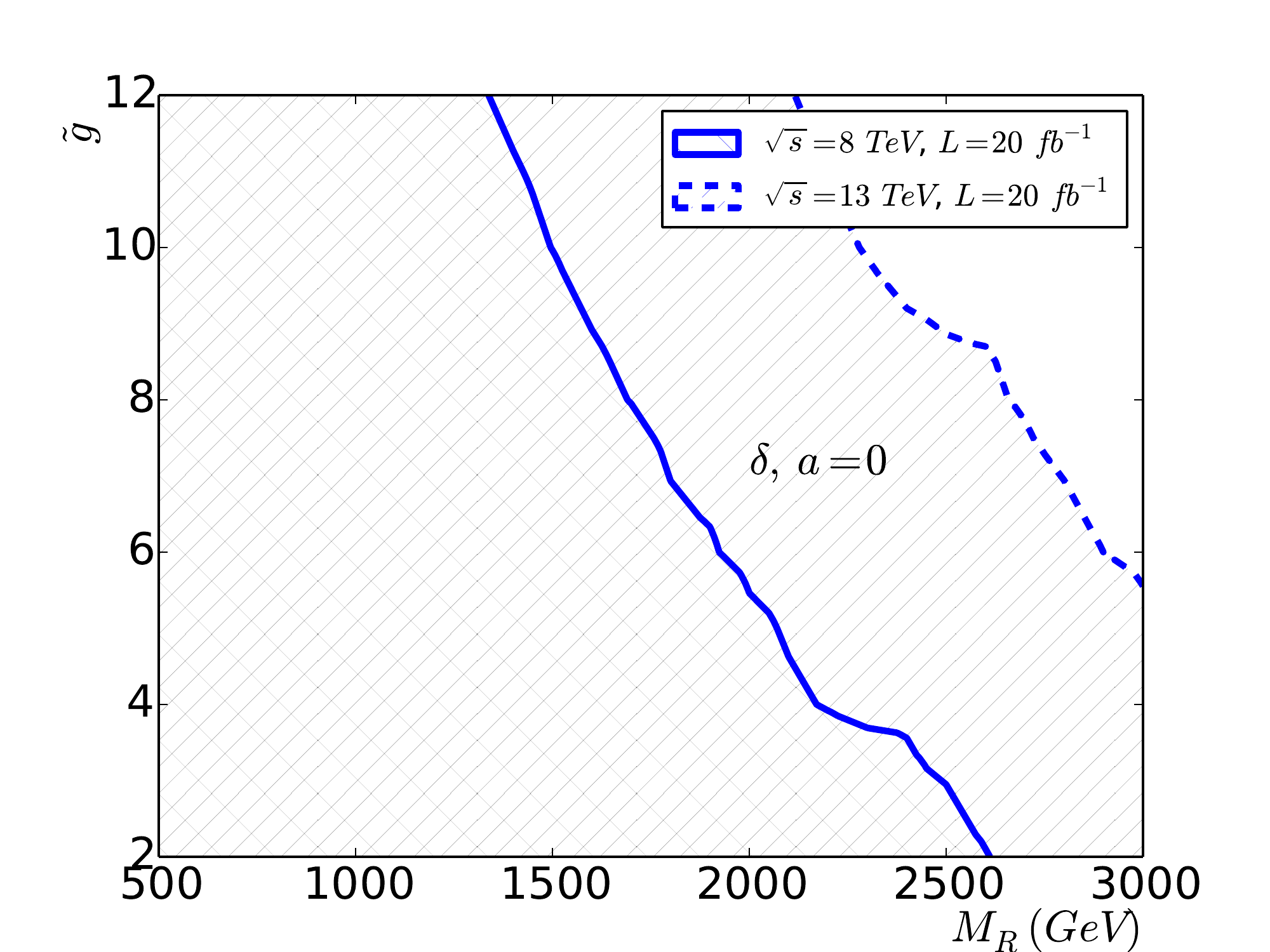}
\caption{Projected discovery reach or exclusion limit of the LHC Run II dilepton resonance search at $\sqrt{s}=13 \TeV$ with $L=20\ifb$ 
in the $(M_R,\,\tilde{g})$ plane (dashed blue). The striped and cross striped regions will be excluded in the absence of discovery. 
Also shown is the equivalently projected exclusion limit for the current run at $\sqrt{s}=8 \TeV$ (solid blue) to assess the validity of the projection. }
\label{fig:Excl95Exp}
\end{center}
\end{figure}
For comparison with the upper left panel of \fig{fig:Excl95delta} we also show the estimated current LHC exclusion curve using the method detailed in appendix~\ref{sec:futurereach}.
Although the computation is simplistic, it compares reasonably well and we therefore expect the projection to be a good guide to the future run. 
 For $a=0$ most of the parameter space will be excluded already with  $L=20\ifb$ while $L=100\ifb$ will be enough to exclude the entire parameter space shown. To exclude the same values of $(M_R, \gt)$ for $|a|\lesssim 10$ the required luminosity is estimated to be $L=200\ifb$.

\section{Summary and Outlook}
\label{sec:conclusions}
In this paper we have presented the Custodial Vector Model (CVM), featuring two new weak triplets of vector resonances in addition to the SM and 3 new parameters determining their interactions. Here we have studied the CVM in its own right but as mentioned in the introduction the model can be interpreted as an effective Lagrangian for several different theories of dynamical EWSB.

We have further discussed the distinct collider phenomenology of the CVM: The presence of two nearly mass degenerate resonances in dilepton final states and a single (dominant) resonance in single charged lepton final states as well as the apparent absence of resonances in the $WW$ and $WZ$ channels. Finally the interactions between the new resonances and the Higgs sector can be probed by the associate Higgs production. 

Despite the simple and distinct pattern of resonances, the identification of the CVM at LHC is challenging because of suppressed couplings to SM fields for $\gt>1$, the narrow spacing of the resonances and, in a significant part of the parameter space, their narrow widths. Given the CVM, Run II of the LHC should be able to discover a signal of new physics from the vector resonances with about 200 inverse femtobarns of luminosity in the parameter space region $M_R<3$ TeV, $\gt <4\pi$ and at least $|a| \lesssim 10$. However to clearly identify the new physics as stemming from the CVM, more luminosity and better resolution will be required. A high energy lepton collider would be ideal to uncover the CVM. 

\section*{Acknowledgements}
We thank Georges Azuelos for careful reading of the manuscript.
DB would also like to thank Stefania De Curtis for discussions.
The CP$^3$-Origins centre is partially funded
by the Danish National Research Foundation, grant number DNRF90. MTF acknowledges a
Sapere Aude Grant no. 11-120829 from the Danish Council for Independent Research.

\newpage
\appendix
\label{sec:appendix}
\section{Mass matrices}
\label{appendix:diag}
The spin-one mass Lagrangian is
\bea
{\cal L}_{\rm mass} =
\begin{pmatrix} \widetilde{W}^-_\mu & A_{L\mu}^- & A_{R\mu}^-  \end{pmatrix} {\cal M}_C^2
\begin{pmatrix} \widetilde{W}^{+\mu} \\ A_L^{+\mu} \\ A_R^{+\mu}  \end{pmatrix} +
\frac{1}{2}\begin{pmatrix} B_\mu & \widetilde{W}^3_\mu & A_{L\mu}^0 & A_{R\mu}^0  \end{pmatrix} {\cal M}_N^2
\begin{pmatrix} B^\mu \\ \widetilde{W}^{3\mu} \\ A_L^{0\mu} \\ A_R^{0\mu}  \end{pmatrix} \ ,
\eea
where
\bea
{\cal M}_C^2 =
\begin{pmatrix}
{\displaystyle g^2\, \frac{f^2+(1+s)v^2}{4}} &
-{\displaystyle  g \tilde{g}\, \frac{f^2+s v^2}{4}} &
{\displaystyle 0} \\
\smallskip \\
-{\displaystyle  g \tilde{g}\, \frac{f^2+s v^2}{4}} &
{\displaystyle \tilde{g}^2\, \frac{f^2+s v^2}{4}} &
0 \\
\smallskip \\
{\displaystyle 0} &
0 &
{\displaystyle \tilde{g}^2\, \frac{f^2+s v^2}{4}}
\end{pmatrix} \ ,
\eea
\bea
{\cal M}_N^2 =
\begin{pmatrix}
{\displaystyle g^{\prime 2}\, \frac{f^2+(1+s)v^2}{4}} &
-{\displaystyle g g^\prime\frac{v^2}{4}} &
{\displaystyle 0} &
-{\displaystyle  g^\prime \tilde{g}\, \frac{f^2+s\, v^2}{4}}  \\
\smallskip \\
-{\displaystyle g g^\prime\frac{v^2}{4}} &
{\displaystyle g^2\, \frac{f^2+(1+s)v^2}{4}} &
-{\displaystyle  g \tilde{g}\, \frac{f^2+s\, v^2}{4}} &
{\displaystyle 0} \\
\smallskip \\
{\displaystyle 0} &
-{\displaystyle  g \tilde{g}\, \frac{f^2+s\, v^2}{4}} &
{\displaystyle \tilde{g}^2\, \frac{f^2+s\, v^2}{4}} &
{\displaystyle 0} \\
\smallskip \\
-{\displaystyle  g^\prime \tilde{g}\, \frac{f^2+s\, v^2}{4}} &
{\displaystyle 0} &
{\displaystyle 0} &
{\displaystyle \tilde{g}^2\, \frac{f^2+s\, v^2}{4}}
\end{pmatrix} \ .
\eea
The charged mass eigenstates are the $W$ boson, $L^\pm$, and $R^\pm$, whereas the neutral mass eigenstates are the photon $A$, the $Z$ boson, $L^0$ and $R^0$. Let ${\cal C}$ and ${\cal N}$ be the charged and neutral rotation matrix, respectively:
\bea
\left(\begin{array}{c} \widetilde{W}_\mu^\pm \\ A_{L\mu}^\pm  \\  A_{R\mu}^\pm \end{array}\right)
= {\cal C}
\left(\begin{array}{c} W_\mu^\pm \\ L_{\mu}^\pm \\ R_{\mu}^\pm \end{array}\right)\ , \quad
\left(\begin{array}{c} \widetilde{B}_\mu \\ \widetilde{W}_\mu^3 \\ A_{L\mu}^3 \\ A_{R\mu}^3 \end{array}\right)
= {\cal N}
\left(\begin{array}{c} A_\mu \\ Z_\mu \\ L_{\mu}^0 \\ R_{\mu}^0 \end{array}\right)\ .
\eea
Diagonalization of the charged-boson mass matrix gives
\bea
\label{eq:chargedmasses}
M_W^2 &=& \frac{g^2+\tilde{g}^2}{\tilde{g}^2}\frac{M_R^2}{2}+\frac{g^2 v^2}{8}
-\sqrt{\left(\frac{g^2+\tilde{g}^2}{\tilde{g}^2}\frac{M_R^2}{2}+\frac{g^2 v^2}{8}\right)^2-\frac{g^2 v^2 M_R^2}{4}}\ , \nonumber \\
M_{R^\pm}^2 &=& M_R^2 \ , \nonumber \\
M_{L^\pm}^2 &=& \frac{g^2+\tilde{g}^2}{\tilde{g}^2}\frac{M_R^2}{2}+\frac{g^2 v^2}{8}
+\sqrt{\left(\frac{g^2+\tilde{g}^2}{\tilde{g}^2}\frac{M_R^2}{2}+\frac{g^2 v^2}{8}\right)^2-\frac{g^2 v^2 M_R^2}{4}} \ ,
\eea
where $M_R$ is defined in \eq{Eq:Massparam}.
\bea
{\cal C} = \left(\begin{array}{ccc} \cos\alpha & -\sin\alpha & 0 \\ \sin\alpha & \cos\alpha & 0 \\ 0 & 0 & 1  \end{array}\right)\ , \quad
\sin\alpha = \sqrt{\frac{M_{L^\pm}^2-M_{R^\pm}^2}{M_{L^\pm}^2 - M_W^2}} = \frac{g}{\gt} + {\cal O}\left(\frac{g^3}{\tilde{g}^3}\right)\ . 
\label{eq:chargedmixing}
\eea
Note that the spin-one charged resonance associated to the $SU(2)_R$ group does not mix with the $W$ boson, and its mass is therefore unaffected, at tree-level, by the electroweak interactions. The $SU(2)_L$ resonance does mix with the $W$ boson, and its mass receives a small and positive contribution.

The $4\times 4$ neutral mass matrix can be diagonalized analytically, because one eigenvalue is the massless photon. However it is more instructive to expand eigenvalues and eigenvectors in powers of $1/\tilde{g}$, assuming that $M_R$ scales as $\tilde{g}$ without a parametric suppression from $f^2+ s v^2$. This gives
\bea
M_Z^2 &=& \frac{g^2+g^{\prime 2}}{4}v^2\left[1-\frac{g^4+g^{\prime 4}}{(g^2+g^{\prime 2})\tilde{g}^2}+{\cal O}\left(\frac{g^4}{\tilde{g}^4}\right)\right]\ , \nonumber \\
M_{R^0}^2 &=& M_R^2\left[1+\frac{g^{\prime 2}}{\tilde{g}^2}+{\cal O}\left(\frac{g^4}{\tilde{g}^4}\right)\right] \nonumber \\
M_{L^0}^2 &=& M_R^2\left[1+\frac{g^2}{\tilde{g}^2}+{\cal O}\left(\frac{g^4}{\tilde{g}^4}\right)\right] \ .
\eea
Note that the $SU(2)_L$ neutral resonance is still heavier than its $SU(2)_R$ counterpart, as $g>g^\prime$. The elements of the neutral boson rotation matrix are 
\bea
&& {\cal N}_{00} = \frac{e}{g^\prime}\ , \quad {\cal N}_{10} = \frac{e}{g}\ , \quad {\cal N}_{20} = \frac{e}{\tilde{g}}\ , \quad {\cal N}_{30} = \frac{e}{\tilde{g}} \ , \nonumber \\
&& {\cal N}_{01} = -\frac{g^\prime}{\sqrt{g^2+g^{\prime 2}}}\left[1+\frac{g^4-2g^2 g^{\prime 2} -g^{\prime 4}}{2(g^2+g^{\prime 2})\tilde{g}^2}
+{\cal O}\left(\frac{g^4}{\tilde{g}^4}\right)\right]\ , \nonumber \\
&& {\cal N}_{11} = \frac{g}{\sqrt{g^2+g^{\prime 2}}}\left[1+\frac{g^{\prime 4}-2g^2 g^{\prime 2}-g^4}{2(g^2+g^{\prime 2})\tilde{g}^2} 
+{\cal O}\left(\frac{g^4}{\tilde{g}^4}\right)\right] \ ,\nonumber \\
&& {\cal N}_{21} = \frac{g^2}{\sqrt{g^2+g^{\prime 2}}\tilde{g}}\left[1+{\cal O}\left(\frac{g^2}{\tilde{g}^2}\right)\right] \ , \quad
{\cal N}_{31} = -\frac{g^{\prime 2}}{\sqrt{g^2+g^{\prime 2}}\tilde{g}}\left[1+{\cal O}\left(\frac{g^2}{\tilde{g}^2}\right)\right]\ , \nonumber \\
&& {\cal N}_{02} = \frac{ v^2 g^{\prime} g^4 }{ 4 \gt M_R^2 ( g^2 - g^{\prime 2} ) }\left[1+{\cal O}\left(\frac{g^2}{\tilde{g}^2}\right)\right] \ , \quad
{\cal N}_{12} = -\frac{g}{\tilde{g}}\left[1+{\cal O}\left(\frac{g^2}{\tilde{g}^2}\right)\right]\ , \nonumber \\
&& {\cal N}_{22} = 1-\frac{g^2}{2\tilde{g}^2} + {\cal O}\left(\frac{g^4}{\tilde{g}^4}\right) \ , \quad
{\cal N}_{32} = -\frac{g^2 g^{\prime 2} v^2}{4(g^2-g^{\prime 2})M_R^2} \left[1+{\cal O}\left(\frac{g^2}{\tilde{g}^2}\right)\right]\ , \nonumber \\
&& {\cal N}_{03} =  -\frac{g^\prime}{\tilde{g}}\left[1+{\cal O}\left(\frac{g^2}{\tilde{g}^2}\right)\right]\ , \quad 
{\cal N}_{13} =  -\frac{ v^2 g'^4 g }{ 4 \gt M_R^2 ( g^2 - g^{\prime 2} ) }\left[1+{\cal O}\left(\frac{g^2}{\tilde{g}^2}\right) \right]\ , \nonumber \\
&& {\cal N}_{23} =  \frac{g^2 g^{\prime 2} v^2}{4(g^2-g^{\prime 2})M_R^2} \left[1+{\cal O}\left(\frac{g^2}{\tilde{g}^2}\right)\right] \ , \quad
{\cal N}_{33} = 1-\frac{g^{\prime 2}}{2\tilde{g}^2} + {\cal O}\left(\frac{g^4}{\tilde{g}^4}\right) \ .
\eea
\section{Couplings}
\label{appendix:couplings}
In order to express the vertices with vectors in a compact form, we define
\bea
\left(\begin{array}{c} {\cal W}_{1\mu}^\pm \\ {\cal W}_{2\mu}^\pm \\ {\cal W}_{3\mu}^\pm \end{array}\right)
\equiv\left(\begin{array}{c} W_\mu^\pm \\ L_{\mu}^\pm \\ R_{\mu}^\pm \end{array}\right)\ , \quad
\left(\begin{array}{c} {\cal Z}_{0\mu} \\ {\cal Z}_{1\mu} \\ {\cal Z}_{2\mu} \\ {\cal Z}_{3\mu} \end{array}\right)
\equiv\left(\begin{array}{c} A_\mu \\ Z_\mu \\ L_{\mu}^0 \\ R_{\mu}^0 \end{array}\right)\ .
\eea
The trilinear spin-one vertices are
\bea
{\cal L}_{{\cal Z W W}} = \sum_{klm}\, g_{klm} \Bigg(
[[{\cal Z}_k\, {\cal W}_l^+\, {\cal W}_m^-]]
+[[{\cal W}_l^+\, {\cal W}_m^-\, {\cal Z}_k]] \Bigg)
\eea
where
\bea
g_{klm} = g\, {\cal N}_{1k} {\cal C}_{1l}  {\cal C}_{1m} 
+ \tilde{g}\left( {\cal N}_{2k} {\cal C}_{2l}  {\cal C}_{2m} + {\cal N}_{3k} {\cal C}_{3l}  {\cal C}_{3m}\right) \ .
\eea

The Higgs vertices with vectors are
\bea
{\cal L}_{H{\cal V}{\cal V}}=\left(2\frac{H}{v}+\frac{H^2}{v^2}\right)\sum_{kl}\left[
\left({\cal C}^T\, \delta{\cal M}_C^2\, {\cal C}\right)_{k l} {\cal W}_{k\mu}^- {\cal W}_l^{+\mu}
+\frac{1}{2}\left({\cal N}^T\, \delta{\cal M}_N^2\, {\cal N}\right)_{k l} {\cal Z}_{k\mu} {\cal Z}_l^{\mu}\right]\ ,
\eea
where $\delta{\cal M}_C^2$ and $\delta{\cal M}_N^2$ are the $v^2$ part of the charged and neutral mass matrices, respectively.
\bea
{\delta\cal M}_C^2 =
\begin{pmatrix}
{\displaystyle g^2\, \frac{(1+s)v^2}{4}} &
-{\displaystyle  g \tilde{g}\, \frac{s v^2}{4}} &
{\displaystyle 0} \\
\smallskip \\
-{\displaystyle  g \tilde{g}\, \frac{s v^2}{4}} &
{\displaystyle \tilde{g}^2\, \frac{s v^2}{4}} &
0 \\
\smallskip \\
{\displaystyle 0} &
0 &
{\displaystyle \tilde{g}^2\, \frac{s v^2}{4}}
\end{pmatrix} \ ,
\eea
\bea
{\delta\cal M}_N^2 =
\begin{pmatrix}
{\displaystyle g^{\prime 2}\, \frac{(1+s)v^2}{4}} &
-{\displaystyle g g^\prime\frac{v^2}{4}} &
{\displaystyle 0} &
-{\displaystyle  g^\prime \tilde{g}\, \frac{s\, v^2}{4}}  \\
\smallskip \\
-{\displaystyle g g^\prime\frac{v^2}{4}} &
{\displaystyle g^2\, \frac{(1+s)v^2}{4}} &
-{\displaystyle  g \tilde{g}\, \frac{s\, v^2}{4}} &
{\displaystyle 0} \\
\smallskip \\
{\displaystyle 0} &
-{\displaystyle  g \tilde{g}\, \frac{s\, v^2}{4}} &
{\displaystyle \tilde{g}^2\, \frac{s\, v^2}{4}} &
{\displaystyle 0} \\
\smallskip \\
-{\displaystyle  g^\prime \tilde{g}\, \frac{s\, v^2}{4}} &
{\displaystyle 0} &
{\displaystyle 0} &
{\displaystyle \tilde{g}^2\, \frac{s\, v^2}{4}}
\end{pmatrix} \ .
\eea

Finally, the SM fermions couple to the spin-one resonances through mixings with the electroweak bosons. This leads to the vertices
\bea
{\cal L}_{{\cal V}ff} &=& \frac{g}{\sqrt{2}}\sum_k \sum_i {\cal C}_{1k} \bar{u}_i\, \slashed{\cal W}_k^+\, P_L\, d_i + {\rm h.c.}  \nonumber \\
&+&\sum_k \sum_f \left(g\, {\cal N}_{1k}-g^\prime\, {\cal N}_{0k}\right) \bar{f} {\cal Z}_k 
\left(T^3_f P_L-\frac{g^{\prime 2}}{g^2+g^{\prime 2}}Q_f\right)f \nonumber \\
&+& e\sum_f \bar{f}\, \slashed{A}\, Q_f\, f\ ,
\eea
where $i$ runs over quark and lepton doublets, with $u_i$ ($d_i$) up-type (down-type) fermion, and $f$ runs over all quark and lepton flavours. 

Here below we list the set of interactions between physical states relevant for the present study in the form presented in \sec{sec:MI}. 
The expansions in $g/\gt$ assume that $M_R$ scales as $\gt$ without parametric suppression from $f^2+sv^2$, the $a$ parameter is or order 1 and $\delta$ scales as $g^4/(\gt^2 M_R^2)$.

The couplings between $L,\,R$ and SM weak bosons are
\bea
&& g_{L W W}^{(1)} = g_{L W W}^{(2)} = g_{211} = \frac{g^4 v^2}{4\tilde{g}M_R^2} +{\cal O}\left(\frac{g^4}{\tilde{g}^4}\right) \nonumber \\
&& g_{R W W}^{(1)} = g_{R W W}^{(2)} = g_{311} = \frac{g^2 g^{\prime 2}v^2}{4M_R^2 \tilde{g}}+{\cal O}\left(\frac{g^4}{\tilde{g}^4}\right) \nonumber \\
&& g_{Z L W}^{(1)} = g_{Z L W}^{(2)} = g_{121} = \frac{g^3\sqrt{g^2+g^{\prime 2}} v^2}{4\tilde{g}M_R^2} +{\cal O}\left(\frac{g^4}{\tilde{g}^4}\right) \nonumber \\
&& g_{Z R W}^{(1)} = g_{Z R W}^{(2)}  = g_{131} = 0 \nonumber \\ 
&& g_{ARR} = e \nonumber \\
&& g_{ZRR} = -\frac{g'^2}{\sqrt{g^2+g'^2}} \nonumber \\ 
&& g_{L Z Z} = g_{L Z\gamma} = g_{R Z Z} = g_{R Z\gamma} = 0\ .
\eea

The  trilinear couplings of two vector fields with the Higgs boson, expressed in terms of $M_Z, v$ and the Weinberg angle $\theta$ defined in \eq{eq:thetadef} as well as $a$ and $\delta$ defined in \eq{def:a} and \eq{eq:deltadef}, are given by 
\bea
\label{Eq:HiggscouplingsZ}
g_{HZZ}
&=& \frac{2}{v}M_Z^2(1-\delta) \nonumber\\
&=& \frac{2}{v}M_Z^2\left\{1 - a\left[ \frac{M_Z^4}{\gt^2 v^2 M_R^2}\Bigl(3+\cos(4\theta)\Bigr) +\mathcal{O}\left(\frac{g^6}{\tilde{g}^6}\right) \right] \right\}\\
\label{Eq:Higgscouplings}
g_{HW^+W^-}
&=& \frac{2}{v}M_W^2\left\{1 - a\left[ \frac{M_Z^4}{\gt^2 v^2 M_R^2}\Bigl(4 \cos^4\theta \Bigr) + \mathcal{O}\left(\frac{g^6}{\tilde{g}^6}\right) \right] \right\} \nonumber\\
&=& \frac{2}{v}M_Z^2 \cos^2\theta \left\{ 1 - \frac{M_Z^4}{\gt^2 v^2 M_R^2}\Bigl( 4 a \cos^4\theta - \sin(2\theta) \tan(2\theta) \Bigr) + \mathcal{O}\left(\frac{g^6}{\tilde{g}^6}\right) \right\} \\
g_{HL^0Z}
&=& -a \frac{ M_Z^3}{\gt\, v^2}\, 4 \cos^2\theta + \mathcal{O}\left(\frac{g^3}{\gt^3}\right) = -\delta\, \frac{ \gt M_R^2}{M_Z} \frac{4 \cos^2\theta}{3+\cos(4\theta)} + \mathcal{O}\left(\frac{g^3}{\gt^3}\right)\\
g_{HR^0Z}
&=& a \frac{ M_Z^3}{\gt\, v^2}\, 4 \sin^2\theta + \mathcal{O}\left(\frac{g^3}{\gt^3}\right) = \delta\, \frac{ \gt M_R^2}{M_Z} \frac{4 \sin^2\theta}{3+\cos(4\theta)} + \mathcal{O}\left(\frac{g^3}{\gt^3}\right) \\
g_{HL^+W^-}
&=& -a\frac{M_Z^3}{\gt\, v^2}4\cos^3\theta + \mathcal{O}\left(\frac{g^3}{\gt^3}\right)  = -\delta \frac{\gt M_R^2}{M_Z}\frac{4\cos^3\theta}{3+\cos (4 \theta )} + \mathcal{O}\left(\frac{g^3}{\gt^3}\right)  \\
\label{Eq:HiggscouplingsLL}
g_{HL^+L^-}
&=&\frac{2 M_R^2}{v}(1-a)\left[1+\frac{4 M_Z^2\cos^2\theta}{\gt^2 v^2}+\mathcal{O}\left(\frac{g^4}{\gt^4}\right)\right] \nonumber\\
&=&\frac{2 M_R^2}{v}\left(1-\delta\frac{\gt^2 M_R^2 v^2}{M_Z^4 (\cos (4 \theta )+3)}\right)\left[1+\frac{4 M_Z^2\cos^2\theta}{\gt^2 v^2}+\mathcal{O}\left(\frac{g^4}{\gt^4}\right)\right] \\
g_{HR^+R^-}
&=&\frac{2 M_R^2}{v}(1-a)
\eea

The couplings between fermions and the vector fields are
\bea
g^{L/R}_{R^0 f}
&=&\frac{g^{\prime 2}}{\gt}\left(T^3_f \delta_L-\frac{g^{\prime 2}}{g^2+g^{\prime 2}}Q_f\right) +\mathcal{O}\left(\frac{g^2}{\gt^2}\right)\\
g^{L/R}_{L^0 f}
&=&-\frac{g^{2}}{\gt}\left(T^3_f \delta_L-\frac{g^{\prime 2}}{g^2+g^{\prime 2}}Q_f\right) +\mathcal{O}\left(\frac{g^2}{\gt^2}\right)\\
g^L_{L^\pm ud}
&=& \frac{g^2}{\sqrt{2}\gt}+\mathcal{O}\left(\frac{g^2}{\gt^2}\right) \\
g^{L/R}_{R^\pm ud}&=&g^R_{L^\pm ud}=0 \,,
\eea
where $\delta_L=1,\,0$ for $L$ and $R$, the \emph{left}-handed and \emph{right}-handed fermions, respectively.

\section{Decay widths}
\label{sec:widths}
Below we give the partial widths of the heavy $\cal R$ resonances, see e.g. \cite{Frandsen:2012rk},
\bea
\Gamma({\cal R}\rightarrow f\bar{f})&=&\frac{m_{\cal R} N_c}{12\pi} 
 \sqrt{1-\frac{4 m_f^2}{m_{\cal R}^2}}  [(g^{V}_{f})^2+(g^{A}_{f})^2 + \frac{m_f^2}{m_{\cal R}^2}(2 (g^{V}_{f})^2 - 4 (g^{A}_{f})^2)  ]  
\eea

\bea
\Gamma({\cal R}\rightarrow W^{+}W^{-})&=&
\frac{1}{192  \pi}m_{\cal R} \left (\frac{m_{\cal R}}{M_W} \right )^4
\left (1-4\frac{M_W^2}{m_{\cal R}^2} \right )^{1/2} \nonumber \\
&\times &\bigg((g_{WW1}^{\cal R})^2\left [4\frac{M_W^2}{m_{\cal R}^2}  -4\frac{M_W^4}{m_{\cal R}^4} - 48\frac{M_W^6}{m_{\cal R}^6} \right ]\nonumber \\
&&+ (g_{WW2}^{\cal R})^2 \left [ 1- 16 \frac{M_W^4}{m_{\cal R}^4} \right ]\nonumber \\
&&+ g_{WW1}^{\cal R} g_{WW2}^{\cal R} \left [12 \frac{M_W^2}{m_{\cal R}^2} - 48\frac{M_W^4}{m_{\cal R}^4}  \right ]\nonumber \\
&&+ (g_{WW3}^{\cal R})^2 \left [4 \frac{M_W^2}{m_{\cal R}^2}- 32 \frac{M_W^4}{m_{\cal R}^4} + 64 \frac{M_W^6}{m_{\cal R}^6}  \right ] \bigg)
\eea

\bea
\Gamma({\cal R}\rightarrow Z Z )&=&
\frac{(g_{ZZ}^{\cal R})^2}{96  \pi}m_{\cal R} \frac{m_{\cal R}^2}{M_Z^2}
\left (1-4\frac{M_Z^2}{m_{\cal R}^2} \right )^{3/2}
\left [1 - 6\frac{M_Z^2}{m_{\cal R}^2} \right ]
\eea

\bea
\Gamma({\cal R}\rightarrow Z \gamma )&=&
\frac{(g_{Z\gamma}^{\cal R})^2}{96  \pi}m_{\cal R} \frac{m_{\cal R}^2}{M_Z^2}
\left (1-\frac{M_Z^2}{m_{\cal R}^2} \right )^{3}
\eea

\bea
\Gamma({\cal R}\rightarrow ZH)&=&\frac{(g^{\cal R}_{ZH})^2}{192\pi M_Z^2} m_{\cal R} 
\sqrt{\lambda(1,x_Z, x_H)}(\lambda(1,x_Z, x_H)+ 12 x_Z) \ , 
\eea
where $x_Z=(M_Z/m_{\cal R})^2$, $x_H=(m_H/m_{\cal R})^2$, and
$\lambda(x,y,z)=x^2+y^2+z^2-2xy-2yz-2zx$. 

\section{Off-diagonal widths}
\label{sec:offdiagonalappendix}

The following is basically a summary of the basic effect from off-diagonal width.
Loop corrections to the vector self-energy can be parametrised as
\bea
\label{eq:selfen}
\Pi_{\mu\nu}=\Pi_T g_{\mu\nu}+\Pi_L p_\mu p_\nu\,.
\eea
The corrected vector-particles propagator can be written as:
\bea
i\Delta_{\mu\nu} = \left(g_{\mu\nu}-\frac{p_\mu p_\nu}{p^2}\right)\frac{-i}{p^2-M_0^2+\Pi_T}+ \frac{p_\mu p_\nu}{p^2}\frac{-i\xi}{p^2-\xi(M_0^2-\Pi_T-p^2\Pi_L)}\,.
\eea
$\Pi_T$ defines the the gauge independent pole mass and width while $\Pi_L$ contributes to the gauge dependent pole and is negligible at the physical pole. Therefore we neglect $\Pi_L$ and adopt the unitary gauge, $\xi\ra \infty$. After diagonalization and renormalization, we get\footnote{Notice that the renormalization of fields and mass parameters allow us to fix the off-diagonal real part one-loop contribution to zero.}
\bea
i\Delta_{\mu\nu} = \left(g_{\mu\nu}-\frac{p_\mu p_\nu}{p^2}\right)(-i)\Delta\,,
\eea
where, in the two-particles case,
\bea
i\Delta=\frac{i}{D}\left(\begin{array}{cc}
  p^2-m_2^2+i\Sigma_{22} & -i\Sigma_{12} \\
  -i\Sigma_{21} & p^2-m_1^2+i\Sigma_{11} 
  \end{array} \right)\,,
\eea
and $\Sigma_{ij} =\Im \Pi_{ij}$ for particle indexes $i,\,j=1,\,2$ and 
\bea
D=(p^2-m_1^2+i\Sigma_{11})(p^2-m_2^2+i\Sigma_{22})+\Sigma_{12}\Sigma_{21}\,.
\eea

\section{Future reach}
\label{sec:futurereach}
Our projected reach estimate is based on the search in the electron channel by the CMS experiment described in \cite{CMS-PAS-EXO-12-061}. 

We assume a constant efficiency of $89\%$ for both background and signal and apply the kinematic cuts:
\bea
|\eta(\ell^\pm)|&<&2.5 \nonumber\\
p_T(\ell^\pm)	&>&25\GeV .
\eea

{\bf Signal:}
The signal cross section, $\sigma_S$, is computed at LO and a mass dependent $K$ factor is applied to account for QCD NNLO corrections. The $K$ factors for resonance masses $M=1\TeV$, $2\TeV$, $3\TeV$ are $1.22$, $1.16$ and $1.16$ respectively. We fix $a=0$ for the projection and expect slightly weaker bounds for non-zero $a$.

{\bf Background:}
The dominant DY background $pp\to Z/\gamma \to e^+e^-$ is computed at LO and the NNLO QCD and NLO EW corrections are incorporated through a mass dependent $K$ factor. The k-factors for $m(\ell\ell)=1\TeV$, $2\TeV$, $3\TeV$  are $1.07$, $1.1$ and $1.14$ respectively.
The sum of other background processes, $t\bar{t}$, $tW$, $WW$, $WZ$, $ZZ$, $\tau\tau$ and jets producing ``fake" electrons have the same exponential fall off as a function of dilepton invariant mass as the DY for  $m(\ell\ell)\gtrsim 200\GeV$. They can therefore be modeled as a number times the DY cross-section. We take this number to be $r=0.24$.

{\bf Statistics:}
We look for a local excess in the mass window $M_R-30\GeV<m(\ell\ell)<M_R+150$, for each value of $M_R$. A Poisson distribution is assumed for the expected number of background events, $N_B$, 
\begin{equation}
P(N;\mu)=\frac{\mu^N}{N!}e^{-\mu} \ , 
\end{equation}
with the predicted cross section times the integrated luminosity as the mean value,
\begin{equation}
\mu=\sigma_B\,L\,.
\end{equation}
We denote the maximum number of events at $95\%$ CL, assuming the background only hypothesis,  by $N_{95}$\footnote{Fractional values of $N_{95}$, assuming uniform probability for each $N$, are necessary to be used for extremelly small values of $\mu$.}. 
\begin{equation}
95\%=\sum_{N=0}^{N_{95}}\frac{\mu^N}{N!}e^{-\mu}\,.
\end{equation}
Cross sections for which $\mu_S=\sigma_S\,L$, is larger than $N_{95}$ are then considered excluded. 
The resulting exclusion limit presented in \fig{fig:Excl95Exp} is slightly stronger than our exclusion limit given in the upper left panel of \fig{fig:Excl95delta}. This is not surprising given the simplicity of the analysis.



\bibliography{Custodial}

\end{document}